\documentclass[11pt]{article}

\renewcommand{\normalsize}{\fontsize{9}{11}\selectfont}
\AtBeginDocument{\normalsize}
\usepackage{newtxtext,newtxmath}
\usepackage{moreverb,url}
\usepackage{authblk} 
\usepackage[colorlinks,bookmarksopen,bookmarksnumbered,citecolor=red,urlcolor=red]{hyperref}
\usepackage{natbib}

\usepackage{array,epsfig,rotating}
\usepackage[]{hyperref}  
\usepackage{sectsty, secdot}
\usepackage{xcolor}
\usepackage{enumitem}

\newcommand\BibTeX{{\rmfamily B\kern-.05em \textsc{i\kern-.025em b}\kern-.08em
T\kern-.1667em\lower.7ex\hbox{E}\kern-.125emX}}

\usepackage{amsmath}
\usepackage{amssymb}
\usepackage{amsfonts}
\usepackage{multirow}
\usepackage{amsthm}

\def\BSigma{\boldsymbol{\Sigma}}
\def\Bnu{\boldsymbol{\nu}}
\def\Bell{\boldsymbol{\ell}}
\def\Bx{\boldsymbol{x}}
\def\BW{\boldsymbol{W}}

\providecommand{\keywords}[1]
{
  \small	
  \textbf{\textit{Keywords---}} #1
}

\newtheorem{theorem}{Theorem}
\newtheorem{lemma}{Lemma}

\theoremstyle{definition}
\newtheorem{definition}{Definition}

\usepackage{fancyhdr}
\newtheorem{condition}{Condition}

\definecolor{zhixin}{rgb}{0.6, 0.2, 0.8}

\DeclareMathOperator*{\argmin}{arg\,min}
\newtheorem{specification}{Specification}
\newtheorem{assumption}{Assumption}
\usepackage{booktabs}   
\usepackage{diagbox}    
\usepackage[flushleft]{threeparttable} 
\usepackage{multirow}
\newcommand{\topone}[1]{\textbf{#1}}
\newcommand{\toptwo}[1]{\underline{#1}}
\usepackage{pdflscape}
\usepackage{graphicx}
\usepackage{xr}
\usepackage{color}

\def\useLocalRef{0} 

\newcommand{\Sref}[2]{%
  \if\useLocalRef1%
    \ref{#1}
  \else
    #2
  \fi
}

\newcommand{\Seqref}[2]{%
  \if\useLocalRef1%
    \eqref{#1}
  \else
    (#2)
  \fi
}

\title{Robust Forest-based Estimation of Heterogeneous Treatment Effect Under High-dimensional Confounding}
\author[1]{Zhixin Qiu}
\author[1]{Xuanxuan Wang}
\author[2]{Xinyuan Song}
\author[3*]{Yinfeng Wang}
\author[1*]{Yanlin Tang}

\affil[1]{East China Normal University}
\affil[2]{The Chinese University of Hong Kong}
\affil[3]{Shanghai Lixin University of Accounting and Finance}

\date{}

\begin{document}
\fancyhead[L]









\maketitle

\begin{abstract}

    Understanding how patients with Alzheimer's disease respond differently to treatment is important for individualized therapeutic assessment. Using observational data from the Alzheimer's Disease Neuroimaging Initiative (ADNI), we investigate how the association between donepezil dosage and 12-month cognitive performance varies across diagnostic stages and hippocampal volumes. This analysis presents several statistical challenges, including potentially nonlinear treatment effect heterogeneity, high-dimensional confounding, and heavy-tailed outcomes with extreme observations. To address these challenges, we develop an Orthogonal Huber Random Forest (OHRF) framework that combines Huber-based robust estimation, Neyman orthogonality, and forest-based adaptive weighting. The proposed method accommodates complex treatment effect heterogeneity while reducing sensitivity to nuisance estimation errors and extreme outcomes. We establish its consistency, asymptotic normality, and finite-sample error bounds. Simulation studies demonstrate favorable estimation accuracy and robustness across a range of error distributions, particularly under heavy-tailed noise and outlier contamination. In the ADNI application, OHRF produces generally narrower pointwise confidence intervals than the conventional Orthogonal Random Forest (ORF) while revealing distinct patterns of treatment effect heterogeneity across diagnostic groups and hippocampal volumes. These findings illustrate the potential of robust, data-adaptive methods for studying heterogeneous treatment responses in complex observational biomedical data.

\end{abstract}


\section{Introduction}

In modern causal inference, estimating treatment effects is a central problem with broad practical relevance. Classical causal inference has primarily focused on average treatment effects, which summarize the population-level impact of a treatment \citep{robins2000marginal, imbens2015causal}. A motivating biomedical problem is how the effects of Donepezil dosage on cognitive outcomes vary across patients. We investigate this problem using data from the Alzheimer's Disease Neuroimaging Initiative (ADNI). In particular, we consider the 12-month
Alzheimer's Disease Assessment Scale-Cognitive Subscale (ADAS-13)
score as the outcome and examine treatment effect heterogeneity
according to baseline diagnostic status and hippocampal volume.
Understanding such heterogeneity may inform further research into
individualized treatment assessment. 
This application illustrates several challenges in estimating
heterogeneous treatment effects (HTEs). First, treatment effects may vary nonlinearly across diagnostic groups and hippocampal volumes. Second, the ADNI data contain a large number of baseline variables, including demographic
characteristics, cognitive assessments, biomarkers, and concurrent
medication use. Accounting for their potential influence on treatment and outcomes requires the estimation of potentially high-dimensional nuisance functions. Third, the empirical distribution of the outcome departs from normality and contains outlying
observations, motivating methods that are less sensitive to extreme
outcomes. 

A large literature studies heterogeneous treatment effects, with the goal of estimating how treated responses vary across individuals or subgroups, including machine learning and deep learning algorithms \citep{bica2020estimating, curth2021nonparametric}, Bayesian methods \citep{alaa2017bayesian,hahn2020bayesian} and representation-learning methods\citep{kazemi2024adversarially}. Among these, random forest \citep{wager2018estimation} has emerged as a dominant approach due to its adaptive nature, providing a powerful data-driven framework for estimating such complex conditional effects \citep{athey2019generalized,starling2021targeted}.

Although these approaches provide flexible tools for modeling
treatment effect heterogeneity, accurate HTE estimation in the
presence of high-dimensional confounders requires addressing the bias introduced by nuisance estimation errors. To mitigate the problem of error propagation, \cite{chernozhukov2018double} proposed the Double Machine Learning (DML) framework, which combines Neyman orthogonal scores with cross-fitting to remove the first-order impact of nuisance estimation error.
\cite{belloni2017program} studied efficient estimators and honest confidence bands for treatment-effect parameters in high dimensional settings, while \cite{oprescu2019orthogonal} developed the Orthogonal Random Forest (ORF) for estimating heterogeneous treatment effects.

Despite these advancements, robustness to heavy-tailed noise and outlier contamination remains a critical challenge in HTE estimation. While standard least-squares-based methods are effective and statistically optimal when the errors follow well-behaved  distributions, they can be sensitive to extreme outcome observations, potentially leading to unstable estimates in the presence of heavy-tailed errors or outlier contamination. Existing approaches often address these scenarios separately and with significant trade-offs. For outlier contamination, a common heuristic is to trim or remove outliers before applying standard least-squares methods; however, such an approach may discard informative data. For intrinsic heavy tails, quantile regression \citep{koenker2005quantile} is a principled robust method and has been extended to causal inference \citep{firpo2007efficient,frolich2013unconditional,giessing2023debiased,cheng2025inverting}. Quantile-based methods target conditional quantile effects,
which generally differ from conditional mean effects. their
relative efficiency also depends on the underlying error
distribution. Under suitable location-shift models, quantile
and mean treatment effects may coincide, but the corresponding
estimators can still exhibit different performance across
error distributions. Huber regression \citep{huber1973robust} offers a compelling alternative that provides a unified solution to both scenarios. 
The Huber loss retains the efficiency of squared loss for small residuals while improving robustness by downweighting large residuals from outliers or heavy-tailed noise.
Recently, \cite{fan2017estimation} studied high dimensional Huber regression, \cite{sun2020adaptive} and \cite{wang2021new} discussed the choice of the robustification parameter for achieving an optimal bias-variance trade-off, while \cite{han2022robust} and \cite{ma2025sparse} further investigated high-dimensional Huber inference.

In this paper, we propose a unified framework, the \textit{Orthogonal Huber Random Forest (OHRF)}. The main innovation is to construct Huber-based Neyman-orthogonal estimating equations within a random forest weighting scheme, so that treatment-effect heterogeneity, high-dimensional nuisance adjustment and robustness to nonstandard errors are handled in a unified estimation procedure. First, we propose a Huber-based orthogonal score for treatment effect estimation with high-dimensional confounders and heavy-tailed or outlier-contaminated outcomes. The orthogonalization reduces the impact of nuisance estimation errors, while the Huber robustification improves stability against extreme outcomes. Second, we incorporate the orthogonal Huber score into forest splitting procedure, allowing the forest to better capture treatment effect heterogeneity while maintaining robustness to outliers and high-dimensional nuisance estimation errors. In contrast to ORF, OHRF is inherently resilient to both heavy-tailed noise and outlier contamination. Simultaneously, unlike quantile-based alternatives, our method is specifically optimized for conditional mean estimation and maintains high efficiency in standard non-heavy-tailed scenarios. By dynamically calibrating the robustness parameter within this synergetic framework of DML and generalized random forest (GRF)\citep{athey2019generalized}, our estimator achieves consistency and asymptotic normality under refined moment conditions. This provides a comprehensive and reliable tool for HTE estimation in the presence of high-dimensional confounders and diverse error structures.

The practical advantages of OHRF are further illustrated through an analysis of observational data from the Alzheimer's Disease Neuroimaging Initiative (ADNI), which motivates our study. The data feature high-dimensional clinical and biological covariates and a markedly right-skewed outcome distribution with extreme observations, making robust estimation particularly relevant. In this application, OHRF identifies heterogeneous patterns in the estimated effects of donepezil dosage on 12-month cognitive performance across diagnostic stages and hippocampal volumes. Compared with the conventional Orthogonal Random Forest (ORF), OHRF yields broadly similar treatment-effect patterns but generally narrower pointwise confidence intervals, suggesting improved estimation stability in the presence of extreme outcomes. The estimates also reveal clinically relevant differences across patient subgroups: the estimated effects become more positive at higher hippocampal volumes in the early mild cognitive impairment (EMCI) group, remain close to zero in the late mild cognitive impairment (LMCI) group, and exhibit a nonmonotonic pattern in the Alzheimer's disease (AD) group. These findings demonstrate how robust estimation can provide more informative assessments of treatment effect heterogeneity in complex observational data, directly addressing the challenges of nonlinear heterogeneity, high-dimensional confounding, and irregular outcome distributions highlighted at the outset.

The remainder of the paper is organized as follows. 
Section \ref{sec:methods} presents the model setup and OHRF construction. Section \ref{sec:theory} presents the theoretical results. Section \ref{sec:simulation} assesses the finite-sample performance of the proposed method and Section \ref{sec:real_data} illustrates its practical utility with ADNI data. The implementation details, additional numerical results and all theoretical proofs are provided in the Supplementary Materials.

\section{Methods}\label{sec:methods}
\subsection{Model setup and motivation}
\label{sec:model_setup}
We consider independent and identically distributed observations\\ $\left\{ \boldsymbol{Z}_i = (Y_i, \boldsymbol{T}_i, \boldsymbol{X}_i, \boldsymbol{W}_i)\right\}_{i=1}^{n}$, generated from the following model:
\begin{align}
    Y_i              & =  \boldsymbol{\theta}_0(\boldsymbol{X}_i)^{\top} \boldsymbol{T}_i+ \boldsymbol{\beta}_0(\boldsymbol{X}_i)^{\top}\boldsymbol{W}_i + \epsilon_i, \quad i=1,2,\dots,n,\label{eq:y_main_eq} \\
    \boldsymbol{T}_i & = \boldsymbol{\ell}_0(\boldsymbol{X}_i)^{\top}\boldsymbol{W}_i + \boldsymbol{\eta}_i,  \label{eq:t_main_eq}
\end{align}
where $Y_i$ denotes the response variable, $\boldsymbol{T}_i\in\mathbb{R}^{d_{\boldsymbol{T}}}$ denotes the treatment variables, $\boldsymbol{W}_i\in\mathbb{R}^{d_{\boldsymbol{W}}}$  denotes the high-dimensional confounding variables that simultaneously influence the response $Y_i$ and treatment $\boldsymbol{T}_i$, and $\boldsymbol{\beta}_0 : \mathbb{R}^{d_{\boldsymbol X}} \to \mathbb{R}^{d_{\boldsymbol W}}$ 
and $\boldsymbol{\ell}_0 : \mathbb{R}^{d_{\boldsymbol X}} \to 
\mathbb{R}^{d_{\boldsymbol W} \times d_{\boldsymbol T}}$ are unknown coefficient functions. The coefficient vector $\boldsymbol{\theta}_0(\boldsymbol{x})$ is our primary object of interest, modeled as an unknown and non-linear function, characterizing the treatment effects that exhibit heterogeneity dependent on the covariates $\boldsymbol{X}_i\in\mathbb{R}^{d_{\boldsymbol{X}}}$. 
We allow the outcome noise $\epsilon_i$ to be heavy-tailed and contain outliers. Additional conditions on the outcome noise $\epsilon_i$ and the treatment residual $\boldsymbol{\eta}_i$ are specified in Assumption \ref{assumption:Non-degeneracy}.

This model captures the three main challenges in HTE estimation. Firstly, the functional form of $\boldsymbol{\theta}_0(\boldsymbol{X}_i)$ represents the complex, non-parametric patterns of heterogeneity we aim to capture. 
Secondly, the inclusion of high-dimensional confounders $\boldsymbol{W}_i$ motivates the use of debiased treatment effect estimation. Lastly, the noise term $\epsilon_i$ allows for heavy-tailed noise or outlier contamination, representing the complex error structures that our robust strategy is designed to handle.

The proposed OHRF is a multi-component framework designed to address the aforementioned challenges. (i) To accommodate heterogeneity, 
we employ a forest-based approach analogous to standard Random Forests 
to construct similarity weights $a_i(\boldsymbol{x})$, which quantify the relevance of each observation $\boldsymbol{X}_i$ to the target point $\boldsymbol{x}$. (ii) To mitigate the bias arising from high-dimensional nuisance functions in \eqref{eq:y_main_eq} and \eqref{eq:t_main_eq}, we utilize the principle of Neyman orthogonality. By constructing a score function whose first-order derivative with respect to the nuisance parameters is zero, we ensure that the resulting estimator for $\boldsymbol{\theta}_0(\boldsymbol{x})$ remains first-order insensitive to estimation errors in the nuisance components. (iii) Crucially, to achieve robustness without sacrificing efficiency, 
we embed an adaptive Huber loss within this orthogonal score framework. 
Unlike fixed-parameter Huber or quantile-based methods, 
the adaptive selection of the robustness parameter allows our method 
to effectively resist heavy-tailed noise while maintaining 
asymptotic unbiasedness and high efficiency 
under Gaussian error settings. The construction of the orthogonal Huber score is presented in Section~\ref{sec:Construction_score}.
A detailed description of each step is provided in Section \ref{sec:het_huber_neyman}. The similarity weights $a_i(\boldsymbol{x})$ are not known a priori, and we present the construction of OHRF and the resulting weights in Section \ref{sec:forest_weights}.

\subsection{Huber score function with Neyman orthogonality}\label{sec:Construction_score}

If we simultaneously estimate $\boldsymbol{\theta}_0(\boldsymbol{x})$ and $\boldsymbol{\beta}_0(\boldsymbol{x})$ by forest-based method based on model (\ref{eq:y_main_eq}) only, the estimation error of high-dimensional $\boldsymbol{\beta}_0(\boldsymbol{x})$ may propagate into the estimator of $\boldsymbol{\theta}_0(\boldsymbol{x})$ and slow down its convergence rate.
We address these limitations by constructing a Neyman orthogonal score function, which prevents nuisance estimation errors from affecting the leading term of the estimator. 

Our framework is designed to be robust to outliers and heavy-tailed noise by constructing the Neyman orthogonal score based on the Huber loss. Let $\boldsymbol{\nu}_0(\boldsymbol{x})=\boldsymbol{\ell}_0(\boldsymbol{x})\boldsymbol{\theta}_0(\boldsymbol{x})+\boldsymbol{\beta}_0(\boldsymbol{x})$, thus the conditional expectation of the outcome in models (\ref{eq:y_main_eq})-(\ref{eq:t_main_eq}) can be expressed as 
$$\mathbb{E}(Y_i \vert \boldsymbol{X}_i, \boldsymbol{W}_i)=\left\{\boldsymbol{\ell}_0(\boldsymbol{X}_i)\boldsymbol{\theta}_0(\boldsymbol{X}_i)\right\}^{\top}\boldsymbol{W}_i+\boldsymbol{\beta}_0(\boldsymbol{X}_i)^{\top}\boldsymbol{W}_i=\boldsymbol{\nu}_0(\boldsymbol{X}_i)^{\top} \boldsymbol{W}_i.$$ 
We construct the orthogonal Huber score function $\boldsymbol{\psi}_{\delta}$ as
\begin{equation}
\begin{split}
    \boldsymbol{\psi}_{\delta_{\boldsymbol{\theta}}}(\boldsymbol{Z}_i; \boldsymbol{\theta}, \boldsymbol{h}) = &\phi_{\delta_{\boldsymbol{\theta}}}\left[Y_i - \boldsymbol{\nu}(\boldsymbol{X}_i)^{\top}\boldsymbol{W}_i - \boldsymbol{\theta}(\boldsymbol{X}_i)^{\top} \left\{\boldsymbol{T}_i - \boldsymbol{\ell}(\boldsymbol{X}_i)^{\top}\boldsymbol{W}_i\right\} \right]\\
    & \cdot \left\{\boldsymbol{T}_i-\boldsymbol{\ell}(\boldsymbol{X}_i)^{\top}\boldsymbol{W}_i\right\}, 
\end{split}
    \label{eq:score_function}
\end{equation}
where $\boldsymbol{h}(\boldsymbol{X}_i) = 
\left(\boldsymbol{\nu}(\boldsymbol{X}_i), \boldsymbol{\ell}(\boldsymbol{X}_i)\right)$ is the vector of nuisance parameters, and
\[
    \phi_{\delta_{\boldsymbol{\theta}}}(u) =  \nabla_{u}\rho_{\delta_{\boldsymbol{\theta}}}(u) = uI(|u|\leq\delta_{\boldsymbol{\theta}})+\delta_{\boldsymbol{\theta}}\text{sign}(u)I(|u|>\delta_{\boldsymbol{\theta}}),
\]
is the derivative of Huber loss
\citep{huber1992robust},
    \[
        \rho_{\delta_{\boldsymbol{\theta}}}(u) = \frac{1}{2} u^{2}I\left(|u|\leq\delta_{\boldsymbol{\theta}}\right)+\delta_{\boldsymbol{\theta}}\left(|u|-\frac{1}{2}\delta_{\boldsymbol{\theta}}\right)I(|u|>\delta_{\boldsymbol{\theta}}).
    \]
Under some mild assumptions, 
$\mathbb{E}\left[\boldsymbol{\psi}_{\delta_{\boldsymbol{\theta}}}\left\{\boldsymbol{Z}_i; \boldsymbol{\theta}_0(\boldsymbol{x}),\boldsymbol{h}_0(\boldsymbol{x})\right\}\mid\boldsymbol{X}_i=\boldsymbol{x}\right]= \boldsymbol{0}.$
Therefore, $\boldsymbol{\theta}_0(\boldsymbol{x})$ can be identified through the orthogonal Huber score.

The score function $\boldsymbol{\psi}_{\delta_{\boldsymbol{\theta}}}(\boldsymbol{Z}; \boldsymbol{\theta}, \boldsymbol{h})$ satisfies the Neyman orthogonality condition with respect to the nuisance functions, meaning that the moment condition is locally insensitive to first-order perturbations of $\boldsymbol{\nu}$ and $\boldsymbol{\ell}$ at the truth. Specifically, 
\begin{equation}
    \nabla_{\boldsymbol{\nu}} \mathbb{E} \big[  \boldsymbol{\psi}_{\delta_{\boldsymbol{\theta}}}\left\{\boldsymbol{Z}_i; \boldsymbol{\theta}_0(\boldsymbol{x}), \left(\boldsymbol{\nu},\boldsymbol{\ell}_0(\boldsymbol{x})\right)\right\} \mid \boldsymbol{X}_{i} = \boldsymbol{x} \big] \big|_{\boldsymbol{\nu}=\boldsymbol{\nu}_0(\boldsymbol{x})} = \boldsymbol{0},
\label{eq:nabla_q}
\end{equation}
and
\begin{equation}
    \nabla_{\boldsymbol{\ell}} \mathbb{E} \big[  \boldsymbol{\psi}_{\delta_{\boldsymbol{\theta}}}\left\{\boldsymbol{Z}_i; \boldsymbol{\theta}_0(\boldsymbol{x}), \left(\boldsymbol{\nu}_0(\boldsymbol{x}),\boldsymbol{\ell}\right)\right\} \mid \boldsymbol{X}_{i} = \boldsymbol{x} \big] \big|_{\boldsymbol{\ell}=\boldsymbol{\ell}_0(\boldsymbol{x})} = \boldsymbol{0}.
\label{eq:nabla_g}
\end{equation}
Consequently, the treatment effect estimator has the same first-order asymptotic behavior as the oracle estimator, provided the nuisance estimators satisfy mild convergence rate conditions. Proofs of \eqref{eq:nabla_q} and \eqref{eq:nabla_g} are provided in  Section \Sref{appendix:proof_treatment}{S.4.1} of the Supplementary Material. This orthogonality property ensures that estimation errors in the nuisance components affect the estimation of the target treatment effect only at second order.

In practice, we estimate the treatment effects at a given target point $\boldsymbol{x}$ by solving the empirical version of the conditional mean $\mathbb{E}\{\boldsymbol{\psi}_{\delta_{\boldsymbol{\theta}}}(\boldsymbol{Z}_i; \boldsymbol{\theta}, \boldsymbol{h}) \vert \boldsymbol{X}_i = \boldsymbol{x}\}$. In particular,
\begin{equation}\label{ee}
    \sum_{i=1}^n a_i(\boldsymbol{x}) \boldsymbol{\psi}_{\delta_{\boldsymbol{\theta}}}\left\{\boldsymbol{Z}_i; \boldsymbol{\theta}, \boldsymbol{h}_0(\boldsymbol{x})\right\} = \boldsymbol{0},
\end{equation}
where $a_i(\boldsymbol{x})$ are weights that quantify the similarity between observation $\boldsymbol{X}_i$ and target covariate $\boldsymbol{x}$. 
\subsection{Robust HTE estimation with Neyman orthogonality}
\label{sec:het_huber_neyman}
To outline the implementation of our estimation procedure, 
we initially assume the similarity weights $a_i(\boldsymbol{x})$ are known.  
The overall procedure follows a two-stage strategy based on sample splitting, which ensures independence between the estimation of nuisance parameters and the target parameter. Specifically, we randomly partition the dataset $\mathcal{D}$ into two disjoint subsets $\mathcal{D}_1$ and $\mathcal{D}_2$. In the first stage, using data from $\mathcal{D}_1$ to estimate nuisance functions $\widehat{\boldsymbol{\ell}}(\boldsymbol{x})$ and $\widehat{\boldsymbol{\nu}}(\boldsymbol{x})$ by performing weighted Lasso regression of $\boldsymbol{T}$ on $\boldsymbol{W}$ and weighted adaptive Huber Lasso of $Y$ on $\boldsymbol{W}$, respectively. In the second stage, we estimate the treatment effects $\widehat{\boldsymbol{\theta}}(\boldsymbol{x})$ by solving a estimating equation using the data in $\mathcal{D}_2$. 
The corresponding estimating equation is given in \eqref{ee}, where the first-stage nuisance estimates are plugged in.

Let $\boldsymbol{\ell}^{(q)}(\boldsymbol{x})$ denote the $q$-th column of
$\boldsymbol{\ell}(\boldsymbol{x})$, $q=1,2,\dots, d_{\boldsymbol{T}}$, and assume that both $\boldsymbol{\nu}(\boldsymbol{x})$ and $\boldsymbol{\ell}^{(q)}(\boldsymbol{x})$ are sparse. The two-stage estimation procedure is summarized as follows. 

\textbf{Estimating the nuisance parameter.}
For $q=1,2,\dots, d_{\boldsymbol{T}}$, we estimate $q$-th column of $\boldsymbol{\ell}(\boldsymbol{x})$ from weighted Lasso estimator,

\begin{equation*} 
\widehat{\boldsymbol{\ell}}^{(q)}(\boldsymbol{x}) = \argmin_{\boldsymbol{\ell}^{(q)}(\boldsymbol{x})} \Biggl\{ \sum_{\{i \in \mathcal{D}_{1}\}} a_i(\boldsymbol{x}) \left\{\boldsymbol{T}_{i}^{(q)}-\boldsymbol{\ell}^{(q)}(\boldsymbol{x})^{\top}\boldsymbol{W}_i \right\}^2 +\lambda_{\ell} \left\|\boldsymbol{\ell}^{(q)}(\boldsymbol{x})\right\|_1 \Biggr\}. 
\end{equation*}

Then we estimate $\boldsymbol{\nu}(\boldsymbol{x})$ from weighted adaptive Huber Lasso,
\begin{equation*}  
    \widehat{\boldsymbol{\nu}}(\boldsymbol{x}) = \argmin_{\boldsymbol{\nu}(\boldsymbol{x})} \Biggl\{ \sum_{\{i \in \mathcal{D}_{1}\}}a_i(\boldsymbol{x}) \rho_{\delta_{\boldsymbol{\nu}}}\left\{Y_i-\boldsymbol{\nu}(\boldsymbol{x})^{\top}\boldsymbol{W}_i\right\}+\lambda_{\nu}\left\|\boldsymbol{\nu}(\boldsymbol{x})\right\|_1 \Biggr\}.
\end{equation*}
Here, $\lambda_{\boldsymbol{\ell}}$ and $\lambda_{\boldsymbol{\nu}}$ are tuning parameters that control the sparsity of the two nuisance estimators, respectively. 
The robustification parameter $\delta_{\boldsymbol{\nu}}$ is used for high-dimensional nuisance estimation \citep{sun2020adaptive} and is chosen separately from $\delta_{\boldsymbol{\theta}}$ in Subsection~\ref{sec:Construction_score}, which is used for estimating the low-dimensional treatment effect.  Details on how we choose these parameters are provided in Section S2.1 of the Supplementary Material.

\textbf{Estimating the treatment effects $\boldsymbol{\theta}(\boldsymbol{x})$.}
To obtain an estimation of $\boldsymbol{\theta}(\boldsymbol{x})$,
we solve the estimating equation~\eqref{ee} as introduced in Section~\ref{sec:Construction_score},
\begin{equation*}
    \sum_{\left\{i \in \mathcal{D}_2\right\}}a_i(\boldsymbol{x}) \boldsymbol{\psi}_{\delta_{\boldsymbol{\theta}}}\left\{\boldsymbol{Z}_i; \boldsymbol{\theta}, \widehat{\boldsymbol{h}}(\boldsymbol{x})\right\} = \boldsymbol{0}, 
\end{equation*}
where $\widehat{\boldsymbol{h}}(\boldsymbol{x})=\left(\widehat{\boldsymbol{\nu}}(\boldsymbol{x}),\widehat{\boldsymbol{\ell}}(\boldsymbol{x})\right)$.
Since the empirical moment condition may not admit an exact zero, we define $\widehat{\boldsymbol{\theta}}(\boldsymbol{x})$ as
\begin{equation*}
    \widehat{\boldsymbol{\theta}}(\boldsymbol{x}) \in \mathop{\arg\min}_{\boldsymbol{\theta} \in \boldsymbol{\Theta}} \bigg\| \sum_{\left\{i \in \mathcal{D}_2\right\}}a_i(\boldsymbol{x}) \boldsymbol{\psi}_{\delta_{\boldsymbol{\theta}}}\left\{\boldsymbol{Z}_i; \boldsymbol{\theta}, \widehat{\boldsymbol{h}}(\boldsymbol{x},\boldsymbol{W}_i)\right\} \bigg\|.
\end{equation*}

The core challenge in solving the above minimization problem stems from its inherently complex structure,
particularly due to the presence of the indicator function in
$\phi_{\delta_{\boldsymbol{\theta}}}(\cdot)$.
This indicator function makes the points of non-differentiability dependent on the unknown parameter $\boldsymbol{\theta}(\boldsymbol{x})$.
Consequently, deriving a closed-form solution for $\boldsymbol{\theta}(\boldsymbol{x})$ is generally not feasible.
Nevertheless, $\phi_{\delta_{\boldsymbol{\theta}}}(\cdot)$ is continuous and piecewise differentiable, making stochastic optimization methods, particularly Stochastic Gradient Descent (SGD), highly suitable for solving this problem. We therefore employ Adam, an adaptive variant of SGD, to obtain a numerical solution.

\subsection{Forest-based similarity weights}
\label{sec:forest_weights}

In practice, the similarity weights are unknown. In this paper, the weights $a_i(\boldsymbol{x})$ are derived from a random forest 
structure following \cite{wager2018estimation}, representing the proximity between the feature vector $\boldsymbol{X}_i$ 
and the target point $\boldsymbol{x}$. For a forest with $B$ trees, 
let $L_b(\boldsymbol{x})$ denote the leaf node containing $\boldsymbol{x}$ in the $b$-th tree. 
The tree-level weight is defined as:
\[
    a_{ib}(\boldsymbol{x}) = \frac{I\{\boldsymbol{X}_i \in L_b(\boldsymbol{x})\}}{|L_b(\boldsymbol{x})|}, 
\]
where $|L_b(\boldsymbol{x})|$ is the number of samples in the leaf. 
The aggregate similarity weight is then computed as the average across the forest, 
\[
    a_i(\boldsymbol{x}) = \frac{1}{B} \sum_{b=1}^B a_{ib}(\boldsymbol{x}).
\]

In this paper, each tree is grown by recursive partitioning in the spirit of GRF splitting algorithm\citep{athey2019generalized}. Within each parent node $P$, we first obtain the local estimators $\widehat{\boldsymbol{\theta}}_{P}$ and $\widehat{\boldsymbol{h}}_{P}$ by solving the Neyman orthogonal estimating equations, following a two-stage process similar to the one described in Section \ref{sec:het_huber_neyman}. The goal is then to select a split that maximizes treatment effect heterogeneity between the two child nodes. Directly recomputing the local treatment effect estimator for every candidate split is computationally expensive, therefore we adopt a GRF-style splitting rule that uses a one-step approximation from the parent-node estimate as a computationally efficient proxy for heterogeneity-based splitting. The splitting procedure is described below, where $n_{P}$ is the sample size in the parent node and $n_{C_j}$ is the sample size in child node $C_j$, $j=1,2$.

\noindent \textbf{Step 1: Estimate the nuisance parameter $\widehat{\boldsymbol{h}}_{P}$.}
The $q$-th $(q=1,2,\dots, d_{\boldsymbol{T}})$ column of $\widehat{\boldsymbol{\ell}}_{P}$ and $\widehat{\boldsymbol{\nu}}_{P}$ can be estimated by
\[
    \widehat{\boldsymbol{\ell}}^{(q)}_{P} = \argmin_{\boldsymbol{\ell}^{(q)}} \left\{ \frac{1}{n_{P}} \sum_{\{i \in P\}}\left\{\boldsymbol{T}_{i}^{(q)}-\boldsymbol{\ell}^{(q){\top}}\boldsymbol{W}_i \right\}^2+\lambda_{\ell}^{(0)} \left\|\boldsymbol{\ell}^{(q)}\right\|_1 \right\},
\]
\[
    \widehat{\boldsymbol{\nu}}_{P} = \argmin_{\boldsymbol{\nu}} \left\{ \frac{1}{n_{P}} \sum_{\{i \in P\}}\rho_{\delta^{(0)}_{\boldsymbol{\nu}}}\left(Y_i-\boldsymbol{\nu}^{\top}\boldsymbol{W}_i\right)+\lambda^{(0)}_{\nu}\left\|\boldsymbol{\nu}\right\|_1 \right\}.
\]
Here, $\lambda_{\boldsymbol{\ell}}^{(0)}$ and $\lambda_{\boldsymbol{\nu}}^{(0)}$ are used for the splitting step and may differ from $\lambda_{\boldsymbol{\ell}}$ and $\lambda_{\boldsymbol{\nu}}$, since the sample size $n_P$ is different.

\noindent\textbf{Step 2: Estimate $\widehat{\boldsymbol{\theta}}_{P}$.} The estimator in the parent node is obtained by
\[
    \widehat{\boldsymbol{\theta}}_{P} \in \argmin_{\boldsymbol{\theta}} \left\| \frac{1}{n_{P}} \sum_{\{i \in P\}} \boldsymbol{\psi}_{\delta_{\theta}}(\boldsymbol{Z}_i; \boldsymbol{\theta}, \widehat{\boldsymbol{h}}_{P}) \right\|.
\]

\noindent\textbf{Step 3: Approximate $\widehat{\boldsymbol{\theta}}_{C_1}$ and $\widehat{\boldsymbol{\theta}}_{C_2}$.}
For computational efficiency, we employ an approximation strategy for the child nodes $C_j$ ($j=1,2$). Denote the residual $\widetilde{Y}_i= Y_i - \widehat{\boldsymbol{\nu}}_{P}^{\top}\boldsymbol{W}_i$ and $\widetilde{\boldsymbol{T}_i}  = \boldsymbol{T}_i - \widehat{\boldsymbol{\ell}}_{P}^{\top}\boldsymbol{W}_i$. The proxy estimate $\widetilde{\boldsymbol{\theta}}_{C_j}$ for each child node is obtained by applying a single Newton step from the parent node's estimate $\widehat{\boldsymbol{\theta}}_{P}$. Specifically,
\[
    \widetilde{\boldsymbol{\theta}}_{C_j} = \widehat{\boldsymbol{\theta}}_{P} - \frac{1}{n_{C_j}} \boldsymbol{A}_{P}^{-1} \sum_{\{i \in C_j\}} \boldsymbol{\psi}_{\delta_{\boldsymbol{\theta}}}(\boldsymbol{Z}_i; \widehat{\boldsymbol{\theta}}_{P}, \widehat{\boldsymbol{h}}_{P}), \quad j=1,2, 
\]
where $\boldsymbol{A}_{P}$ is the empirical gradient of the score function with respect to $\boldsymbol{\theta}$, evaluated at the parent node, defined as
\[
    \boldsymbol{A}_{P} = \frac{1}{n_{P}} \sum_{\{i \in P\}} \nabla_{\boldsymbol{\theta}}\boldsymbol{\psi}_{\delta_{\boldsymbol{\theta}}}(\boldsymbol{Z}_i; \widehat{\boldsymbol{\theta}}_{P}, \widehat{\boldsymbol{h}}_{P}),
\]
in which the derivative of $\boldsymbol{\psi}_{\delta_{\boldsymbol{\theta}}}$ with respect to $\boldsymbol{\theta}$ is given by
\[
    \nabla_{\boldsymbol{\theta}}\boldsymbol{\psi}_{\delta_{\boldsymbol{\theta}}}(\boldsymbol{Z}_i; \widehat{\boldsymbol{\theta}}_{P}, \widehat{\boldsymbol{h}}_{P}) = -\widetilde{\boldsymbol{T}}_i\widetilde{\boldsymbol{T}}_i^{\top} I\left(\left| \widetilde{Y}_i - \widehat{\boldsymbol{\theta}}_{P}^{\top} \widetilde{\boldsymbol{T}}_i \right| \leq \delta_{\boldsymbol{\theta}} \right).
\]

\noindent\textbf{Step 4: Split $P$ via maximizing the heterogeneity score.}
We first define $\boldsymbol{\boldsymbol{V}}_i = \boldsymbol{A}_{P}^{-1} \boldsymbol{\psi}_{\delta_{\boldsymbol{\theta}}}(\boldsymbol{Z_i};\widehat{\boldsymbol{\theta}}_{P}, \widehat{\boldsymbol{h}}_{P})$.
For any candidate child nodes $C_1$ and $C_2$, define $\widetilde{\Xi}_q(C_1, C_2)$ as
\[
    \widetilde{\Xi}_q(C_1, C_2) = \sum_{j=1}^2 \frac{1}{n_{C_j}}\left(\sum_{\{i \in C_j\}} \boldsymbol{\boldsymbol{V}}_{i,q}\right)^2, \quad q=1,2,\dots, d_{\boldsymbol{T}},
\]
where $\boldsymbol{\boldsymbol{V}}_{i,q}$ is the $q$-th component of $\boldsymbol{A}_{P}^{-1} \boldsymbol{\psi}_{\delta_{\boldsymbol{\theta}}}(\boldsymbol{Z_i};\widehat{\boldsymbol{\theta}}_{P}, \widehat{\boldsymbol{h}}_{P})$.
Define the heterogeneity score $\Xi^{*}(C_1, C_2)$ as a convex combination of the maximum and the average of $\widetilde{\Xi}_q(C_1, C_2)$ over $q$. Specifically, 
\[
    \Xi^{*}(C_1, C_2) = u \max_{q} \widetilde{\Xi}_q(C_1, C_2) + (1-u) \frac{1}{d_{\boldsymbol{T}}} \sum_{q=1}^{d_{\boldsymbol{T}}} \widetilde{\Xi}_q(C_1, C_2), 
\]
where $u$ is drawn from $U(0, 1)$. Split parent node $P$ into two children $C_1$ and $C_2$ to maximize the heterogeneity score $\Xi^{*}(C_1, C_2)$ \citep{oprescu2019orthogonal}.


\section{Theoretical Properties}\label{sec:theory}

Our forest construction follows the framework of \cite{wager2018estimation}. Each tree is grown on a subsample $\mathcal{J}$ of size $s=s_n$, where $s_n/n\to0$ and $s_n\to\infty$ as $n\to\infty$. 
We impose standard regularity conditions on the tree construction, including honesty, balanced splitting, minimum leaf-size and random feature selection at each split; see detailed specification of the forest construction in Section S.1 of the Supplementary Material.


\label{sec:theory_treatment}
To rigorously analyze the statistical properties for $\widehat{\boldsymbol{\theta}}(\boldsymbol{x})$, specifically its consistency, convergence rates, and asymptotic normality,
we introduce the following foundational assumptions.

\begin{assumption}\label{assumption:TE_Lipschitz}
    The covariate vector $\boldsymbol{X}$ has non-vanishing density function across its support $[0,1]^{d_{\boldsymbol{X}}}$. The treatment effect $\boldsymbol{\theta}(\boldsymbol{x})$ is $ O(1)$-Lipschitz in $\boldsymbol{x}$. That is, there exists an absolute constant $L=O(1)$, $\left\|\boldsymbol{\theta}(\boldsymbol{x}_1)-\boldsymbol{\theta}(\boldsymbol{x}_2)\right\|\leq L\left\|\boldsymbol{x}_1-\boldsymbol{x}_2\right\|$ for all $\boldsymbol{x}_1,\boldsymbol{x}_2\in[0,1]^{d_{\boldsymbol{X}}}$.
\end{assumption}

\begin{assumption}\label{assumption:Non-degeneracy}

The outcome error $\epsilon_i$ follows a symmetric distribution given $\boldsymbol{X}_i, \boldsymbol{W}_i, \boldsymbol{T}_i$, while satisfying the conditional moment condition $$\mathbb{E}\left\{|\epsilon_i|^{1+\tau} \vert \boldsymbol{X}_i, \boldsymbol{W}_i, \boldsymbol{T}_i\right\}< \infty$$ with $\tau > 1/3$. Furthermore, the density function $f_{\epsilon}(u)$ is bounded and continuous in neighborhoods of $u=\pm\delta_{\boldsymbol{\theta}}$. 
For the treatment residual $\boldsymbol{\eta}_i$, we assume $\mathbb{E}(\boldsymbol{\eta}_i\mid \boldsymbol{X}_i,\boldsymbol{W}_i)=\boldsymbol{0}$ and $\boldsymbol{\eta}_i$ is conditionally independent with $\epsilon_i$  given $\boldsymbol{X}_i, \boldsymbol{W}_i$. The matrix $\mathbb{E}\left\{\boldsymbol{\eta}\boldsymbol{\eta}^{\top}I(\vert \epsilon \vert \leq \delta_{\boldsymbol{\theta}})\right\}$ is uniformly positive definite, with its minimum eigenvalue bounded below by a positive constant $\varpi^{\star}$. Moreover, there exists a constant $K_{\boldsymbol\eta}>0$ such that, almost surely,
$$\sup_{\substack{\boldsymbol u\in\mathbb R^{d_{\boldsymbol T}}\\ \|\boldsymbol u\|_2=1}}
\mathbb E\left[\left. \exp\left\{\frac{(\boldsymbol u^\top\boldsymbol\eta)^2}{K_{\boldsymbol\eta}^2}\right\}\right|\boldsymbol{X},\boldsymbol W\right]\leq 2.$$
\end{assumption}

\begin{assumption}\label{assumption:nuisance error}
Let $\boldsymbol{h}_0=\left(\boldsymbol{\nu}_0,\boldsymbol{\ell}_0\right)$ and $\widehat{\boldsymbol{h}}=\left(\widehat{\boldsymbol{\nu}},\widehat{\boldsymbol{\ell}}\right)$. We measure the estimation error of the nuisance functions by
 $$\mathcal{E}(\widehat{\boldsymbol{h}})=\sqrt{\mathbb{E}\left\{\left\|\widehat{\boldsymbol{\nu}}(\boldsymbol{x})-\boldsymbol{\nu}_0(\boldsymbol{x})\right\|^2+\left\|\widehat{\boldsymbol{\ell}}(\boldsymbol{x})-\boldsymbol{\ell}_0(\boldsymbol{x})\right\|^2\right\}}.$$
Assume the nuisance estimator $\widehat{\boldsymbol{h}}$ satisfies the following two conditions.
 \begin{enumerate}
        \item For some fixed integer $p\ge 1$,
        $\mathbb E\left\{\mathcal{E}(\widehat{\boldsymbol h})^{2p}\right\}^{1/p}=o\left(\sqrt{\frac{s}{n}}\right)$. 

        \item  For every $t\ge 0$, there exists a $\mathcal{R}_{n,s,t}$ satisfying $\mathbb{P}\left(\mathcal{E}(\widehat{\boldsymbol{h}})\leq\mathcal{R}_{n,s,t}\right)\geq 1-e^{-t}$ with $\mathcal{R}_{n,s,t}=O\left(\sqrt{\frac{k^2s\left(t+\log d_{\boldsymbol{W}}\right)}{n}}\right)=o\left(s^{1/4}/n^{1/4}\right)$.
\end{enumerate}
   
\end{assumption}

Assumption~\ref{assumption:TE_Lipschitz} specifies a commonly used smoothness condition on the treatment effect. Assumption~\ref{assumption:Non-degeneracy} specifies the conditional error structure required for the Huber moment restriction and Neyman orthogonality. It accommodates heavy-tailed outcomes by requiring only a finite $(1+\tau)$-th moment of the outcome error term. Following \cite{huber1973robust}, we assume that the conditional distribution of $\epsilon_i$ given $\boldsymbol{X}_i$, $\boldsymbol{W}_i$ and $\boldsymbol{T}_i$ is symmetric. The density condition provides regularity around the Huber thresholds. The conditional curvature condition prevents local degeneracy and provides local identification of the treatment effect. Assumption~\ref{assumption:nuisance error} is a high-level assumption, which quantifies the accuracy required from the first-stage nuisance estimators. We prove our proposed nuisance estimators satisfy Assumption~\ref{assumption:nuisance error} in Section S.3 of Supplementary Materials. The first part is is used to establish the asymptotic properties of the treatment effect estimator. The second part gives a high-probability bound for the nuisance error and is used when deriving finite-sample error bounds for the treatment effect estimator. We next establish the main theoretical properties of the proposed estimator, including consistency, $L_p$ error bound and asymptotic normality. 

\noindent \textbf{Remark:}
The conditional symmetry assumption on outcome error $\epsilon_i$ in Assumption~\ref{assumption:Non-degeneracy} allows us to work under relatively weak moment conditions on the outcome error. Existing studies can accommodate asymmetric outcome errors, but typically require stronger moment conditions; for example, \cite{ma2025sparse} establish their theoretical results under a finite third-moment condition on $\varepsilon_i$. In contrast,
our framework accommodates more severely heavy-tailed outcome errors, including distributions with infinite variance. Together with the other conditions stated in Assumption~\ref{assumption:Non-degeneracy}, conditional symmetry ensures the exact Neyman orthogonality of
the proposed Huber score.

\begin{theorem}(Asymptotic properties for $\widehat{\boldsymbol{\theta}}$)
\label{thm:Asymptotic properties}
Under Assumptions \ref{assumption:TE_Lipschitz}--\ref{assumption:nuisance error}, suppose that $B \gtrsim n/s$, $\widehat{\boldsymbol{\theta}}(\boldsymbol{x})$ satisfies the following properties.
\begin{enumerate}
        \item[(i)] \textbf{Consistency.} For any covariate $\boldsymbol{x}$, our proposed estimator is consistent for the true treatment effect, that is, 
        $\left\|\widehat{\boldsymbol{\theta}}(\boldsymbol{x})-\boldsymbol{\theta}_0(\boldsymbol{x})\right\| = o_p(1)$.

        \item[(ii)] \textbf{$L_p$ error bound.}  We obtain a bound on the $L_p$ norm of the estimation error. For constant integer $p\ge 1$ defined in Assumption~\ref{assumption:nuisance error},
           $\left[\mathbb{E}\left\{\| \widehat{\boldsymbol{\theta}}(\boldsymbol{x}) - \boldsymbol{\theta}_0(\boldsymbol{x}) \|^p\right\}\right]^{1/p}
        = O\left(
        s^{-\frac{1}{2\alpha d_{\boldsymbol{X}}}}
        + \sqrt{\frac{s\log(n/s)}{n}}
        + \sqrt{\frac{s}{n}}
        \right)$.

    \item[(iii)] \textbf{Asymptotic normality.} 
        Suppose in addition that the subsample size $s$ is chosen such that  $s^{-1/(2\alpha d_{\boldsymbol{X}})}=o\left((s/n)^{1/2-\varepsilon}\right)$ for any $\varepsilon>0$. We further define $\boldsymbol{M}_{\delta_{\boldsymbol{\theta}}} \equiv -\nabla_{\boldsymbol{\theta}} \mathbb{E}\left[\boldsymbol{\psi}_{\delta_{\boldsymbol{\theta}}}\left\{\boldsymbol{Z}_i; \boldsymbol{\theta}, \boldsymbol{h}_0(\boldsymbol{x})\right\}\mid \boldsymbol X=\boldsymbol x\right]\mid_{\boldsymbol{\theta}=\boldsymbol{\theta}_0(\boldsymbol{x})} = \mathbb{E}\left\{\boldsymbol{\eta} \boldsymbol{\eta}^{\top} I(\vert \epsilon \vert \leq \delta_{\boldsymbol{\theta}}) \mid \boldsymbol{X} = \boldsymbol{x}\right\}$.
        Then for any $\boldsymbol{b}\in\mathbb{R}^{d_{\boldsymbol{T}}}$ with $\|\boldsymbol{b}\|\le 1$,
        \[
        \mathrm{Var}\!\left[
        \boldsymbol{b}^{\top}\boldsymbol{M}_{\delta_{\boldsymbol{\theta}}}^{-1}
        \boldsymbol{\psi}_{\delta_{\boldsymbol{\theta}}}\left\{\boldsymbol{Z};\boldsymbol{\theta}_0,\boldsymbol{h}(\boldsymbol{X},\boldsymbol{W})\right\}
        \,\middle|\, \boldsymbol{X}=\boldsymbol{x}'
        \right]
        >0
        \]
        for all $\boldsymbol{x}'\in[0,1]^{d_{\boldsymbol{X}}}$.
        Then there exists a sequence $\sigma_n=O\!\left(\sqrt{\frac{\mathrm{polylog}(n/s)^{-1}s}{n}}\right)$,
        such that
        \begin{equation}\label{theorem:asymptotic_normality}
             \sigma_n^{-1}
        \left\langle\boldsymbol{b},
        \widehat{\boldsymbol{\theta}}(\boldsymbol{x})-\boldsymbol{\theta}_0(\boldsymbol{x})
        \right\rangle
        \overset{d}{\longrightarrow}
        \mathcal{N}(0,1),
        \end{equation}
        where $\mathrm{polylog}(n/s)$ is bounded away from $0$ and grows at most polynomially in $\log(n/s)$. 
    \end{enumerate}
\end{theorem}

Theorem~\ref{thm:Asymptotic properties} provides a complete asymptotic characterization of the proposed treatment effect estimator.
Part (i) guarantees that the estimated treatment effect converges to the true target parameter, establishing the basic reliability of the method. Part (ii) provides the $L_p$ error bound of our proposed estimator. 
Part (iii) further shows that the estimator admits an asymptotically Gaussian distribution after appropriate normalization. By using the Huber loss, we obtain these asymptotic results without assuming finite error variance, thereby extending the applicability of the method to a much broader class of heavy-tailed or outlier-contaminated distributions.

We next provide a non-asymptotic error bound for the proposed estimator by deriving a high-probability bound for $\left\|\widehat{\boldsymbol{\theta}} - \boldsymbol{\theta}_0 \right\|$.
\begin{theorem}(High Probability Error Bound)
    \label{theorem:high_probability_error_bound}
      Under Assumptions \ref{assumption:TE_Lipschitz}--\ref{assumption:nuisance error} and when the subsample size $s$ satisfies  $s^{-1/(2\alpha d_{\boldsymbol{X}})}=o\left((s/n)^{1/2-\varepsilon}\right)$, for any $t>0$, with probability at least $1-2e^{-t}$, we have:
    $$\left\| \widehat{\boldsymbol{\theta}} - \boldsymbol{\theta}_0 \right\| =\frac{O\left(s^{-\frac{1}{2\alpha d_{\boldsymbol{X}}}} + \sqrt{\frac{s\{\log(n/s)+t\}}{n}}+\mathcal{R}_{n,s,t}^2\right)}{\varpi^{\star}-O\left(\mathcal{R}_{n,s,t}\right)}.$$
\end{theorem}
\noindent Theorem \ref{theorem:high_probability_error_bound} decomposes the estimation error into three terms: the localization bias $s^{-1/(2\alpha d_{\boldsymbol X})}$, the empirical fluctuation term $\sqrt{s\left\{\log(n/s)+t\right\}/n}$ and the second-order contribution of nuisance estimation $\mathcal{R}_{n,s,t}^2$. The empirical fluctuation term refers to the randomness of the localized estimating equation around its population counterpart. Among the three terms, the empirical fluctuation term determines the leading rate. To see this, the localization bias is negligible under the condition on $s$, while the nuisance term is negligible by Assumption~\ref{assumption:nuisance error}. Therefore, the leading rate is of order $O\left(\sqrt{\frac{s\{\log(n/s)+t\}}{n}}\right)$.

\section{Simulation}

\label{sec:simulation}
\subsection{Baseline simulation study}\label{subsec:baseline simulation}

We focus on one-dimensional treatment variable $T_i$ here, and data are generated from the following model,
\begin{align}
    Y_i              & = T_i\theta_0(\boldsymbol{X}_i) + \boldsymbol{\beta}_0(\boldsymbol{X}_i)^{\top}\boldsymbol{W}_i  + \epsilon_i, \quad i=1,2,\dots,n, \\
    T_i & = \boldsymbol{\ell}_0(\boldsymbol{X}_i)^{\top}\boldsymbol{W}_i + \eta_i, 
\end{align}
where $\boldsymbol{W}_i \sim \mathcal{N}(\boldsymbol{0}, \boldsymbol{I}_{d_{\boldsymbol{W}}})$, 
$\eta_i \sim U(-1, 1)$. 
We set the sample size $n$ to be $5000$ and the confounder dimension $d_{\boldsymbol{W}}$ to be $500$. 
We vary the support size $k$ of $\boldsymbol{\beta}_0(\boldsymbol{x})$ and $\boldsymbol{\ell}_0(\boldsymbol{x})$ from $\{1, 5\}$.
We evaluate the performance of the estimated heterogeneous treatment effects.  

\paragraph{Setting 1: Two-dimensional effect modifiers $\boldsymbol{X}$.} 
In this setting, the modifier is two-dimensional, with $\boldsymbol{X}_i=(X_i^{(1)},X_i^{(2)})^\top$, with
$X_i^{(1)}\sim U(0,1)$ and
$X_i^{(2)}\sim\mathrm{Bernoulli}(0.5)$. 
The nuisance functions are defined as:
    $\boldsymbol{\beta}_0(x) = (\boldsymbol{1}^{\top}_{\lceil k/2 \rceil}, 
    -\boldsymbol{1}^{\top}_{\lfloor k/2 \rfloor}, \boldsymbol{0}^{\top})^{\top}$,
    $\boldsymbol{\ell}_0(x) = (\boldsymbol{0}^{\top}_k,\boldsymbol{1}^{\top}_{\lceil k/2 \rceil}, 
    -\boldsymbol{1}^{\top}_{\lfloor k/2 \rfloor}, \boldsymbol{0}^{\top})^{\top}$.
The heterogeneous treatment effect is
\[
    \theta_0(\boldsymbol{x})
    =
    \theta_1(x^{(1)})\mathbb{I}\{x^{(2)}=0\}
    +
    \theta_2(x^{(1)})\mathbb{I}\{x^{(2)}=1\},
\]
where
\[
    \theta_1(x)=
    \begin{cases}
        x+1,       & x\leq 0.3,\\
        8x-1.1,    & 0.3<x\leq 0.7,\\
        -4x+7.3,   & x>0.7,
    \end{cases}
\]
and
\[
    \theta_2(x)=
    \begin{cases}
        x^2+2.54,  & x\leq 0.4,\\
        3.25x+1.4, & 0.4<x\leq 0.8,\\
        -x^2+4.64, & x>0.8.
    \end{cases}
\]
This setting allows the treatment-effect function to vary
nonlinearly with the continuous modifier and to differ across the
two levels of the binary modifier.

We also consider two settings with a one-dimensional modifier
$X_i\sim U(0,1)$. In both settings, the true treatment effect is set as $\theta_1(x)$, corresponding to the treatment effect
in Setting 1 when $x^{(2)}=0$. Setting 2 uses nuisance coefficient
functions that are constant in $x$, whereas Setting 3 allows these
coefficients to vary with $x$. The detailed specifications and
results for these two settings are provided in the Supplementary
Material.

\paragraph{Error Distributions.}
To assess robustness to noise, we generate the error term $\epsilon_i$ from five different distributions: 
(i) uniform distribution $U(-1, 1)$,
(ii) normal distribution $\mathcal{N}(0, 1)$,
(iii) mixture of normals $\boldsymbol{z}\mathcal{N}(0, 1) + (1-\boldsymbol{z})\mathcal{N}(0, 5^2), \boldsymbol{z} \sim \mathcal{B}ernoulli(0.9)$,
(iv) $t$-distribution with 2.1 degrees of freedom and (v) $t$-distribution with 1.5 degrees of freedom.


We compare the proposed OHRF with several baseline methods. The competing methods are detailed as follows:
(1) \textbf{OQRF}: the Orthogonal Convolution-type Quantile Random Forest \citep{qiu2025heterogeneous}, which estimates heterogeneous quantile treatment effects using a quantile-based orthogonal score together with random forests.
(2) \textbf{ORF}: the Orthogonal Random Forest \citep{oprescu2019orthogonal}, which serves as a primary benchmark for orthogonalization-based forest methods without the specific robustification introduced in our approach.
(3) \textbf{GRF-Res-Lasso}: use cross-validated Lasso to residualize the treatment and outcome with respect to $\boldsymbol{X}$ and $\boldsymbol{W}$, then apply GRF to the residuals for conditional treatment effect estimation.
(4) \textbf{DML-Lasso}: the standard DML framework of \cite{chernozhukov2018double}, with a polynomial basis expansion for the heterogeneous treatment effect and cross-validated Lasso for nuisance estimation.
(5) \textbf{DML-RF}: a non-parametric variant of the DML algorithm, with random forests used for both nuisance estimation and conditional treatment effect estimation.

For each simulation setting, we conduct 100 Monte Carlo replications. 
The performance of each estimator is evaluated on a fixed test grid set defined as $\mathcal{X}_{\text{test}}=\{ (x_1,x_2) : x_1 \in \{ x^{(1)}, \dots, x^{(100)} \}, x_2 \in \{ 0,1 \} \}$, where $\left\lbrace x^{(1)},x^{(2)},\dots,x^{(100)}\right\rbrace$ are equally spaced in  $[0, 1]$. The test grid set contains 200 test points. We assess the estimation performance by computing the Integrated Mean Squared error (IMSE) as $\frac{1}{100} \sum_{r=1}^{100}\frac{1}{|\mathcal{X}_{\text{test}}|}\sum_{x_j\in \mathcal{X}_{\text{test}}} \left\{ \widehat{\theta}^{(r)}(x_j)-\theta(x_j) \right\}^{2}$.
The reported results are aggregated across $100$ replications to assess the overall performance and stability of each method. The results for Setting 1 are reported in Table~\ref{tab:simulation_results_setting3} and the results for the remaining simulation settings are provided in Section S.2.2 of the Supplementary Material. The bold and underlined values represent the best and the second-best results, respectively.

\begin{table}[t!]
    \centering
    \caption{Integrated Mean Square Error $(\times 10^{-2})$ of Setting 1}
    \label{tab:simulation_results_setting3}
    \scriptsize
    \setlength{\tabcolsep}{3.5pt}
    \renewcommand{\arraystretch}{0.7}
    \begin{tabular}{lllcccccc}
        \toprule
        & \multicolumn{2}{l}{\diagbox[width=2cm]{\tiny Error}{\tiny Methods}} & OHRF & OQRF & ORF & GRF-Res-Lasso & DML-Lasso & DML-RF \\
        \midrule
        
\multirow{5}{*}{$k=1$} & $U(-1,1)$ & & \topone{1.71}(0.04) & \toptwo{1.78}(0.05) & 3.31(0.10) & 8.07(0.15) & 4.33(0.03) & 4.78(0.04) \\
 & $\mathcal{N}(0,1)$ & & \topone{2.21}(0.05) & \toptwo{2.94}(0.09) & 4.16(0.11) & 11.21(0.17) & 4.54(0.03) & 5.11(0.07) \\
 & $Mix\text{-}\mathcal{N}$ & & \topone{3.17}(0.09) & \toptwo{3.40}(0.10) & 6.12(0.22) & 24.31(0.32) & 5.64(0.11) & 6.12(0.12) \\
 & $t_{2.1}$ & & \topone{3.98}(0.11) & \toptwo{4.30}(0.15) & 16.35(6.34) & 52.22(5.96) & 7.70(0.36) & 8.27(0.32) \\
 & $t_{1.5}$ & & \toptwo{14.68}(7.46) & \topone{5.04}(0.16) & 152.44(40.15) & 772.82(204.79) & 81.63(25.12) & 69.12(17.33) \\
\cmidrule{1-9}
\multirow{5}{*}{$k=5$} & $U(-1,1)$ & & \topone{2.07}(0.06) & \toptwo{2.53}(0.09) & 4.03(0.11) & 39.53(0.60) & 8.14(0.14) & 10.45(0.19) \\
 & $\mathcal{N}(0,1)$ & & \topone{2.63}(0.07) & \toptwo{3.46}(0.11) & 4.69(0.14) & 42.63(0.74) & 8.28(0.16) & 10.62(0.21) \\
 & $Mix\text{-}\mathcal{N}$ & & \topone{3.86}(0.10) & \toptwo{3.96}(0.13) & 6.42(0.22) & 53.99(0.93) & 9.43(0.25) & 10.87(0.27) \\
 & $t_{2.1}$ & & \topone{4.83}(0.15) & \toptwo{4.95}(0.17) & 13.08(2.21) & 81.68(6.57) & 11.74(0.50) & 10.99(0.27) \\
 & $t_{1.5}$ & & \toptwo{11.41}(2.85) & \topone{5.04}(0.17) & 146.27(43.07) & 872.75(252.95) & 95.20(29.44) & 22.22(3.78) \\
        
        \bottomrule
    \end{tabular}
        \begin{tablenotes}
            \footnotesize
            \item \textit{Note:} The standard errors the integrated squared errors are shown in parentheses. The bold and underlined values indicate the best and second-best results, respectively.
    \end{tablenotes}
\end{table}

\begin{figure}[htbp]
\centering
\includegraphics[width=\linewidth]{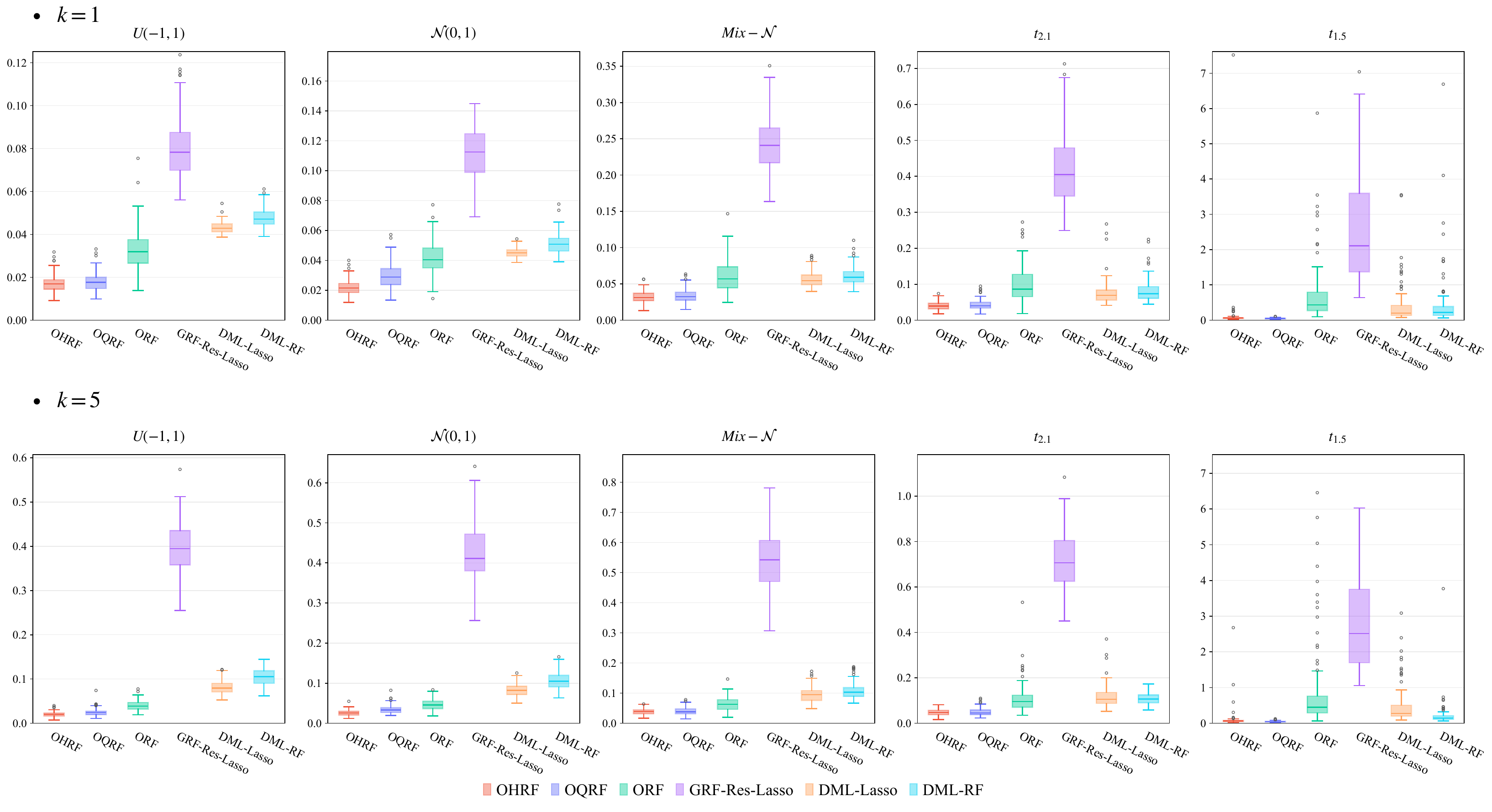}
\caption{Boxplots of the integrated squared errors for all methods across different error distributions in Setting 1.}
\label{fig:boxplot_setting3}
\end{figure}

To illustrate the stability and distributional properties of the estimators, we provide detailed boxplots of the integrated squared errors from the 100 Monte Carlo replications in Figure \ref{fig:boxplot_setting3}. The boxplot summarizes the distribution of the IMSE values obtained for a given 
method and error distribution. 

The empirical results demonstrate that the proposed Orthogonal Huber Random Forest (OHRF) achieves a superior balance between estimation efficiency and robustness. 
While ORF performs well under non-heavy-tailed distributions such as $U(-1,1)$ and $\mathcal{N}(0,1)$, 
its performance deteriorates rapidly as the error variance becomes larger ($t_{2.1}$ and $t_{1.5}$).
In contrast, OHRF maintains remarkable stability and consistently low estimation errors 
across all distribution types, effectively mitigating the influence of outliers where 
standard squared-loss methods fail. 
Furthermore, in non-heavy-tailed settings, including $U(-1,1)$, $\mathcal{N}(0,1)$, and $Mix\text{-}\mathcal{N}$, 
OHRF consistently outperforms OQRF, avoiding the efficiency loss often associated  with quantile-based approaches when data are relatively well-behaved. 
Although OQRF exhibits advantage in the most extreme heavy-tailed scenario ($t_{1.5}$)—an expected result 
given the theoretical invariance of quantiles to outliers—OHRF remains highly competitive and significantly more robust than non-robust baselines. In Settings 2 and 3, OHRF continues to achieve the best overall performance. Overall, this evidence suggests that OHRF is particularly well-suited for scenarios involving complex nuisance parameters and multi-dimensional heterogeneity.

\subsection{Simulation mimicking the real data setting}

To further evaluate the performance of the proposed method in a setting that aligns to our real data application, we consider an additional simulation study with a sample size of $n=1200$ and $75$ confounders. The effect modifiers $\boldsymbol{X}$ are two-dimensional, consisting of one continuous variable and one binary variable, as in Setting 1 described in Subsection~\ref{subsec:baseline simulation}. These choices are motivated by the sample size and covariate structure of our real data. Moreover, since the estimated outcome nuisance parameter in the real data analysis contains approximately $10$ nonzero coefficients, we consider two sparsity levels, $k=5$ and $k=10$. 

\begin{table}[t!]
    \centering
    \caption{Integrated Mean Squared Error under the Real-Data-Mimicking Simulation Setting}
    \label{tab:simulation_results_mimic}
    \scriptsize
    \setlength{\tabcolsep}{2pt}
    \renewcommand{\arraystretch}{0.9}
    \begin{tabular}{lllcccccc}
        \toprule
        & \multicolumn{2}{l}{\diagbox[width=2cm]{\tiny Error}{\tiny Methods}} & OHRF & OQRF & ORF & GRF-Res-Lasso & DML-Lasso & DML-RF  \\
        \midrule
        
\multirow{5}{*}{$k=5$} & $U(-1,1)$ & & \topone{0.087}(0.003) & 0.135(0.092) & \toptwo{0.113}(0.004) & 0.442(0.015) & 0.128(0.004) & 0.515(0.010)\\
 & $\mathcal{N}(0,1)$ & & \topone{0.106}(0.004) & 0.160(0.014) & \toptwo{0.129}(0.005)& 0.499(0.017) & 0.138(0.005) & 0.537(0.010) \\
 & $Mix\text{-}\mathcal{N}$ & & \topone{0.146}(0.005) & 0.228(0.210) & \toptwo{0.189}(0.007)& 0.610(0.017) & 0.201(0.009) & 0.659(0.013) \\
 & $t_{2.1}$ & & \topone{0.169}(0.006) & \toptwo{0.229}(0.042) & 0.311(0.042) & 1.135(0.261) & 0.348(0.070)& 0.920(0.042)  \\
 & $t_{1.5}$ & & \toptwo{0.434}(0.155) & \topone{0.241}(0.166)  & 3.402(1.499) & 5.406(1.535) & 2.831(1.275) & 3.308(0.943) \\
\cmidrule{1-9}
\multirow{5}{*}{$k=10$} & $U(-1,1)$ & & \toptwo{0.184}(0.006) & 0.214(0.010) & \topone{0.164}(0.006) & 0.986(0.033) & 0.248(0.009) & 0.280(0.005)\\
 & $\mathcal{N}(0,1)$ & & \toptwo{0.184}(0.006) & 0.231(0.013) & \topone{0.173}(0.006) & 1.022(0.029) & 0.263(0.011) & 0.299(0.005)  \\
 & $Mix\text{-}\mathcal{N}$ & & \topone{0.226}(0.009) & 0.259(0.016) & \toptwo{0.254}(0.010) & 1.146(0.033) & 0.300(0.014) & 0.342(0.007) \\
 & $t_{2.1}$ & & \topone{0.244}(0.008) & \toptwo{0.268}(0.014) & 0.366(0.021) & 1.620(0.230) & 0.480(0.079) & 0.403(0.012)  \\
 & $t_{1.5}$ & & \toptwo{0.568}(0.198)  & \topone{0.269}(0.233) & 4.625(2.100) & 5.253(1.043) & 2.561(0.886) & 1.873(0.591)  \\
        
        \bottomrule
    \end{tabular}
        \begin{tablenotes}
            \footnotesize
            \item \textit{Note:} The standard errors of the integrated squared errors are shown in parentheses. The bold and underlined values indicate the best and second-best results, respectively.
    \end{tablenotes}
\end{table}

\begin{figure}[htbp]
\centering
\includegraphics[width=\linewidth]{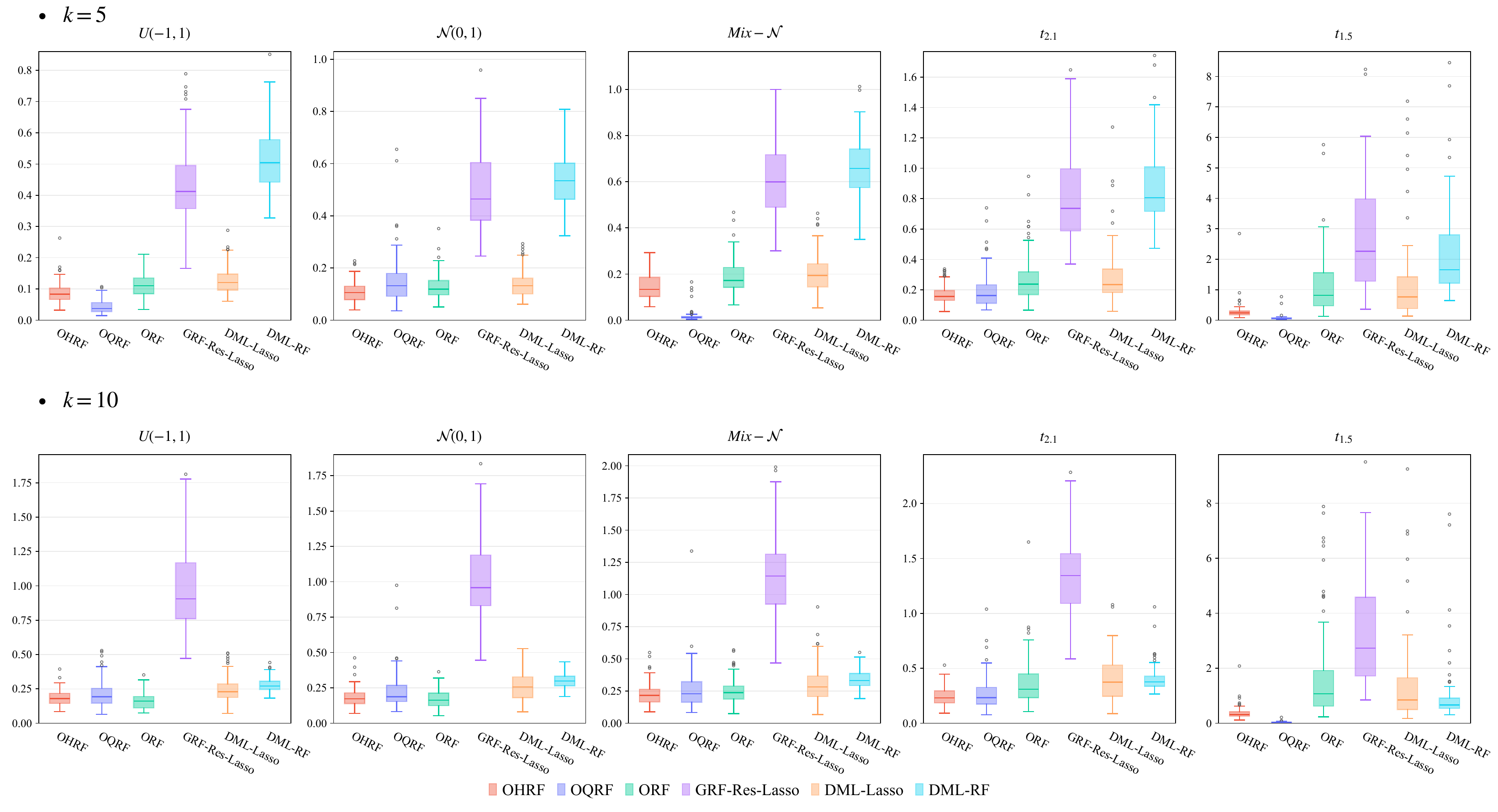}
\caption{Boxplots of the integrated squared errors for all methods across different error distributions under the Real-Data-Mimicking Simulation Setting.}
\label{fig:mimic}
\end{figure}

Table~\ref{tab:simulation_results_mimic} reports the results of the real-data-mimicking simulation. The results of the real-data-mimicking simulation are broadly consistent with those of Setting 1 reported in Subsection~\ref{subsec:baseline simulation}. 
Despite the smaller sample size, OHRF remains competitive across the error distributions and nuisance sparsity levels. Its advantage over ORF is particularly evident under heavy-tailed or contaminated errors, while ORF can be slightly more accurate under uniform and standard normal errors. As the nuisance structure becomes less sparse, the IMSE of OHRF increases, but the method remains competitive. In contrast, the adjusted OQRF results show relatively large Monte Carlo standard errors in several scenarios, reflecting substantial variation in replicate-level squared errors and sensitivity to occasional extreme estimates. Such variability may reflect the difficulty of estimating local quantile effects and nuisance functions from fewer informative observations, as well as the influence of smoothing and regularization choices. Under the most severe heavy-tailed errors, OQRF has the lowest reported IMSE; nevertheless, Monte Carlo variability is considerable for both OQRF and OHRF.

\section{A Real Data Example}

\label{sec:real_data}

\subsection{Background and clinical motivation}
Alzheimer's disease (AD) remains a leading cause of dementia worldwide, 
characterized by progressive cognitive decline and neurodegeneration. 
The reduction of Acetylcholine (ACh) synthesis is widely considered 
the most critical biochemical alteration underlying AD. 
This cholinergic deficit, coupled with the progressive degeneration of 
cholinergic neurons and the subsequent loss of effective neurotransmission, 
has been identified as a principal mechanism driving the cognitive impairments 
characteristic of the disease \citep{sharma2019cholinesterase}. 
In clinical practice and research, the Alzheimer's Disease Assessment Scale-Cognitive 
Subscale (ADAS-13) \citep{rosen1984new} serves as a canonical instrument for monitoring 
the cognitive trajectory of patients, where higher scores reflect 
greater levels of cognitive impairment. 
As a standard longitudinal measure, changes in ADAS-13 scores provide 
critical insights into disease progression and the relative efficacy 
of therapeutic interventions.

Among the existing therapeutic options, Donepezil is one of the few 
medications authorized by the FDA for the symptomatic management of 
mild to severe AD dementia \citep{adlimoghaddam2018review}. As a highly selective and reversible 
acetylcholinesterase inhibitor (AChEI), Donepezil has demonstrated 
pharmacological benefits in cognitive improvement and functional stabilization. 
However, while widely prescribed, its clinical efficacy exhibits 
substantial heterogeneity across the patient population. 
Evidence from prior randomized controlled trials \citep{doody2009donepezil, birks1996donepezil} suggests that 
Donepezil offers negligible cognitive benefits, or even induces 
adverse effects, in individuals at the prodromal stages of 
Mild Cognitive Impairment (MCI). 
Furthermore, since the therapeutic mechanism of Donepezil relies on 
the preservation of cholinergic neurons, structural neurodegeneration 
may severely limit its efficacy. Clinical studies \citep{roman2010randomized} indicate 
that hippocampal atrophy—a hallmark of neurodegeneration—is strongly 
associated with poor response to AChEIs, suggesting that 
drug benefits are conditional on brain structural integrity.

\subsection{Data description and experimental setup}
To investigate the heterogeneity in treatment response conditional on 
neuroanatomical and diagnostic markers, we analyzed a real-world dataset 
from the Alzheimer's Disease Neuroimaging Initiative (ADNI) database (adni.loni.usc.edu). The study cohort consists of 1,241 participants diagnosed with Early Mild Cognitive Impairment (EMCI), Late Mild Cognitive Impairment (LMCI), or Alzheimer's Disease (AD) at baseline. All participants underwent standardized cognitive assessments and MRI scans at both baseline and a 12-month follow-up.

Our primary objective is to evaluate the heterogeneous treatment effect of Donepezil dosage ($T$) on cognitive performance. 
The response variable ($Y$) is defined as the ADAS-13 score measured at the 12-month follow-up. The heterogeneity is modeled conditional on two primary modifiers ($\boldsymbol{X}$): 
the baseline diagnostic status and the standardized hippocampal volume, the latter calculated as $100 \times (\text{Hippocampal Volume} / \text{Intracranial Volume})$. To account for measured differences across participants, we
adjusted for $d_W=77$ baseline covariates ($\boldsymbol{W}$),
including sociodemographics (age, gender, education, APOE4 genotype), 
clinical and functional staging (CDR-SB, MMSE, MoCA), 
and an extensive neuropsychological battery (RAVLT, Trail Making Test). 
We further controlled for neuropathological biomarkers (CSF Abeta, Tau, Amyloid PET), 
additional regional brain volumes, technical covariates 
(MRI field strength), and the concurrent use of other AD-related medications. 
Crucially, the baseline ADAS-13 score was included in $\boldsymbol{W}$ to account for initial disease severity. Since treatment assignment in ADNI is observational, a causal interpretation of the estimated treatment effects requires additional assumptions including conditional exchangeability given the effect modifiers and observed confounders (i.e., no unmeasured confounding) and sufficient overlap across dosage levels. Under these assumptions and the model assumption in Subsection~\ref{sec:model_setup}, $\theta_0(\boldsymbol{x})$ represents the conditional mean treatment effect per unit increase in the recorded donepezil dosage. These assumptions cannot be fully verified from the observed data and the estimated treatment effects should therefore be interpreted with appropriate caution.

In addition, \cite{roman2010randomized}
reported differences in cognitive trajectories according to
hippocampal atrophy status in a trial of patients with vascular
dementia. More recently, an observational
ADNI study reported an association between cholinesterase inhibitor
use and faster cognitive decline in patients with MCI
\citep{liu2025impact}, suggesting that hippocampal structure may be a useful
characteristic to investigate in studies of treatment heterogeneity. When examining treatment heterogeneity across diagnostic stages, we account for differences in the observed distributions of standardized hippocampal volume. To ensure the reliability of our estimates and avoid extrapolation bias in regions with sparse data, 
we estimate the HTE within a localized range, 
specifically between the 0.1 and 0.9 quantiles of the 
standardized hippocampal volume observed within each diagnostic group. 
This adaptive grid ensures that the reported effects are supported by 
sufficient local observations in each subpopulation.

\subsection{Results and discussion}

\begin{figure*}
    \centering
    \includegraphics[width=1\textwidth]{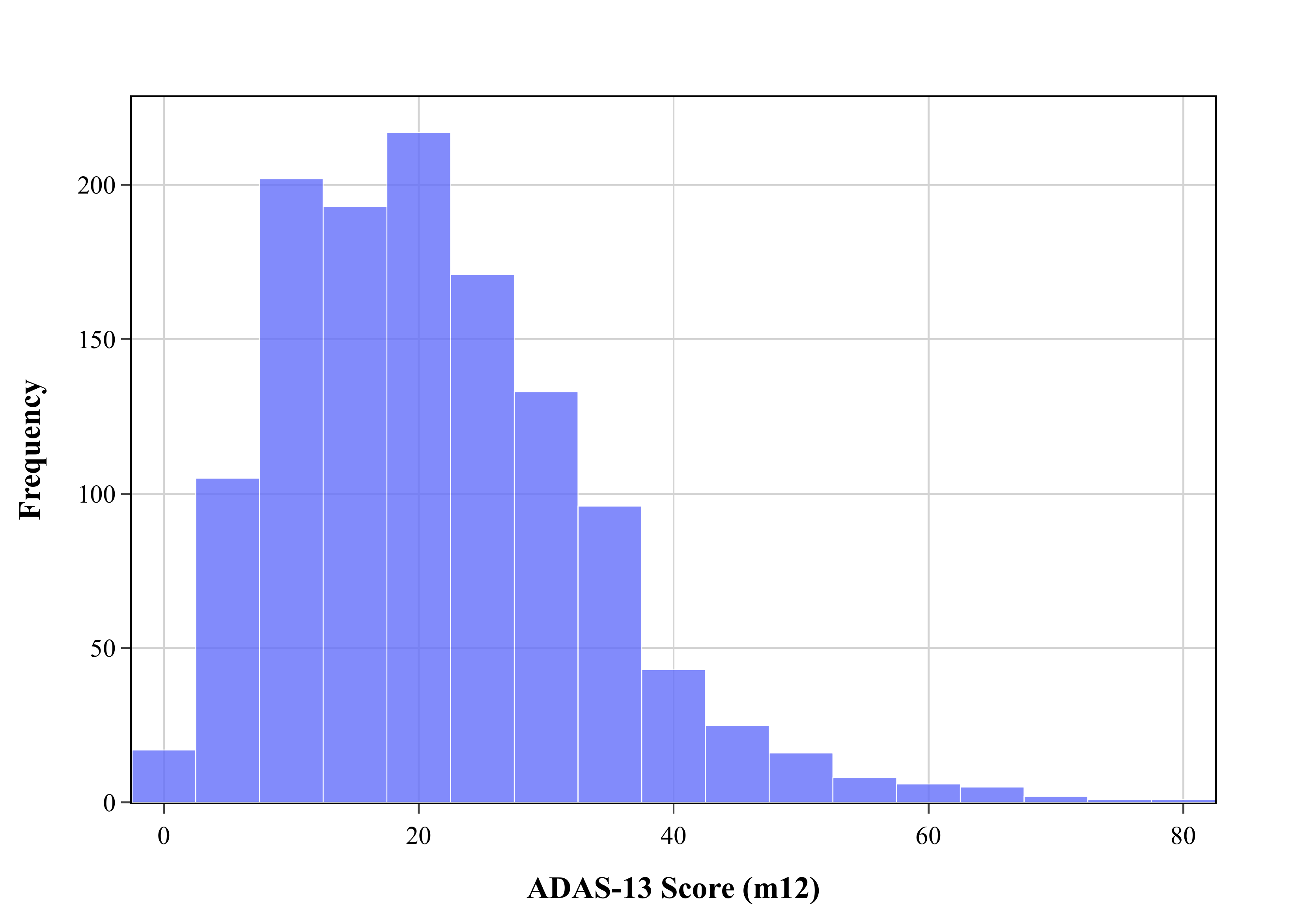}
    \caption{The distribution of ADAS-13 scores at month 12.}
    \label{fig:adas13_m12_histogram}
\end{figure*}
\begin{figure}
    \centering
    \makebox[\textwidth][c]{
        \includegraphics[width=1\textwidth]{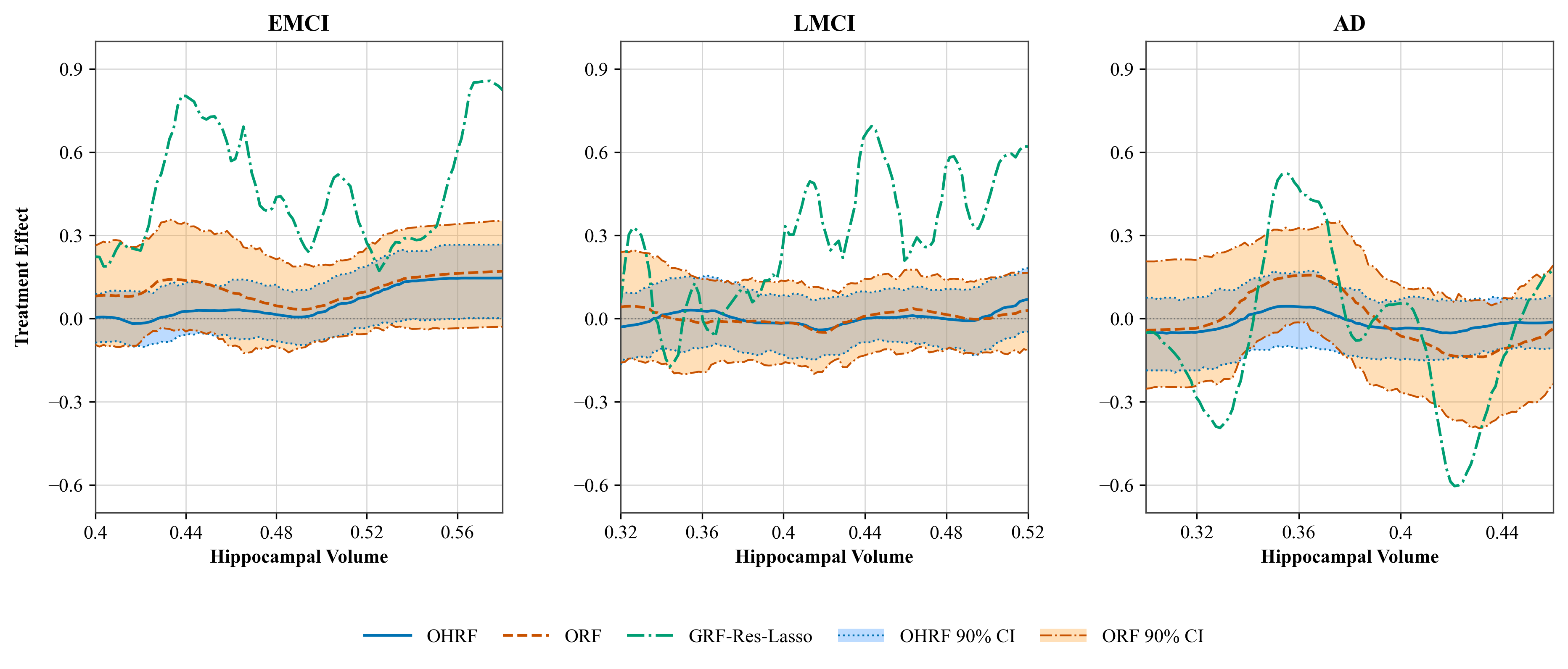}
    }
    \caption{Estimated treatment effects of donepezil dose on 12-month ADAS-13 scores conditional on baseline hippocampal volume, shown separately for different diagnostic status. The shaded region represents the 90\% pointwise bootstrap confidence intervals for OHRF and ORF.}
    \label{fig:bootstrap_treatment_effects}
\end{figure}

As illustrated in Figure \ref{fig:adas13_m12_histogram}, the distribution of the 
12-month ADAS-13 scores exhibits a markedly heavy-tailed characteristic. 
This statistical property, likely driven by the varying rates of 
cognitive deterioration in the AD population, motivates the application 
of our proposed Orthogonal Huber Random Forest (OHRF) to achieve 
robust estimation. The estimated treatment effects, along with 
90\% confidence intervals obtained via 100 bootstrap replicates, 
are presented in Figure \ref{fig:bootstrap_treatment_effects}.
Note that the x-axis in each panel represents the stage-specific 
inter-quantile range (from 0.1 to 0.9) of the standardized hippocampal volume, 
as previously described. 

OHRF and ORF obtain similar treatment-effect patterns across the three different diagnostic groups. Compared to ORF, OHRF yields generally narrower confidence intervals. These results suggest that OHRF provides more stable estimates in this application, highlighting the robustness of our proposed Huber-based approach. In contrast, GRF-Res-Lasso exhibits more pronounced fluctuations and wider bootstrap intervals, its confidence intervals are omitted from the figure for readability.

The estimated treatment effects suggest that the effect of
donepezil dosage on 12-month ADAS-13 scores may vary with
hippocampal volume across diagnostic groups. Specifically, in the EMCI group, the OHRF and ORF estimates become increasingly positive toward the upper end of the hippocampal-volume range. Since higher ADAS-13 scores indicate poorer cognition, the EMCI pattern raises the possibility of limited benefit from dose escalation in patients with relatively preserved hippocampal volume. In the LMCI group, both methods yield estimates
generally close to zero. In the AD group, the estimated curves are
nonmonotonic, with positive estimates at intermediate hippocampal volumes and modestly negative estimates over part of the upper range. These findings are supported by previous evidence regarding the clinical benefits of donepezil in MCI. \cite{petersen2005vitamin} found a lower rate of progression from MCI to AD during the first 12 months of donepezil treatment, whereas the difference was not sustained over the full three-year follow-up. Moreover, \cite{doody2009donepezil} found a small improvement in cognitive performance in a 48-week randomized trial, but no corresponding improvement in global function. This suggests that the clinical benefits of donepezil in MCI may depend on the outcome considered and provides context for the near-zero estimates observed in the LMCI group. Since donepezil is a cholinesterase inhibitor (ChEI), studies examining ChEIs can also provide relevant clinical conclusion. Using observational ADNI data, \cite{liu2025impact} and \cite{pyun2024cholinesterase} reported associations between ChEI use and faster cognitive decline in patients with MCI. These findings provide context for the positive estimates observed over part of the EMCI range.


\section{Discussion}
In this paper, we propose Orthogonal Huber Random Forest for estimating heterogeneous treatment effects in the presence of high-dimensional confounders. By integrating double machine learning algorithm with Huber loss and forest-based local weighting, our method achieves robustness against outliers and heavy-tailed noise while retaining high estimation efficiency when the noise distribution is close to Gaussian. This makes the proposed method especially appealing in practice, where the tail behavior of the noise is often unknown. On the theoretical side, we establish rigorous guarantees for the proposed estimator, including finite-sample error control and asymptotic normality.

This work can be extended or improved in several directions. First, causal inference with unobserved confounders remains a common yet challenging problem. It would be interest to investigate whether the proposed framework can be combined with recent methods designed to address hidden confounding \citep{cevid2020spectral,ouyang2023high}. Second, developing alternative strategies that accommodate stronger dependence structures is an important topic for future research. Finally, although statistical inference for forest-based estimators has been studied in prior work \citep{mentch2016quantifying}, extending these inference frameworks to the proposed orthogonal Huber forest setting is nontrivial. Developing valid inference procedures for our estimator remains an interesting direction for future research.


\section*{Supplementary Materials}
The online supplementary materials provide implementation details, additional simulation results and proofs of all theoretical results in this paper.


\section*{Acknowledgements}
 Generative AI tool ChatGPT (GPT-5) was occasionally used for language editing and proofreading.

\section*{Funding statement}
This work was partially supported by the National Key R$\&$D Program of China, No. 2023YFA1008700 and 2023YFA1008703; National Natural Science Foundation of China, No. 12371265; Natural Science Foundation of Shanghai, No. 24ZR1455200; the General Research Funds from Research Grant Council of the Hong Kong Special Administration Region, No. 14303622, 14300425.

\section*{Declaration of conflicting interest}
The author(s) declared no potential conflicts of interest with respect to the research, authorship, and/or publication of this article.

\section*{Ethical considerations}
Ethical approval was not required.

\section*{Consent to participate}
Not applicable.

\section*{Consent for publication}
Not applicable.

\section*{Data availability statement}
Data used in the preparation of this article were obtained from the Alzheimer's Disease Neuroimaging Initiative (ADNI) database (adni.loni.usc.edu). The ADNI was launched in 2003 as a public-private partnership, led by Principal Investigator Michael W. Weiner, MD. The original goal of ADNI was to test whether serial magnetic resonance imaging (MRI), positron emission tomography (PET), other biological markers, and clinical and neuropsychological assessment can be combined to measure the progression of mild cognitive impairment (MCI) and early Alzheimer's disease (AD). The current goals include validating biomarkers for clinical trials, improving the generalizability of ADNI data by increasing diversity in the participant cohort, and to provide data concerning the diagnosis and progression of Alzheimer's disease to the scientific community. For up-to-date information, see adni.loni.usc.edu.


\bibhang=1.7pc
\bibsep=2pt
\fontsize{9}{14pt plus.8pt minus .6pt}\selectfont
\renewcommand\bibname{\large \bf References}
\expandafter\ifx\csname
natexlab\endcsname\relax\def\natexlab#1{#1}\fi
\expandafter\ifx\csname url\endcsname\relax
  \def\url#1{\texttt{#1}}\fi
\expandafter\ifx\csname urlprefix\endcsname\relax\def\urlprefix{URL}\fi

\bibliographystyle{elsarticle-harv}      
\bibliography{references.bib}

\begin{thebibliography}{40}
\expandafter\ifx\csname natexlab\endcsname\relax\def\natexlab#1{#1}\fi
\providecommand{\url}[1]{\texttt{#1}}
\providecommand{\href}[2]{#2}
\providecommand{\path}[1]{#1}
\providecommand{\DOIprefix}{doi:}
\providecommand{\ArXivprefix}{arXiv:}
\providecommand{\URLprefix}{URL: }
\providecommand{\Pubmedprefix}{pmid:}
\providecommand{\doi}[1]{\href{http://dx.doi.org/#1}{\path{#1}}}
\providecommand{\Pubmed}[1]{\href{pmid:#1}{\path{#1}}}
\providecommand{\bibinfo}[2]{#2}
\ifx\xfnm\relax \def\xfnm[#1]{\unskip,\space#1}\fi
\bibitem[{Adlimoghaddam et~al.(2018)Adlimoghaddam, Neuendorff, Roy and
  Albensi}]{adlimoghaddam2018review}
\bibinfo{author}{Adlimoghaddam, A.}, \bibinfo{author}{Neuendorff, M.},
  \bibinfo{author}{Roy, B.}, \bibinfo{author}{Albensi, B.C.},
  \bibinfo{year}{2018}.
\newblock \bibinfo{title}{{A Review of Clinical Treatment Considerations of
  Donepezil in Severe Alzheimer's Disease}}.
\newblock \bibinfo{journal}{CNS neuroscience \& therapeutics}
  \bibinfo{volume}{24}, \bibinfo{pages}{876--888}.
\bibitem[{Alaa and Van Der~Schaar(2017)}]{alaa2017bayesian}
\bibinfo{author}{Alaa, A.M.}, \bibinfo{author}{Van Der~Schaar, M.},
  \bibinfo{year}{2017}.
\newblock \bibinfo{title}{Bayesian inference of individualized treatment
  effects using multi-task gaussian processes}, in:
  \bibinfo{booktitle}{Advances in Neural Information Processing Systems}, p.
  \bibinfo{pages}{3427–3435}.
\bibitem[{Athey et~al.(2019)Athey, Tibshirani and Wager}]{athey2019generalized}
\bibinfo{author}{Athey, S.}, \bibinfo{author}{Tibshirani, J.},
  \bibinfo{author}{Wager, S.}, \bibinfo{year}{2019}.
\newblock \bibinfo{title}{{Generalized Random Forests}}.
\newblock \bibinfo{journal}{The Annals of Statistics} \bibinfo{volume}{47},
  \bibinfo{pages}{1148--1178}.
\bibitem[{Belloni et~al.(2017)Belloni, Chernozhukov, Fern{\'a}ndez-Val and
  Hansen}]{belloni2017program}
\bibinfo{author}{Belloni, A.}, \bibinfo{author}{Chernozhukov, V.},
  \bibinfo{author}{Fern{\'a}ndez-Val, I.}, \bibinfo{author}{Hansen, C.},
  \bibinfo{year}{2017}.
\newblock \bibinfo{title}{Program evaluation and causal inference with
  high-dimensional data}.
\newblock \bibinfo{journal}{Econometrica} \bibinfo{volume}{85},
  \bibinfo{pages}{233--298}.
\bibitem[{Bica et~al.(2020)Bica, Jordon and van~der
  Schaar}]{bica2020estimating}
\bibinfo{author}{Bica, I.}, \bibinfo{author}{Jordon, J.},
  \bibinfo{author}{van~der Schaar, M.}, \bibinfo{year}{2020}.
\newblock \bibinfo{title}{{Estimating the Effects of Continuous-Valued
  Interventions Using Generative Adversarial Networks}}, in:
  \bibinfo{booktitle}{Advances in Neural Information Processing Systems}, pp.
  \bibinfo{pages}{16434--16445}.
\bibitem[{Bickel et~al.(2009)Bickel, Ritov and
  Tsybakov}]{bickel2009simultaneous}
\bibinfo{author}{Bickel, P.J.}, \bibinfo{author}{Ritov, Y.},
  \bibinfo{author}{Tsybakov, A.B.}, \bibinfo{year}{2009}.
\newblock \bibinfo{title}{{Simultaneous Analysis of Lasso and Dantzig
  Selector}}.
\newblock \bibinfo{journal}{Annals of Statistics} \bibinfo{volume}{37},
  \bibinfo{pages}{1705–1732}.
\bibitem[{Birks et~al.(2006)Birks, Flicker, Dementia and
  Group}]{birks1996donepezil}
\bibinfo{author}{Birks, J.}, \bibinfo{author}{Flicker, L.},
  \bibinfo{author}{Dementia, C.}, \bibinfo{author}{Group, C.I.},
  \bibinfo{year}{2006}.
\newblock \bibinfo{title}{{Donepezil for Mild Cognitive Impairment}}.
\newblock \bibinfo{journal}{Cochrane Database of Systematic Reviews}
  \bibinfo{volume}{2006}.
\bibitem[{{\'C}evid et~al.(2020){\'C}evid, B{\"u}hlmann and
  Meinshausen}]{cevid2020spectral}
\bibinfo{author}{{\'C}evid, D.}, \bibinfo{author}{B{\"u}hlmann, P.},
  \bibinfo{author}{Meinshausen, N.}, \bibinfo{year}{2020}.
\newblock \bibinfo{title}{{Spectral Deconfounding via Perturbed Sparse Linear
  Models}}.
\newblock \bibinfo{journal}{Journal of Machine Learning Research}
  \bibinfo{volume}{21}, \bibinfo{pages}{1--41}.
\bibitem[{Cheng and Li(2025)}]{cheng2025inverting}
\bibinfo{author}{Cheng, C.}, \bibinfo{author}{Li, F.}, \bibinfo{year}{2025}.
\newblock \bibinfo{title}{{Inverting Estimating Equations for Causal Inference
  on Quantiles}}.
\newblock \bibinfo{journal}{Biometrika} \bibinfo{volume}{112},
  \bibinfo{pages}{asae058}.
\bibitem[{Chernozhukov et~al.(2018)Chernozhukov, Chetverikov, Demirer, Duflo,
  Hansen, Newey et~al.}]{chernozhukov2018double}
\bibinfo{author}{Chernozhukov, V.}, \bibinfo{author}{Chetverikov, D.},
  \bibinfo{author}{Demirer, M.}, \bibinfo{author}{Duflo, E.},
  \bibinfo{author}{Hansen, C.}, \bibinfo{author}{Newey, W.}, et~al.,
  \bibinfo{year}{2018}.
\newblock \bibinfo{title}{{Double/Debiased Machine Learning for Treatment and
  Structural Parameters}}.
\newblock \bibinfo{journal}{The Econometrics Journal} \bibinfo{volume}{21},
  \bibinfo{pages}{1--68}.
\bibitem[{Curth and Van~der Schaar(2021)}]{curth2021nonparametric}
\bibinfo{author}{Curth, A.}, \bibinfo{author}{Van~der Schaar, M.},
  \bibinfo{year}{2021}.
\newblock \bibinfo{title}{{Nonparametric Estimation of Heterogeneous Treatment
  Effects: From Theory to Learning Algorithms}}, in:
  \bibinfo{booktitle}{International Conference on Artificial Intelligence and
  Statistics}, \bibinfo{organization}{PMLR}. pp. \bibinfo{pages}{1810--1818}.
\bibitem[{Doody et~al.(2009)Doody, Ferris, Salloway, Sun, Goldman, Watkins
  et~al.}]{doody2009donepezil}
\bibinfo{author}{Doody, R.}, \bibinfo{author}{Ferris, S.},
  \bibinfo{author}{Salloway, S.}, \bibinfo{author}{Sun, Y.},
  \bibinfo{author}{Goldman, R.}, \bibinfo{author}{Watkins, W.}, et~al.,
  \bibinfo{year}{2009}.
\newblock \bibinfo{title}{{Donepezil Treatment of Patients with MCI: A 48-Week
  Randomized, Placebo-Controlled Trial}}.
\newblock \bibinfo{journal}{Neurology} \bibinfo{volume}{72},
  \bibinfo{pages}{1555--1561}.
\bibitem[{Fan et~al.(2017)Fan, Li and Wang}]{fan2017estimation}
\bibinfo{author}{Fan, J.}, \bibinfo{author}{Li, Q.}, \bibinfo{author}{Wang,
  Y.}, \bibinfo{year}{2017}.
\newblock \bibinfo{title}{{Estimation of High Dimensional Mean Regression in
  the Absence of Symmetry and Light Tail Assumptions}}.
\newblock \bibinfo{journal}{Journal of the Royal Statistical Society Series B:
  Statistical Methodology} \bibinfo{volume}{79}, \bibinfo{pages}{247--265}.
\bibitem[{Firpo(2007)}]{firpo2007efficient}
\bibinfo{author}{Firpo, S.}, \bibinfo{year}{2007}.
\newblock \bibinfo{title}{{Efficient Semiparametric Estimation of Quantile
  Treatment Effects}}.
\newblock \bibinfo{journal}{Econometrica} \bibinfo{volume}{75},
  \bibinfo{pages}{259--276}.
\bibitem[{Fr{\"o}lich and Melly(2013)}]{frolich2013unconditional}
\bibinfo{author}{Fr{\"o}lich, M.}, \bibinfo{author}{Melly, B.},
  \bibinfo{year}{2013}.
\newblock \bibinfo{title}{{Unconditional Quantile Treatment Effects Under
  Endogeneity}}.
\newblock \bibinfo{journal}{Journal of business \& economic statistics}
  \bibinfo{volume}{31}, \bibinfo{pages}{346--357}.
\bibitem[{Giessing and Wang(2023)}]{giessing2023debiased}
\bibinfo{author}{Giessing, A.}, \bibinfo{author}{Wang, J.},
  \bibinfo{year}{2023}.
\newblock \bibinfo{title}{{Debiased Inference on Heterogeneous Quantile
  Treatment Effects with Regression Rank Scores}}.
\newblock \bibinfo{journal}{Journal of the Royal Statistical Society Series B:
  Statistical Methodology} \bibinfo{volume}{85}, \bibinfo{pages}{1561--1588}.
\bibitem[{Hahn et~al.(2020)Hahn, Murray and Carvalho}]{hahn2020bayesian}
\bibinfo{author}{Hahn, P.R.}, \bibinfo{author}{Murray, J.S.},
  \bibinfo{author}{Carvalho, C.M.}, \bibinfo{year}{2020}.
\newblock \bibinfo{title}{{Bayesian Regression Tree Models for Causal
  Inference: Regularization, Confounding, and Heterogeneous Effects (With
  Discussion)}}.
\newblock \bibinfo{journal}{Bayesian Analysis} \bibinfo{volume}{15},
  \bibinfo{pages}{965--1056}.
\bibitem[{Han et~al.(2022)Han, Huang, Lin and Shen}]{han2022robust}
\bibinfo{author}{Han, D.}, \bibinfo{author}{Huang, J.}, \bibinfo{author}{Lin,
  Y.}, \bibinfo{author}{Shen, G.}, \bibinfo{year}{2022}.
\newblock \bibinfo{title}{Robust post-selection inference of high-dimensional
  mean regression with heavy-tailed asymmetric or heteroskedastic errors}.
\newblock \bibinfo{journal}{Journal of econometrics} \bibinfo{volume}{230},
  \bibinfo{pages}{416--431}.
\bibitem[{Huber(1973)}]{huber1973robust}
\bibinfo{author}{Huber, P.J.}, \bibinfo{year}{1973}.
\newblock \bibinfo{title}{{Robust Regression: Asymptotics, Conjectures and
  Monte Carlo}}.
\newblock \bibinfo{journal}{The annals of statistics} \bibinfo{volume}{1},
  \bibinfo{pages}{799--821}.
\bibitem[{Huber(1992)}]{huber1992robust}
\bibinfo{author}{Huber, P.J.}, \bibinfo{year}{1992}.
\newblock \bibinfo{title}{{Robust Estimation of a Location Parameter}}, in:
  \bibinfo{booktitle}{Breakthroughs in statistics: Methodology and
  distribution}. \bibinfo{publisher}{Springer}, pp. \bibinfo{pages}{492--518}.
\bibitem[{Imbens and Rubin(2015)}]{imbens2015causal}
\bibinfo{author}{Imbens, G.W.}, \bibinfo{author}{Rubin, D.B.},
  \bibinfo{year}{2015}.
\newblock \bibinfo{title}{Causal inference in statistics, social, and
  biomedical sciences}.
\newblock \bibinfo{publisher}{Cambridge university press}.
\bibitem[{Kazemi and Ester(2024)}]{kazemi2024adversarially}
\bibinfo{author}{Kazemi, A.}, \bibinfo{author}{Ester, M.},
  \bibinfo{year}{2024}.
\newblock \bibinfo{title}{Adversarially balanced representation for continuous
  treatment effect estimation}, in: \bibinfo{booktitle}{Proceedings of the AAAI
  Conference on Artificial Intelligence}, pp. \bibinfo{pages}{13085--13093}.
\bibitem[{Koenker(2005)}]{koenker2005quantile}
\bibinfo{author}{Koenker, R.}, \bibinfo{year}{2005}.
\newblock \bibinfo{title}{{Quantile Regression}}. volume~\bibinfo{volume}{38}.
\newblock \bibinfo{publisher}{Cambridge university press}.
\bibitem[{Liu et~al.(2025)Liu, Li, Qin, Wang, Li, Li, Initiative and
  Jia}]{liu2025impact}
\bibinfo{author}{Liu, W.}, \bibinfo{author}{Li, Y.}, \bibinfo{author}{Qin, W.},
  \bibinfo{author}{Wang, X.}, \bibinfo{author}{Li, W.}, \bibinfo{author}{Li,
  Y.}, \bibinfo{author}{Initiative, A.D.N.}, \bibinfo{author}{Jia, J.},
  \bibinfo{year}{2025}.
\newblock \bibinfo{title}{The impact of cholinesterase inhibitors on cognitive
  trajectories in mild cognitive impairment patients based on amyloid beta
  status}.
\newblock \bibinfo{journal}{Alzheimer's \& Dementia} \bibinfo{volume}{21},
  \bibinfo{pages}{e70193}.
\bibitem[{Ma et~al.(2025)Ma, Gao, Wang and Lian}]{ma2025sparse}
\bibinfo{author}{Ma, W.}, \bibinfo{author}{Gao, J.}, \bibinfo{author}{Wang,
  L.}, \bibinfo{author}{Lian, H.}, \bibinfo{year}{2025}.
\newblock \bibinfo{title}{Sparse and debiased adaptive huber regression in
  distributed data: aggregated and communication-efficient approaches}.
\newblock \bibinfo{journal}{Statistica Sinica} \bibinfo{volume}{35},
  \bibinfo{pages}{1--20}.
\bibitem[{Meinshausen and Yu(2009)}]{meinshausen2009lasso}
\bibinfo{author}{Meinshausen, N.}, \bibinfo{author}{Yu, B.},
  \bibinfo{year}{2009}.
\newblock \bibinfo{title}{{Lasso-Type Recovery of Sparse Representations for
  High-Dimensional Data}}.
\newblock \bibinfo{journal}{Annals of Statistics} \bibinfo{volume}{37},
  \bibinfo{pages}{246–270}.
\bibitem[{Mentch and Hooker(2016)}]{mentch2016quantifying}
\bibinfo{author}{Mentch, L.}, \bibinfo{author}{Hooker, G.},
  \bibinfo{year}{2016}.
\newblock \bibinfo{title}{{Quantifying Uncertainty in Random Forests via
  Confidence Intervals and Hypothesis Tests}}.
\newblock \bibinfo{journal}{Journal of Machine Learning Research}
  \bibinfo{volume}{17}, \bibinfo{pages}{1--41}.
\bibitem[{Oprescu et~al.(2019)Oprescu, Syrgkanis and
  Wu}]{oprescu2019orthogonal}
\bibinfo{author}{Oprescu, M.}, \bibinfo{author}{Syrgkanis, V.},
  \bibinfo{author}{Wu, Z.S.}, \bibinfo{year}{2019}.
\newblock \bibinfo{title}{{Orthogonal Random Forest for Causal Inference}}, in:
  \bibinfo{booktitle}{International Conference on Machine Learning},
  \bibinfo{organization}{PMLR}. pp. \bibinfo{pages}{4932--4941}.
\bibitem[{Ouyang et~al.(2023)Ouyang, Tan and Xu}]{ouyang2023high}
\bibinfo{author}{Ouyang, J.}, \bibinfo{author}{Tan, K.M.}, \bibinfo{author}{Xu,
  G.}, \bibinfo{year}{2023}.
\newblock \bibinfo{title}{{High-Dimensional Inference for Generalized Linear
  Models with Hidden Confounding}}.
\newblock \bibinfo{journal}{Journal of Machine Learning Research}
  \bibinfo{volume}{24}, \bibinfo{pages}{1--61}.
\bibitem[{Petersen et~al.(2005)Petersen, Thomas, Grundman, Bennett, Doody,
  Ferris, Galasko, Jin, Kaye, Levey et~al.}]{petersen2005vitamin}
\bibinfo{author}{Petersen, R.C.}, \bibinfo{author}{Thomas, R.G.},
  \bibinfo{author}{Grundman, M.}, \bibinfo{author}{Bennett, D.},
  \bibinfo{author}{Doody, R.}, \bibinfo{author}{Ferris, S.},
  \bibinfo{author}{Galasko, D.}, \bibinfo{author}{Jin, S.},
  \bibinfo{author}{Kaye, J.}, \bibinfo{author}{Levey, A.}, et~al.,
  \bibinfo{year}{2005}.
\newblock \bibinfo{title}{Vitamin e and donepezil for the treatment of mild
  cognitive impairment}.
\newblock \bibinfo{journal}{New England Journal of Medicine}
  \bibinfo{volume}{352}, \bibinfo{pages}{2379--2388}.
\bibitem[{Pyun et~al.(2024)Pyun, Park, Kang and Kim}]{pyun2024cholinesterase}
\bibinfo{author}{Pyun, J.M.}, \bibinfo{author}{Park, Y.H.},
  \bibinfo{author}{Kang, M.J.}, \bibinfo{author}{Kim, S.},
  \bibinfo{year}{2024}.
\newblock \bibinfo{title}{Cholinesterase inhibitor use in amyloid pet-negative
  mild cognitive impairment and cognitive changes}.
\newblock \bibinfo{journal}{Alzheimer's Research \& Therapy}
  \bibinfo{volume}{16}, \bibinfo{pages}{210}.
\bibitem[{Qiu et~al.(2025)Qiu, Zhu, Wang and Tang}]{qiu2025heterogeneous}
\bibinfo{author}{Qiu, Z.}, \bibinfo{author}{Zhu, H.}, \bibinfo{author}{Wang,
  W.}, \bibinfo{author}{Tang, Y.}, \bibinfo{year}{2025}.
\newblock \bibinfo{title}{Heterogeneous quantile treatment effect estimation
  for longitudinal data with high-dimensional confounding}.
\newblock \bibinfo{journal}{arXiv preprint arXiv:2508.16326} .
\bibitem[{Robins et~al.(2000)Robins, Hernan and Brumback}]{robins2000marginal}
\bibinfo{author}{Robins, J.M.}, \bibinfo{author}{Hernan, M.A.},
  \bibinfo{author}{Brumback, B.}, \bibinfo{year}{2000}.
\newblock \bibinfo{title}{Marginal structural models and causal inference in
  epidemiology}.
\newblock \bibinfo{journal}{Epidemiology} \bibinfo{volume}{11},
  \bibinfo{pages}{550--560}.
\bibitem[{Rom{\'a}n et~al.(2010)Rom{\'a}n, Salloway, Black, Royall, DeCarli,
  Weiner et~al.}]{roman2010randomized}
\bibinfo{author}{Rom{\'a}n, G.C.}, \bibinfo{author}{Salloway, S.},
  \bibinfo{author}{Black, S.E.}, \bibinfo{author}{Royall, D.R.},
  \bibinfo{author}{DeCarli, C.}, \bibinfo{author}{Weiner, M.W.}, et~al.,
  \bibinfo{year}{2010}.
\newblock \bibinfo{title}{{Randomized, Placebo-Controlled, Clinical Trial of
  Donepezil in Vascular Dementia: Differential Effects by Hippocampal Size}}.
\newblock \bibinfo{journal}{Stroke} \bibinfo{volume}{41},
  \bibinfo{pages}{1213--1221}.
\bibitem[{Rosen et~al.(1984)Rosen, Mohs and Davis}]{rosen1984new}
\bibinfo{author}{Rosen, W.G.}, \bibinfo{author}{Mohs, R.C.},
  \bibinfo{author}{Davis, K.L.}, \bibinfo{year}{1984}.
\newblock \bibinfo{title}{{A New Rating Scale for Alzheimer's Disease.}}
\newblock \bibinfo{journal}{The American journal of psychiatry}
  \bibinfo{volume}{141}, \bibinfo{pages}{1356--1364}.
\bibitem[{Sharma(2019)}]{sharma2019cholinesterase}
\bibinfo{author}{Sharma, K.}, \bibinfo{year}{2019}.
\newblock \bibinfo{title}{{Cholinesterase Inhibitors as Alzheimer's
  Therapeutics}}.
\newblock \bibinfo{journal}{Molecular medicine reports} \bibinfo{volume}{20},
  \bibinfo{pages}{1479--1487}.
\bibitem[{Starling et~al.(2021)Starling, Murray, Lohr, Aiken, Carvalho and
  Scott}]{starling2021targeted}
\bibinfo{author}{Starling, J.E.}, \bibinfo{author}{Murray, J.S.},
  \bibinfo{author}{Lohr, P.A.}, \bibinfo{author}{Aiken, A.R.},
  \bibinfo{author}{Carvalho, C.M.}, \bibinfo{author}{Scott, J.G.},
  \bibinfo{year}{2021}.
\newblock \bibinfo{title}{{Targeted Smooth Bayesian Causal Forests: An Analysis
  of Heterogeneous Treatment Effects for Simultaneous vs. Interval Medical
  Abortion Regimens Over Gestation}}.
\newblock \bibinfo{journal}{The Annals of Applied Statistics}
  \bibinfo{volume}{15}, \bibinfo{pages}{1194--1219}.
\bibitem[{Sun et~al.(2020)Sun, Zhou and Fan}]{sun2020adaptive}
\bibinfo{author}{Sun, Q.}, \bibinfo{author}{Zhou, W.X.}, \bibinfo{author}{Fan,
  J.}, \bibinfo{year}{2020}.
\newblock \bibinfo{title}{{Adaptive Huber Regression}}.
\newblock \bibinfo{journal}{Journal of the American Statistical Association}
  \bibinfo{volume}{115}, \bibinfo{pages}{254--265}.
\bibitem[{Wager and Athey(2018)}]{wager2018estimation}
\bibinfo{author}{Wager, S.}, \bibinfo{author}{Athey, S.}, \bibinfo{year}{2018}.
\newblock \bibinfo{title}{{Estimation and Inference of Heterogeneous Treatment
  Effects Using Random Forests}}.
\newblock \bibinfo{journal}{Journal of the American Statistical Association}
  \bibinfo{volume}{113}, \bibinfo{pages}{1228--1242}.
\bibitem[{Wang et~al.(2021)Wang, Zheng, Zhou and Zhou}]{wang2021new}
\bibinfo{author}{Wang, L.}, \bibinfo{author}{Zheng, C.}, \bibinfo{author}{Zhou,
  W.}, \bibinfo{author}{Zhou, W.X.}, \bibinfo{year}{2021}.
\newblock \bibinfo{title}{A new principle for tuning-free huber regression}.
\newblock \bibinfo{journal}{Statistica Sinica} \bibinfo{volume}{31},
  \bibinfo{pages}{2153--2177}.

\end{thebibliography}

\newpage
\appendix

\fontsize{12}{14pt plus.8pt minus .6pt}\selectfont \vspace{0.8pc}
\centerline{\large\bf Supplementary for "Robust Forest-based
Estimation of Heterogeneous}
\vspace{2pt} 
\centerline{\large\bf  Treatment Effect Under High-dimensional Confounding"}
\vspace{.4cm} 

\def\thefigure{\arabic{figure}}
\def\thetable{\arabic{table}}

\renewcommand{\theequation}{\thesection.\arabic{equation}}

\fontsize{12}{14pt plus.8pt minus .6pt}\selectfont

The Online Supplementary Materials are organized as follows. Section \ref{appendix:rf_construction} provides additional details on the random forest construction. Section~\ref{appendix:additional simulation} presents additional simulation results. Section \ref{appendix:proof_nuisance} presents the assumptions and theoretical results for the proposed nuisance estimators and the corresponding technical proofs.
Section \ref{appendix:proof_treatment} contains the technical proofs for 
the consistency and asymptotic normality of the treatment effect 
estimators introduced in Section \Sref{sec:theory_treatment}3 of main paper. 

\label{appendix:proofs}
\renewcommand{\thesection}{S.\arabic{section}}
\setcounter{section}{0}
\renewcommand{\thelemma}{S.\arabic{lemma}} 
\setcounter{lemma}{0}
\renewcommand{\thecondition}{S.\arabic{condition}} 
\setcounter{condition}{0}
\renewcommand{\thedefinition}{S.\arabic{definition}} 
\setcounter{definition}{0}
\renewcommand{\theequation}{S.\arabic{equation}} 
\setcounter{equation}{0}

\renewcommand{\thefigure}{S.\arabic{figure}} 
\setcounter{figure}{0}
\renewcommand{\thetable}{S.\arabic{table}} 
\setcounter{table}{0}

\renewcommand{\theassumption}{S.\arabic{assumption}} 
\setcounter{assumption}{0}
\renewcommand{\thetheorem}{S.\arabic{theorem}} 
\setcounter{theorem}{0}


\section{Forest Construction Details}\label{appendix:rf_construction}

\begin{specification}(Forest Regularity)
    \label{specification:forest_regularity}
    The construction of each tree adheres to the following structural requirements.
    
        \item\textbf{(i) Honesty}. The training data $\mathcal{J}$ is randomly partitioned 
              into two disjoint sets, $\mathcal{J}_{1}$ and $\mathcal{J}_{2}$. 
              The first set $\mathcal{J}_{1}$ is used to determine the tree structure, 
              while $\mathcal{J}_{2}$ is reserved for calculating the leaf estimates.
        
        \item \textbf{(ii) $\Upsilon$-balanced}. Every split must be balanced such that each resulting child node contains at least a fraction $\Upsilon = 0.2$ of the observations from $\mathcal{J}_{2}$ available in the parent node.
        
        \item \textbf{(iii) Minimum leaf size $n_{\text{leaf}}$}.
        Splitting stops when the sample size from $\mathcal{J}_{2}$ in a terminal leaf is between $n_{\text{leaf}}$ and $2n_{\text{leaf}}-1$ for some $n_{\text{leaf}}\geq1$.

        \item \textbf{(iv) $\pi$-random-split}. At each node, every feature $j \in \{1, \dots, d_{\boldsymbol{X}}\}$ 
              has a probability of at least $\pi/d_{\boldsymbol{X}}$ of being selected for the split, 
              marginalized over the internal randomness of the learner, for some $0 < \pi \leq 1$.
  
\end{specification}

To ensure the validity of our estimation procedure and enable honest estimation, for each tree, we apply the sample-splitting mechanism. More specifically, the input sample $\mathcal{J}$ is randomly partitioned into two disjoint subsets: $\mathcal{J}_1$ and $\mathcal{J}_2$. The subset $\mathcal{J}_1$ is utilized exclusively for determining the tree structure, including the identification of optimal split points and features. In contrast, $\mathcal{J}_2$ is reserved for local parameter estimation within the terminal leaves of the constructed trees. This estimation utilizes the orthogonal Huber score developed in Section \Sref{sec:Construction_score}{2.2} and follows the comprehensive procedure outlined in Section \Sref{sec:het_huber_neyman}{2.3} of main paper.
During the tree-building process, several critical constraints are imposed to ensure robust and consistent estimation. Each split is enforced to be $\Upsilon$-balanced, meaning that every split partitions the observations from $\mathcal{J}_2$ in the parent node such that each resulting child node contains at least a fraction $\Upsilon$ of these observations. Following the main literature, we set $\Upsilon = 0.2$ to maintain structural balance. 
Furthermore, to prevent over-fitting, we impose a minimum leaf size of $n_{\text{leaf}}$ samples from $\mathcal{J}_2$. A node is allowed to split as long as its $\mathcal{J}_2$ sample size is at least $2n_{\text{leaf}}$ samples from $\mathcal{J}_2$, thereby preventing overly small leaves and excessive tree depth. Consequently, each terminal leaf of the tree is required to contain between $n_{\text{leaf}}$ and $2n_{\text{leaf}}-1$ samples from $\mathcal{J}_2$. In our implementation, we set $n_{\text{leaf}}=20$ as a balance between local adaptivity and estimation precision.
A crucial aspect for establishing theoretical guarantees is the $\pi$-random-split condition. This property ensures that at every splitting step, after marginalizing over the internal randomness of the splitting rule, the probability that the next split occurs along any specific feature $j$ is at least $\pi / d_{\boldsymbol{X}}$ for all $j \in \{1, \ldots, d_{\boldsymbol{X}}\}$. 
In practice, we implement this condition by drawing candidate splits uniformly at random from the set of all possible features within a node. Therefore, the lower-bound condition holds with $\pi=1$, which is the largest possible value of $\pi$.
Finally, the complete forest is constructed by aggregating a multitude of individual trees. In this section, we assume that the covariates $\boldsymbol{X}$ take values in $[0, 1]^{d_{\boldsymbol{X}}}$, which can be ensured by rescaling each coordinate, as in \cite{athey2019generalized} and \cite{oprescu2019orthogonal}. 

The following lemma, adapted from the proof of Theorem 3 in \cite{wager2018estimation}, shows that under the structural requirements defined in Specification~\ref{specification:forest_regularity}, 
the tree weights $a_{ib}(\boldsymbol{x})$ exhibit a localization property, 
remaining non-zero only within a small neighborhood of the target point $\boldsymbol{x}$. 

\begin{lemma}(Kernel shrinkage, \citealp{wager2018estimation})
    \label{lemma:kernel_shrinkage}
    Suppose the minimum leaf size parameter $n_{\text{leaf}}=O(1)$,
    the tree is $\Upsilon$-balanced and $\pi$-random-split
    and the distribution of $\boldsymbol{X}$ admits a density in $[0, 1]^{d_{\boldsymbol{X}}}$
    that is bounded away from zero and infinity.
    Then the tree weights satisfy
    \[
        \mathbb{E}\left[\sup\left\{\left\|\boldsymbol{X}_i - \boldsymbol{x} \right\|:a_{ib}(\boldsymbol{x}) > 0\right\}\right] = O\left(s^{-1/(2\alpha d_{\boldsymbol{X}})}\right),
    \]
    where $\alpha = \frac{\log(\Upsilon^{-1})}{\pi \log\left\{(1-\Upsilon)^{-1}\right\}}$ and $s$ is the size of the subsamples.
\end{lemma}

\section{Additional Numerical Results}\label{appendix:additional simulation}

\subsection{Implementation Details}

We implement the Orthogonal Huber Random Forest (OHRF) following the procedure described in Section 2 and the details of the estimation procedure are given below.

As in the main paper, we consider the model
\begin{align}
    Y_i              & =  T_i\theta_0(\boldsymbol{X}_i) + \boldsymbol{\beta}_0(\boldsymbol{X}_i)^{\top}\boldsymbol{W}_i + \epsilon_i, \quad i=1,2,\dots,n,\\
    T_i & = \boldsymbol{\ell}_0(\boldsymbol{X}_i)^{\top}\boldsymbol{W}_i + \eta_i, 
\end{align}
Here, $\theta_0(\boldsymbol{x})$ is the treatment effect and the nuisance parameters are denoted by $\boldsymbol{h}_0(\boldsymbol{x})=\left\{\boldsymbol{\nu}_0(\boldsymbol{x}),\boldsymbol{\ell}_0(\boldsymbol{x})\right\}$ with $\boldsymbol{\nu}_0(\boldsymbol{x})=\theta_0(\boldsymbol{x})\boldsymbol{\ell}_0(\boldsymbol{x})+\boldsymbol{\beta}_0(\boldsymbol{x})$.


\noindent \textbf{Node Splitting.} 
For each parent node, we need to obtain local estimation $\widehat{\boldsymbol{\nu}}_P$ and $\widehat{\boldsymbol{\ell}}_P$. Following the setting of \cite{oprescu2019orthogonal}, we estimate the $q$-th $(q=1,2,\dots, d_{\boldsymbol{T}})$ column of $\boldsymbol{\ell}_P$ by 
\[
    \widehat{\boldsymbol{\ell}}^{(q)}_{P} = \argmin_{\boldsymbol{\ell}^{(q)}} \left\{ \frac{1}{n_{P}} \sum_{\{i \in P\}}\left\{\boldsymbol{T}_{i}^{(q)}-\boldsymbol{\ell}^{(q){\top}}\boldsymbol{W}_i \right\}^2+\lambda_{\boldsymbol{\ell}}^{(0)}\left\|\boldsymbol{\ell}^{(q)}\right\|_1 \right\},
\]
with the regularization parameter defined as
    $\lambda_{\boldsymbol{\ell}}^{(0)} = \frac{1}{20}\sqrt{\frac{\log d_{\boldsymbol{W}}}{n_P}}$.
For $\boldsymbol{\nu}$, we estimate it by
\[
    \widehat{\boldsymbol{\nu}}_{P} = \argmin_{\boldsymbol{\nu}} \left\{ \frac{1}{n_{P}} \sum_{\{i \in P\}}\rho_{\delta_{\boldsymbol{\nu}}^{(0)}}\left(Y_i-\boldsymbol{\nu}^{\top}\boldsymbol{W}_i\right)+\lambda_{\boldsymbol{\nu}}^{(0)} \left\|\boldsymbol{\nu}\right\|_1 \right\}.
\]
Following the theoretical results in \cite{sun2020adaptive}, the robustness parameter $\delta_{\boldsymbol{\nu}}^{(0)}$ and regularization parameter $\lambda_{\nu}^{(0)}$ are set as
    $\delta_{\boldsymbol{\nu}}^{(0)} = \widehat{\sigma}\sqrt{\frac{n_p}{\log d_{\boldsymbol{W}}}}$ and $\lambda_{\boldsymbol{\nu}}^{(0)} = \frac{\widehat{\sigma}}{50} \sqrt{\frac{\log d_{\boldsymbol{W}}}{n_P}}$,
where $\widehat{\sigma}$ is the sample standard deviation of $\boldsymbol{Y}$.

\noindent \textbf{Nuisance Parameter Estimation.} 
The nuisance $\boldsymbol{\ell}(\boldsymbol{x})$ is estimated using a weighted Lasso regression, inheriting the regularization parameter $\lambda_{\boldsymbol{\ell}}$ from the splitting. 
Similarly, we estimate $\boldsymbol{\nu}(\boldsymbol{x})$ by a weighted Huber regression with an $L_1$-penalty. The robustness parameter $\delta_{\boldsymbol{\nu}}$ and regularization parameter $\lambda_{\boldsymbol{\nu}}$ used in the nuisance estimation step are selected via grid search.   Specifically, we set $\delta_{\boldsymbol{\nu}} = c_{\delta} \widehat{\sigma}\sqrt{\frac{n}{s\log d_{\boldsymbol{W}}}}$ and $\lambda_{\boldsymbol{\nu}} = \frac{c_{\lambda}\widehat{\sigma}}{50} \sqrt{\frac{s\log d_{\boldsymbol{W}}}{n}}$, where the scaling factors $c_{\delta}$ and $c_{\lambda}$ are selected from the set $\{0.25, 0.5, 0.75, 1\}$ separately to minimize the empirical loss.

\noindent \textbf{Orthogonal Score Construction.} 
 The robustness parameter $\delta_{\boldsymbol{\theta}}$ is calibrated once using the full sample before the forest construction. Specifically, we first compute the fitted values for the outcome as $\widehat{Y}_i = \widehat{\boldsymbol{\nu}}^\top \boldsymbol{W}_i$, where $\widehat{\boldsymbol{\nu}}$ is obtained by regressing $Y$ on $\boldsymbol{W}$ with an $L_1$-penalty with all data. 
We then define the corresponding outcome residuals $\widetilde{Y}_i = Y_i - \widehat{Y}_i$. Based on these full-sample residuals, we set the orthogonal-score robustness parameter as $\delta_{\boldsymbol{\theta}} = \widehat{\sigma}_{\widetilde{Y}} \sqrt{\frac{n}{s}}$, where $\widehat{\sigma}_{\widetilde{Y}}$ denotes the sample standard deviation of $\widetilde{Y}_i$ and $s$ is the subsample size used to grow each tree. 

\noindent{\it Note:} The robustness parameter $\delta_{\boldsymbol{\theta}}$ for the orthogonal score function $\boldsymbol{\psi}_{\delta}$ is chosen separately from the first-stage robustness parameter $\delta_{\boldsymbol{\nu}}$, because $\delta_{\boldsymbol{\theta}}$ is used for estimating the low-dimensional treatment effect rather than the high-dimensional nuisance coefficients. \cite{sun2020adaptive} emphasize that, in high-dimensional settings, the robustness parameter should adapt not only to the sample size but also to the dimensionality. The extra logarithmic factor arises because the high-dimensional nuisance estimator requires uniform control over all covariates. 

It is also important to note that $\delta$ is fixed before the local nuisance estimation and forest-based treatment effect estimation. In particular, we do not first estimate the local nuisance function $\widehat{\boldsymbol{\nu}}(\boldsymbol{X}_i)$ and then choose $\delta_{\boldsymbol{\theta}}$ based on residuals of the form $Y_i-\widehat{\boldsymbol{\nu}}(\boldsymbol{X}_i)^\top \boldsymbol{W}_i$. 
Instead, we use a global preliminary regression to obtain a stable residual scale. This avoids making the robustness parameter depend on local forest weights or on the estimated nuisance functions themselves, which could introduce additional randomness and circularity into the tuning procedure.

\noindent \textbf{Forest Architecture.} 
For structural hyperparameters of the random forest, we set the number of trees to $200$, the maximum tree depth to $10$ and the minimum leaf size to $20$.
The subsample size $s$ is adapted based on the dimensionality of the covariate vector $\boldsymbol{X}$: $s = 0.7n$ for the two-dimensional case (Setting 1) and $s = 0.5n$ for one-dimensional $\boldsymbol{X}$ (Settings 2 and 3). 

\subsection{Additional Numerical Results}
This section provides additional simulation results for Settings 2 and 3. These settings consider homogeneous nuisance and heterogeneous nuisance parameters with one-dimensional modifier $\boldsymbol{X}$.

\paragraph{Setting 2: Homogeneous Nuisance with One-dimensional Modifier $\boldsymbol{X}$.} 
In this setting, the modifier is one-dimensional, with $X_i \sim U(0, 1)$. 
The nuisance functions are defined as:
    $\boldsymbol{\beta}_0(x) = (\boldsymbol{1}^{\top}_{\lceil k/2 \rceil}, 
    -\boldsymbol{1}^{\top}_{\lfloor k/2 \rfloor}, \boldsymbol{0}^{\top})^{\top}$,
    $\boldsymbol{\ell}_0(x) = (\boldsymbol{0}^{\top}_k,\boldsymbol{1}^{\top}_{\lceil k/2 \rceil}, 
    -\boldsymbol{1}^{\top}_{\lfloor k/2 \rfloor}, \boldsymbol{0}^{\top})^{\top}$.
The treatment effect $\theta_0(x)$ follows a piecewise linear structure:
$$\theta_0(x)=
\left\{
\begin{array}{ll}
x+1, & x \le 0.3,\\
8x-1.1, & 0.3 < x \le 0.7,\\
-4x+7.3, & x > 0.7.
\end{array}
\right.$$

\paragraph{Setting 3: Heterogeneous Nuisance Parameters.} 
Setting 3 maintains the same $\theta_0(x)$ and $X_i$ as in Setting 2 but introduces heterogeneity for the nuisance parameters. 
We define the base vectors $\widetilde{\boldsymbol{\beta}}(x) $ 
and $\widetilde{\boldsymbol{\ell}}(x)$ as
    $$\widetilde{\boldsymbol{\beta}}(x)= \left(\cos(\tfrac{\pi}{2} x), 
    \tfrac{1+2x^3}{3}, (1-\tfrac{x}{3})^2, -1, -1\right)^{\top},$$ $$\widetilde{\boldsymbol{\ell}}(x)= \left(\sin(\tfrac{\pi}{2} x), 
    1-x^2, \tfrac{1}{2-x^2}, -1, -1\right)^{\top}.$$ For a given sparsity level $k$, the parameters are constructed as $\boldsymbol{\beta}_0(x) = (\widetilde{\boldsymbol{\beta}}_{1:k}(x)^{\top}, \boldsymbol{0}^{\top})^{\top}$, $\boldsymbol{\ell}_0(x) = (\boldsymbol{0}^{\top}_k, \widetilde{\boldsymbol{\ell}}_{1:k}(x)^{\top}, \boldsymbol{0}^{\top})^{\top}$,
where $\widetilde{\boldsymbol{\beta}}_{1:k}$ and $\widetilde{\boldsymbol{\ell}}_{1:k}$ denote the sub-vectors of $\widetilde{\boldsymbol{\beta}}(x)$ and $\widetilde{\boldsymbol{\ell}}(x)$ respectively, formed by their first first $k$ elements. We consider $k=1$ and $k=5$.

\begin{figure}[htbp]
\centering
\includegraphics[width=\linewidth]{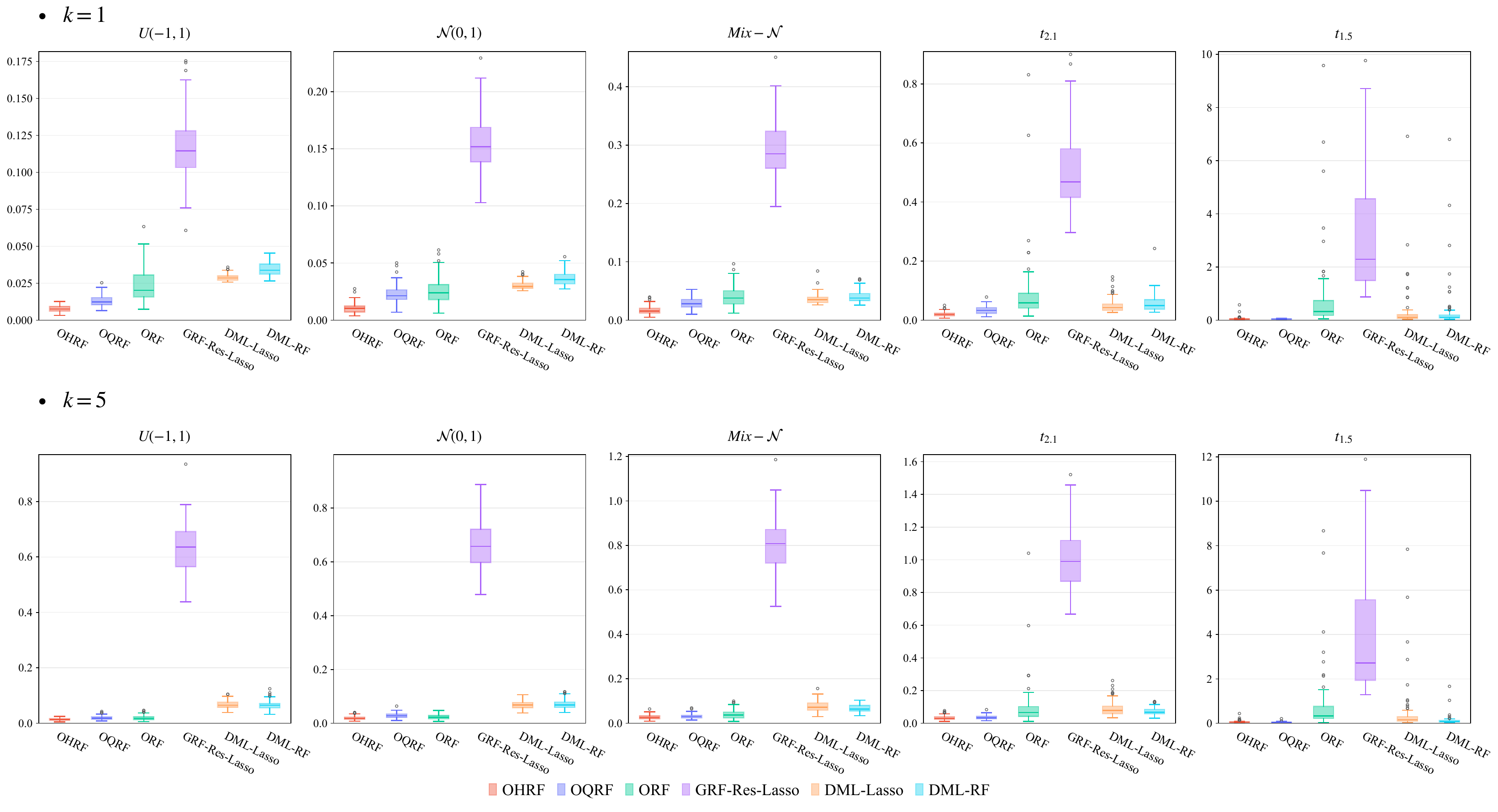}
\caption{Boxplots of the integrated squared errors for all methods across different error distributions in Setting 2.}
\label{fig:boxplot_setting1}
\end{figure}

\begin{figure}[htbp]
\centering
\includegraphics[width=\linewidth]{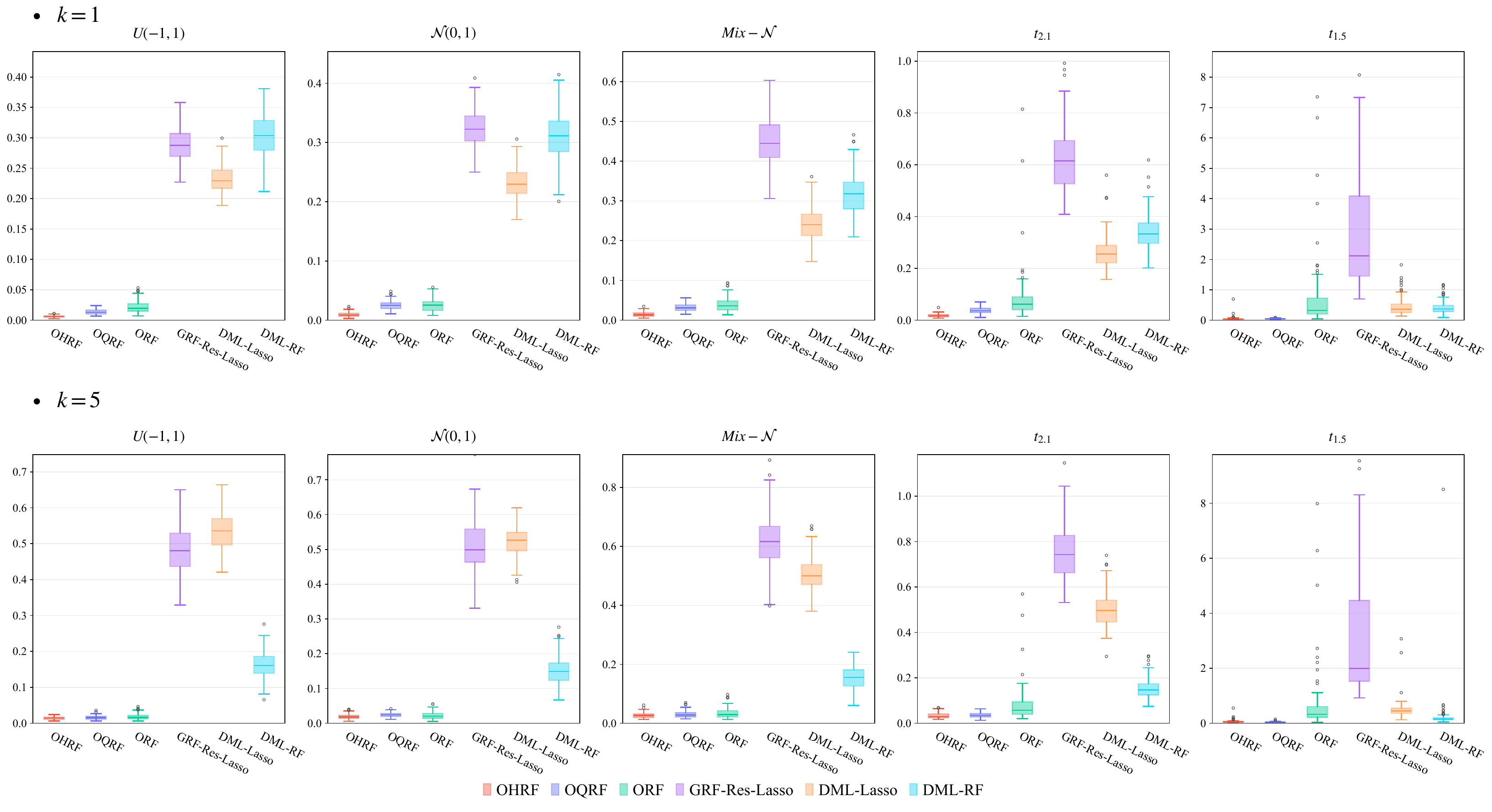}
\caption{Boxplots of the integrated squared errors for all methods across different error distributions in Setting 3.}
\label{fig:boxplot_setting2}
\end{figure}

For the one-dimensional settings, we define the test grid set as $\mathcal{X}_{\text{test}}=\left\lbrace x^{(1)},x^{(2)},\dots,x^{(100)}\right\rbrace$,  where the points are equally spaced in the domain of $X_i \in [0, 1]$. 
For Settings 2 and 3, we compute the Integrated Mean Squared Error as $\frac{1}{100} \sum_{r=1}^{100}\frac{1}{|\mathcal{X}_{\text{test}}|}\sum_{x_j\in \mathcal{X}_{\text{test}}} \left\{ \widehat{\theta}^{(r)}(x_j)-\theta(x_j) \right\}^{2}$ and the IMSE  for these two settings are reported in Tables~\ref{tab:simulation_results_setting1} and \ref{tab:simulation_results_setting2}, respectively.
Figures~\ref{fig:boxplot_setting1} and \ref{fig:boxplot_setting2} present the corresponding boxplots of the integrated squared errors. 

\begin{table}[t!]
    \centering
    \caption{Integrated Mean Square Error $(\times 10^{-2})$ of Setting 2}
    \label{tab:simulation_results_setting1}
    \scriptsize
    \setlength{\tabcolsep}{2pt}
    \renewcommand{\arraystretch}{0.9}
    \begin{tabular}{lllcccccc}
        \toprule
        & \multicolumn{2}{l}{\diagbox[width=2cm]{\tiny Error}{\tiny Methods}} & OHRF & OQRF & ORF & GRF-Res-Lasso & DML-Lasso & DML-RF \\
        \midrule
        
\multirow{5}{*}{$k=1$} & $U(-1,1)$ & & \topone{0.78}(0.02) & \toptwo{1.32}(0.04) & 2.34(0.11) & 11.58(0.20) & 2.88(0.02) & 3.45(0.05) \\
 & $\mathcal{N}(0,1)$ & & \topone{1.04}(0.04) & \toptwo{2.29}(0.07) & 2.55(0.11) & 15.43(0.25) & 3.05(0.03) & 3.65(0.06) \\
 & $Mix\text{-}\mathcal{N}$ & & \topone{1.69}(0.07) & \toptwo{2.92}(0.09) & 4.06(0.18) & 29.20(0.51) & 3.62(0.09) & 4.05(0.10) \\
 & $t_{2.1}$ & & \topone{2.01}(0.08) & \toptwo{3.41}(0.13) & 8.52(1.04) & 61.22(7.76) & 4.91(0.23) & 5.73(0.29) \\
 & $t_{1.5}$ & & \toptwo{5.22}(0.64) & \topone{4.18}(0.14) & 126.74(36.83) & 751.73(226.96) & 29.46(7.78) & 32.24(8.53) \\
\cmidrule{1-9}
\multirow{5}{*}{$k=5$} & $U(-1,1)$ & & \topone{1.39}(0.04) & \toptwo{1.91}(0.06) & 1.91(0.08) & 63.45(0.90) & 6.64(0.13) & 6.53(0.15) \\
 & $\mathcal{N}(0,1)$ & & \topone{1.96}(0.07) & 2.82(0.08) & \toptwo{2.39}(0.09) & 66.55(0.91) & 6.91(0.16) & 7.09(0.16) \\
 & $Mix\text{-}\mathcal{N}$ & & \topone{2.74}(0.10) & \toptwo{3.13}(0.11) & 4.00(0.20) & 80.46(1.17) & 7.51(0.24) & 6.75(0.17) \\
 & $t_{2.1}$ & & \topone{3.35}(0.13) & \toptwo{3.60}(0.12) & 9.11(1.20) & 113.09(8.26) & 8.99(0.43) & 7.30(0.23) \\
 & $t_{1.5}$ & & \toptwo{6.69}(0.50) & \topone{4.44}(0.24) & 122.51(32.60) & 663.45(136.92) & 42.90(10.47) & 12.44(1.89) \\
        
        \bottomrule
    \end{tabular}
        \begin{tablenotes}
            \footnotesize
            \item \textit{Note:} The standard errors of the integrated squared errors are shown in parentheses. The bold and underlined values indicate the best and second-best results, respectively.
    \end{tablenotes}
\end{table}

\begin{table}[t!]
    \centering
    \caption{Integrated Mean Square Error $(\times 10^{-2})$ of Setting 3}
    \label{tab:simulation_results_setting2}
    \scriptsize
    \setlength{\tabcolsep}{3.5pt}
    \renewcommand{\arraystretch}{0.7}
    \begin{tabular}{lllcccccc}
        \toprule
        & \multicolumn{2}{l}{\diagbox[width=2cm]{\tiny Error}{\tiny Methods}} & OHRF & OQRF & ORF & GRF-Res-Lasso & DML-Lasso & DML-RF \\
        \midrule
        
\multirow{5}{*}{$k=1$} & $U(-1,1)$ & & \topone{0.63}(0.02) & \toptwo{1.35}(0.04) & 2.25(0.10) & 28.98(0.25) & 23.25(0.22) & 30.52(0.37) \\
 & $\mathcal{N}(0,1)$ & & \topone{0.95}(0.04) & \toptwo{2.51}(0.07) & 2.52(0.10) & 32.32(0.33) & 23.22(0.28) & 31.12(0.41) \\
 & $Mix\text{-}\mathcal{N}$ & & \topone{1.50}(0.06) & \toptwo{3.12}(0.09) & 3.88(0.19) & 44.71(0.57) & 24.14(0.41) & 31.84(0.52) \\
 & $t_{2.1}$ & & \topone{1.82}(0.07) & \toptwo{3.71}(0.13) & 8.44(1.03) & 82.44(13.32) & 26.51(0.64) & 34.03(0.68) \\
 & $t_{1.5}$ & & \toptwo{4.98}(0.71) & \topone{4.33}(0.15) & 96.67(19.26) & 402.70(57.27) & 44.91(2.99) & 42.78(2.25) \\
\cmidrule{1-9}
\multirow{5}{*}{$k=5$} & $U(-1,1)$ & & \topone{1.42}(0.04) & \toptwo{1.57}(0.05) & 1.85(0.09) & 48.25(0.66) & 53.54(0.55) & 16.30(0.37) \\
 & $\mathcal{N}(0,1)$ & & \topone{1.97}(0.07) & 2.46(0.07) & \toptwo{2.18}(0.09) & 51.13(0.75) & 52.25(0.47) & 15.22(0.42) \\
 & $Mix\text{-}\mathcal{N}$ & & \topone{2.66}(0.09) & \toptwo{3.01}(0.11) & 3.48(0.17) & 61.99(0.91) & 50.93(0.62) & 15.25(0.36) \\
 & $t_{2.1}$ & & \topone{3.35}(0.12) & \toptwo{3.62}(0.11) & 8.08(0.80) & 85.71(8.99) & 50.05(0.77) & 15.40(0.46) \\
 & $t_{1.5}$ & & \toptwo{6.77}(0.60) & \topone{4.31}(0.19) & 112.95(30.41) & 666.30(215.26) & 66.90(17.23) & 26.57(8.39) \\
        
        \bottomrule
    \end{tabular}
        \begin{tablenotes}
            \footnotesize
            \item \textit{Note:} The standard errors of the integrated squared errors are shown in parentheses. The bold and underlined values indicate the best and second-best results, respectively.
    \end{tablenotes}
\end{table}

The results for Settings 2 and 3 provide additional evidence on the performance of OHRF with a one-dimensional effect modifier. 
The findings in Setting 2 are consistent with those in Setting 1, further confirming the robustness of OHRF. In Setting 3, when the nuisance structure becomes more complex, the advantages of the first three methods, OHRF, OQRF, and ORF, become more pronounced. These methods rely on generalized random forest procedures and can therefore better exploit the local structure of the effect modifiers to flexibly capture heterogeneous treatment effects. In contrast, DML-Lasso and DML-RF deteriorate more noticeably in these settings as they treat $\boldsymbol{W}$ and $\boldsymbol{X}$ jointly as generic covariates and fail to exploit the structural information. GRF-Res-Lasso continues to perform poorly because it does not correct for the bias induced by high-dimensional confounders. These results highlight the robustness of OHRF to heavy-tailed errors while accommodating heterogeneous nuisance functions.

\section{Theoretical Results for Estimated Nuisance Parameters}
\label{appendix:proof_nuisance}
This section presents the theoretical results for the nuisance parameter estimators. Section~\ref{S3.1} introduces the assumptions needed for the theoretical analysis. Section~\ref{Nuisance proof} states and proves the main theoretical results for nuisance estimation. Section~\ref{nuisance lemma} provides the auxiliary lemmas used in Section~\ref{Nuisance proof}, together with their proofs.

\subsection{Assumptions and Main Results for Nuisance Estimation}\label{S3.1}

To rigorously establish the theoretical properties and performance guarantees for our nuisance estimators, we introduce a set of fundamental assumptions concerning the underlying data generating process and the characteristics of the nuisance functions.

\begin{assumption}
    \label{assumption:sparse_nuisance}
    For any $\boldsymbol{x} \in[0, 1]^{d_{\boldsymbol{X}}}$, the true nuisance coefficient vectors $\boldsymbol{\ell}_0( \boldsymbol{x} )$ and $\boldsymbol{\nu}_0( \boldsymbol{x} )$ are $k$-sparse and $L$-Lipschitz continuous in $\boldsymbol{x}$.
\end{assumption}

\begin{assumption}
    \label{assumption:re_nu}
    The confounders satisfies $\|\boldsymbol W_i\|_\infty \le L_{\boldsymbol{W}}$.
    For a given $\boldsymbol{x}$, let $\mathcal{K}(\boldsymbol{\ell}_0,\boldsymbol{x})$ and $\mathcal{K}(\boldsymbol{\nu}_0,\boldsymbol{x})$ denote the support set (i.e., the set of indices corresponding to the non-zero coefficients of a vector) of the true nuisance coefficient vectors $\boldsymbol{\ell}_0(\boldsymbol{x})$ and $\boldsymbol{\nu}_0(\boldsymbol{x})$, respectively. Let $\BSigma_i = \boldsymbol{\boldsymbol{W}}_i\boldsymbol{\boldsymbol{W}}_i^{\top}$, it satisfies the restricted eigenvalue condition. Specifically, there exists a positive constant $\varpi_{lo}$ such that 
    \[
        \boldsymbol{u}( \boldsymbol{x} )^{\top} \mathbb{E} \left(\BSigma_i  \right) \boldsymbol{u}( \boldsymbol{x} ) \geq 2\varpi_{lo} \left\|\boldsymbol{u}( \boldsymbol{x} )\right\|_{2}^{2}
    \]
    holds for all $\boldsymbol{u}( \boldsymbol{x} ) \in \{\boldsymbol{u}( \boldsymbol{x} )\in\mathbb{R}^{d_{\boldsymbol{W}}} : \left\| \boldsymbol{u}( \boldsymbol{x} )_{\mathcal{A}^{c}} \right\|_{1} \leq3 \left\| \boldsymbol{u}( \boldsymbol{x} )_{\mathcal{A}} \right\|_{1} \}\doteq \mathcal{C} ( \mathcal{A}; 3 )$ for both $\mathcal{A}=\mathcal{K}(\boldsymbol{\ell}_0,\boldsymbol{x})$ and $\mathcal{A}=\mathcal{K}(\boldsymbol{\nu}_0,\boldsymbol{x})$.
\end{assumption}

\begin{assumption}
    \label{assumption:scaling}
    The subsample size $s$, the dimension of the high-dimensional confounders $d_{\boldsymbol{W}}$ and the sparsity level $k$ of the true nuisance parameters satisfy the following conditions: 
    (i) $s \to \infty$ and $s = o(n)$; 
    (ii) $\log d_{\boldsymbol{W}} = o(n/s)$; and 
    (iii) $k = o\left( (n/s)^{\min\left(\frac{\tau}{1+\tau}, 1/2\right) - 1/4} \right)$. 
\end{assumption}

Assumption \ref{assumption:sparse_nuisance} posits that the true nuisance parameters are $k$-sparse and $L$-Lipschitz continuous in $\boldsymbol{x}$.
Complementarily, Assumption \ref{assumption:re_nu} is a standard restricted eigenvalue condition in high-dimensional statistics,
asserting that the covariance matrix of $\boldsymbol{W}$ 
is well-behaved on a specific sparse cone \citep{bickel2009simultaneous,meinshausen2009lasso}.
 Assumption \ref{assumption:scaling} formalizes the admissible scaling relationships among the total sample size $n$, subsample size $s$, covariate dimension $d_{\boldsymbol{W}}$ and the sparsity level $k$.
Under Assumptions \ref{assumption:sparse_nuisance}-\ref{assumption:scaling}, we next establish convergence rates of our nuisance estimators.

\begin{theorem}(Properties of the Nuisance Estimators)
    \label{thm:nuisance_properties}
    Under Assumptions \ref{assumption:sparse_nuisance} - \ref{assumption:scaling}, the following statements hold.

    \begin{enumerate}
        \item[(i)] \textbf{Forest Lasso bound.}
        For any $t>0$, if $\lambda_{\boldsymbol{\ell}} \gtrsim s^{-1/(2\alpha d_{\boldsymbol X})}
        + \sqrt{\frac{s\log \left(d_{\boldsymbol W}/t\right)}{n}}$, then we have,
         with probability at least $1-e^{-t}$,
         $\left\| \widehat{\boldsymbol{\ell}}(\boldsymbol{x})-\boldsymbol{\ell}_0(\boldsymbol{x}) \right\|_1\le\frac{\lambda_{\boldsymbol{\ell}} k}{\varpi_{lo}-16k\sqrt{s\log (d_{\boldsymbol W}/t)/n}}$.

        \item[(ii)] \textbf{Forest Huber-Lasso bound.}
         Let $\mu_\tau<\infty$ and $\delta_0$ be two constants satisfying $\delta_0 \ge (\mu_\tau)^{\max\{\frac{1}{1+\tau},\frac{1}{2}\}}$. For any $t>0$, as long as $\delta_{\boldsymbol{\nu}} \ge \delta_0
        \left\{\frac{n}{s\log (d_{\boldsymbol W}/t)}\right\}^{\max\left(\frac{1}{1+\tau},\frac{1}{2}\right)}$, 
        $
        \lambda_{\boldsymbol{\nu}} \ge 4\delta_0L_{\boldsymbol{W}}
        \left\{\frac{s\log (d_{\boldsymbol W}/t)}{n}\right\}^{\min\left(\frac{\tau}{1+\tau},\frac{1}{2}\right)}$, 
        then with probability at least $1-e^{-t}$, we have
        \[
        \left\| \widehat{\boldsymbol{\nu}}(\boldsymbol{x})-\boldsymbol{\nu}_0(\boldsymbol{x}) \right\|_1
        \le 12k\lambda_{\boldsymbol{\nu}} \varpi_{lo}^{-1}.
        \]
        In particular, taking $t=c\log d_{\boldsymbol W}$ for some $c>0$, then with probability at least $1-d_{\boldsymbol W}^{-c}$,   
        $\left\| \widehat{\boldsymbol{\nu}}(\boldsymbol{x})-\boldsymbol{\nu}_0(\boldsymbol{x}) \right\|_1
        \lesssim
        k\varpi_{lo}^{-1}\delta_0L_{\boldsymbol{W}}
        \left\{\frac{s(1+c)\log d_{\boldsymbol W}}{n}\right\}^{\min\{\tau/(1+\tau),\,1/2\}}$.

        \item[(iii)] \textbf{Nuisance function convergence.} Define
        $$\mathcal{E}(\widehat{\boldsymbol{h}})=\sqrt{\mathbb{E}\left\{\left\|\widehat{\boldsymbol{\nu}}(\boldsymbol{x})-\boldsymbol{\nu}_0(\boldsymbol{x})\right\|^2+\left\|\widehat{\boldsymbol{\ell}}(\boldsymbol{x})-\boldsymbol{\ell}_0(\boldsymbol{x})\right\|^2\right\}}.$$ Then, for any fixed integer $p\ge 1$,
        $$\mathbb E\left\{\mathcal{E}(\widehat{\boldsymbol h})^{p}\right\}^{1/p}=o\left(\left(\frac{s}{n}\right)^{1/4}\right).$$
    \end{enumerate}
\end{theorem}
Theorem~\ref{thm:nuisance_properties} summarizes the theoretical results for the nuisance estimators. First, part (i) provides a high-probability error bound for $\widehat{\boldsymbol{\ell}}(\boldsymbol{x})$, with the proof follows directly from the proof of Theorem 5.2 of \cite{oprescu2019orthogonal}. For $\widehat{\boldsymbol{\nu}}(\boldsymbol{x})$, part (ii) establishes a high-probability error bound when $\delta_{\boldsymbol{\nu}}$ and $\lambda_{\boldsymbol{\nu}}$ are properly chosen. 
By Neyman orthogonality of the score function, nuisance estimation errors affect the treatment effect estimator only through second-order remainder terms. 
To this end, we bound the conditional mean-squared error $\mathcal{E}(\widehat{\boldsymbol h})$ in part (iii). These results establish that Assumption 3 is satisfied by the proposed nuisance estimators.

\subsection{Proofs of the Nuisance Estimation Results}\label{Nuisance proof}

To prove the main results for the nuisance estimator $\widehat{\boldsymbol\nu}(\boldsymbol{x})$, we first introduce the notation needed to state the theorem:
\begin{itemize}
    \item $e\left\{\boldsymbol{Z}_i; \boldsymbol{\nu}(\boldsymbol{x})\right\} = Y_i - \boldsymbol{\nu}(\boldsymbol{x})^{\top}\boldsymbol{W}_i$,
    \item $\rho_{\delta}\left\{\boldsymbol{Z}_i; \boldsymbol{\nu}(\boldsymbol{x})\right\} = \phi_{\delta}\left[e\left\{\boldsymbol{Z}_i; \boldsymbol{\nu}(\boldsymbol{x})\right\}\right]$,
    \item $\nabla^{2}_{\boldsymbol{\nu}} \rho_{\delta}\left\{\boldsymbol{Z}_i; \boldsymbol{\nu}(\boldsymbol{x})\right\} = \boldsymbol{\boldsymbol{W}}_i\boldsymbol{\boldsymbol{W}}_i^\top I\left[|e\left\{\boldsymbol{Z}_i; \boldsymbol{\nu}(\boldsymbol{x})\right\}|\leq\delta\right]$,
    \item $\mathcal{L}_{\delta}\left\{\boldsymbol{\nu}(\boldsymbol{x})\right\} = \sum_{i=1}^n a_i \rho_{\delta}\left\{\boldsymbol{Z}_i,\boldsymbol{\nu}(\boldsymbol{x})\right\}$,
    \item $\boldsymbol{H}_\delta\left\{\boldsymbol{\nu}(\boldsymbol{x})\right\} = \nabla^2_{\nu} \mathcal{L}_{\delta}\left\{\boldsymbol{\nu}(\boldsymbol{x})\right\} = \sum_{i=1}^n a_i \boldsymbol{\boldsymbol{W}}_i \boldsymbol{\boldsymbol{W}}_i^\top I\left[|e\left\{\boldsymbol{Z}_i; \boldsymbol{\nu}(\boldsymbol{x})\right\}| \leq \delta\right]$,
    \item $\BSigma_i = \boldsymbol{\boldsymbol{W}}_i \boldsymbol{\boldsymbol{W}}_i^\top$,
    \item $\BSigma_a = \sum_{i=1}^n a_i \BSigma_i$,
    \item $\phi_{\delta}\left[e\left\{\boldsymbol{Z}_i; \boldsymbol{\nu}(\boldsymbol{x})\right\}\right] =  \nabla_{e}\rho_{\delta}\left\{\boldsymbol{Z}_i; \boldsymbol{\nu}(\boldsymbol{x})\right\}$,
    \item $\phi_{1}\left[e\left\{\boldsymbol{Z}_i; \boldsymbol{\nu}(\boldsymbol{x})\right\}\right] = \frac{1}{\delta}\phi_{\delta}\left[\delta e\left\{\boldsymbol{Z}_i; \boldsymbol{\nu}(\boldsymbol{x})\right\}\right]$,
    \item $\mu_{i,\tau} = \mathbb{E}\left( |\epsilon_i|^{1+\tau}\right)$, $\mu_{\tau} = \sum_{i=1}^n a_i \mu_{i,\tau}$.

\end{itemize}
  For the remainder of this section, we {\bf write $\delta$ in place of $\delta_{\boldsymbol{\nu}}$ for notational simplicity}. We first establish the high-probability $L_1$-error bound for the forest Huber-Lasso estimator in part (ii).

\noindent\textbf{Proof of Theorem~\ref{thm:nuisance_properties} (ii).}
\begin{proof}
The proof consists of four steps. First, we show that the estimation error lies in the restricted cone. Second, we verify that the weighted empirical loss satisfies a restricted strong convexity condition with high probability. Finally, since the Hessian of the Huber loss depends on the nuisance parameter, we construct an intermediate estimator on the line segment between the true nuisance parameter and the estimated one, so that the corresponding error direction lies in the localized restricted eigenvalue region. Combining the basic inequality with the localized restricted strong convexity condition yields the desired $L_1$-error bound for the original estimator.

\noindent \textbf{Step 1: Cone condition for the nuisance estimation error.}
We first prove that the deviation between $\boldsymbol{\nu}_0(\boldsymbol{x})$ and $\widehat{\boldsymbol{\nu}}(\boldsymbol{x})$ lies in the cone $\mathcal{C} (\mathcal{K}, 3)$. By the convexity of $\widehat{\boldsymbol{\nu}}(\boldsymbol{x})$, we have
    \[
        \sum_{i=1}^{n}a_i \rho_{\delta}\left\{\boldsymbol{Z}_i;\widehat{\boldsymbol{\nu}}(\boldsymbol{x})\right\}+\lambda_{\boldsymbol{\nu}} \left\|\widehat{\boldsymbol{\nu}}\right\|_1 \leq \sum_{i=1}^{n}a_i \rho_{\delta}(\boldsymbol{Z}_i;\boldsymbol{\nu})+\lambda_{\boldsymbol{\nu}} \left\|\boldsymbol{\nu}\right\|_1,
    \]
    and
    \begin{align*}
         &\; \lambda_{\boldsymbol{\nu}}\left\{\left\|\boldsymbol{\nu}_0(\boldsymbol{x})\right\|_1 - \left\|\widehat{\boldsymbol{\nu}}(\boldsymbol{x})\right\|_1\right\}                                                                                        \\
          \geq  &\;\sum_{i=1}^{n}a_i \rho_{\delta}\left\{\boldsymbol{Z}_i;\widehat{\boldsymbol{\nu}}(\boldsymbol{x})\right\} - \sum_{i=1}^{n}a_i \rho_{\delta}\left\{\boldsymbol{Z}_i;\boldsymbol{\nu}_0(\boldsymbol{x})\right\}                                         \\
          \geq  &\;\sum_{i=1}^{n}a_i \langle \nabla_{\nu} \rho_{\delta}\left\{\boldsymbol{Z}_i;\boldsymbol{\nu}_0(\boldsymbol{x})\right\},\widehat{\boldsymbol{\nu}}(\boldsymbol{x})-\boldsymbol{\nu}_0(\boldsymbol{x}) \rangle                            \\
          \geq  &\;-\left\|\sum_{i=1}^{n}a_i \nabla_{\nu} \rho_{\delta}\left\{\boldsymbol{Z}_i;\boldsymbol{\nu}_0(\boldsymbol{x})\right\} \right\|_{\infty} \left\|\widehat{\boldsymbol{\nu}}(\boldsymbol{x})-\boldsymbol{\nu}_0(\boldsymbol{x})\right\|_1.\\
    \end{align*}
    Lemma~\ref{lemma:gradient} in Subsection~\ref{nuisance lemma} indicates that $-\left\|\sum_{i=1}^{n}a_i \nabla_{\nu} \rho_{\delta}\left\{\boldsymbol{Z}_i;\boldsymbol{\nu}_0(\boldsymbol{x})\right\} \right\|_{\infty}\geq-\frac{\lambda_{\boldsymbol{\nu}}}{2}$, then
    \begin{equation}\label{eq1}
        \lambda_{\boldsymbol{\nu}}\left\{\left\|\boldsymbol{\nu}_0(\boldsymbol{x})\right\|_1 - \left\|\widehat{\boldsymbol{\nu}}(\boldsymbol{x})\right\|_1\right\}
        \geq -\frac{\lambda_{\boldsymbol{\nu}}}{2}\left\|\widehat{\boldsymbol{\nu}}(\boldsymbol{x})-\boldsymbol{\nu}_0(\boldsymbol{x})\right\|_1.
   \end{equation}   
    Let $\boldsymbol{\zeta}(\boldsymbol{x}) = \widehat{\boldsymbol{\nu}}(\boldsymbol{x})-\boldsymbol{\nu}_0(\boldsymbol{x})$. It follows that
    \begin{equation}\label{eq2}
    \begin{aligned}
         &\; \left\|\boldsymbol{\nu}_0(\boldsymbol{x})\right\|_1 - \left\|\widehat{\boldsymbol{\nu}}(\boldsymbol{x})\right\|_1\\
        =&\;\left\|\boldsymbol{\nu}_0(\boldsymbol{x})_{\mathcal{K}}\right\|_1+\left\|\boldsymbol{\nu}_0(\boldsymbol{x})_{\mathcal{K}^c}\right\|_1 - \left\|\widehat{\boldsymbol{\nu}}(\boldsymbol{x})_{\mathcal{K}}\right\|_1 - \left\|\widehat{\boldsymbol{\nu}}(\boldsymbol{x})_{\mathcal{K}^c}\right\|_1 \\
          = &\;\left\|\boldsymbol{\nu}_0(\boldsymbol{x})_{\mathcal{K}}\right\|_1 - \left\|\widehat{\boldsymbol{\nu}}(\boldsymbol{x})_{\mathcal{K}}\right\|_1 - \left\|\widehat{\boldsymbol{\nu}}(\boldsymbol{x})_{\mathcal{K}^c}\right\|_1                                                           \\
          = &\;\left\|\boldsymbol{\nu}_0(\boldsymbol{x})_{\mathcal{K}}\right\|_1 - \left\|\widehat{\boldsymbol{\nu}}(\boldsymbol{x})_{\mathcal{K}}\right\|_1 - \left\|\widehat{\boldsymbol{\nu}}(\boldsymbol{x})_{\mathcal{K}^c} - \boldsymbol{\nu}_0(\boldsymbol{x})_{\mathcal{K}^c}\right\|_1                \\
          \leq &\;\left\|\boldsymbol{\nu}_0(\boldsymbol{x})_{\mathcal{K}} - \widehat{\boldsymbol{\nu}}(\boldsymbol{x})_{\mathcal{K}}\right\|_1 - \left\|\widehat{\boldsymbol{\nu}}(\boldsymbol{x})_{\mathcal{K}^c} - \boldsymbol{\nu}_0(\boldsymbol{x})_{\mathcal{K}^c}\right\|_1                              \\
          \leq &\;\left\|\boldsymbol{\zeta}(\boldsymbol{x})_{\mathcal{K}}\right\|_1 - \left\|\boldsymbol{\zeta}(\boldsymbol{x})_{\mathcal{K}^c}\right\|_1. \\
    \end{aligned}
    \end{equation}
Combining Equation~\eqref{eq1} with Equation~\eqref{eq2} yields
    \[
        \lambda_{\boldsymbol{\nu}}\left\{\left\|\boldsymbol{\zeta}(\boldsymbol{x})_{\mathcal{K}}\right\|_1 - \left\|\boldsymbol{\zeta}(\boldsymbol{x})_{\mathcal{K}^c}\right\|_1\right\} \geq -\frac{\lambda_{\boldsymbol{\nu}}}{2}\left\|\boldsymbol{\zeta}(\boldsymbol{x})\right\|_1,
    \]
    then
    \[
        \left\|\boldsymbol{\zeta}(\boldsymbol{x})\right\|_1 \geq 2\left\{\left\|\boldsymbol{\zeta}(\boldsymbol{x})_{\mathcal{K}}\right\|_1 - \left\|\boldsymbol{\zeta}(\boldsymbol{x})_{\mathcal{K}^c}\right\|_1\right\}.
    \]
    Moreover, since
    \[
        \left\|\boldsymbol{\zeta}(\boldsymbol{x})\right\|_1 = \left\|\widehat{\boldsymbol{\nu}}(\boldsymbol{x})_{\mathcal{K}} - \boldsymbol{\nu}_0(\boldsymbol{x})_{\mathcal{K}} + \widehat{\boldsymbol{\nu}}(\boldsymbol{x})_{\mathcal{K}^c} - \boldsymbol{\nu}_0(\boldsymbol{x})_{\mathcal{K}^c}\right\|_1
        \leq \left\|\boldsymbol{\zeta}(\boldsymbol{x})_{\mathcal{K}}\right\|_1 + \left\|\boldsymbol{\zeta}(\boldsymbol{x})_{\mathcal{K}^c}\right\|_1,
    \]
   we have $3\left\|\boldsymbol{\zeta}(\boldsymbol{x})_{\mathcal{K}}\right\|_1 \geq \left\|\boldsymbol{\zeta}(\boldsymbol{x})_{\mathcal{K}^c}\right\|_1$,
    i.e. $\boldsymbol{\zeta}(\boldsymbol{x}) \in \mathcal{C} (\mathcal{K}, 3).$

\noindent\textbf{Step 2: Restricted strong convexity of the weighted empirical loss.}
    We now show that the weighted emperical loss function satisfies a conditional
    restricted strong convexity property with high probability. The Hessian matrix $\BSigma_a = \sum_{i=1}^{n} a_i \BSigma_i = \frac{1}{B}\sum_{b}\sum_{i \in b} a_{ib}\BSigma_i$
    is lower bounded by a matrix whose entries corresponding to a Monte-Carlo approximation of the U-statistic:
    \[
        \boldsymbol{U}(n,s)=\frac{1} {\binom{n} {s}} \sum_{S \subseteq[ n ] : | S |=s} \sum_{i \in S} \mathbb{E}_{\omega} \left\{ a_{i} ( S, \omega) \BSigma_i \right\}.
    \]
    Hence, for sufficiently large $B$, by a U-statistic concentration inequality and a union bound,
    each entry will concentrate around the expected value of U-statistic with probability at least $1-e^{-t}$:
    \[
        \left\| \frac{1}{B}\sum_{b}\sum_{i \in b} a_{ib}\BSigma_i - \mathbb{E} \left\{ \boldsymbol{U}(n,s) \right\} \right\|_{\infty} \leq 2\sqrt{\frac{s\left(t+\log{d_{\boldsymbol{W}}}\right)}{n}}.
    \]
    Moreover, by the tower law of expectation and the honesty of ORF trees, we can write:
    \[
        \mathbb{E} \left\{ \boldsymbol{U}(n,s) \right\} = \mathbb{E} \left\{\frac{1} {\binom{n} {s}} \sum_{S \subseteq[ n ] : | S |=s} \sum_{i \in S} a_{i} ( S, \omega) \mathbb{E} \left(\BSigma_i|\boldsymbol{x} \right)\right\}.
    \]
    Since each $\mathbb{E} \left( \BSigma_i|\boldsymbol{x} \right)$ satisties the restricted eigenvalue condition,
    we can conclude that $\mathbb{E}\left\{\boldsymbol{U}(n,s)\right\}$ also satisfies the restricted eigenvalue condition.     
    
    \noindent \textbf{Step 3: Derivation of the final nuisance error bound.}
    We next combine the cone condition with the localized restricted strong convexity condition. Since the latter is only guaranteed in a local neighborhood, we use a proof by contradiction to show that the nuisance estimator must lie within this neighborhood, yielding the desired $L_1$-error bound.

    Define a constant $r$ such that $r \ge 12k\lambda_{\boldsymbol{\nu}}\varpi_{lo}^{-1}$. We then construct an intermediate estimator $\widehat{\boldsymbol{\nu}}_{\varsigma}(\boldsymbol{x}) = \boldsymbol{\nu}_0(\boldsymbol{x}) + \varsigma\left\{\widehat{\boldsymbol{\nu}}(\boldsymbol{x})-\boldsymbol{\nu}_0(\boldsymbol{x})\right\}$
    such that $\| \widehat{\boldsymbol{\nu}}(\boldsymbol{x})-\boldsymbol{\nu}_0(\boldsymbol{x}) \|_1 \leq r$ for $r > 0$ to be specified.
    We take $\varsigma=1$ if $\| \widehat{\boldsymbol{\nu}}(\boldsymbol{x})-\boldsymbol{\nu}_0(\boldsymbol{x}) \|_1 \leq r$
    and choose $\varsigma \in (0,1)$ so that $\| \widehat{\boldsymbol{\nu}}_{\varsigma}(\boldsymbol{x})-\boldsymbol{\nu}_0(\boldsymbol{x}) \|_1 = r$ otherwise.
    Define a local $L_1$ cone $$\mathcal{C} ( \mathcal{K}, \gamma, r ) :=\left\{( \boldsymbol{u},\boldsymbol{\nu}) \in\mathbb{S}^{d_{\boldsymbol{W}}-1} \times \mathbb{R}^{d_{\boldsymbol{W}}} : \left\|\boldsymbol{u}_{\mathcal{K}^{c}} \right\|_{1} \leq\gamma\left\| \boldsymbol{u}_{\mathcal{K}} \right\|_{1}, \left\| \boldsymbol{\nu}-\boldsymbol{\nu}_0 \right\|_{1} \leq r \right\},$$ where $\mathbb{S}^{d_{\boldsymbol{W}}-1} = \{ \boldsymbol{u} \in \mathbb{R}^{d_{\boldsymbol{W}}} : \|\boldsymbol{u}\|_2 = 1 \}$ 
    denotes the $(d_{\boldsymbol{W}}-1)$-dimensional unit sphere in $\mathbb{R}^{d_{\boldsymbol{W}}}$.
    By the definition of $\widehat{\boldsymbol{\nu}}_{\varsigma}(\boldsymbol{x})$, we have $\boldsymbol{\zeta}_{\varsigma}(\boldsymbol{x})=\widehat{\boldsymbol{\nu}}_{\varsigma}(\boldsymbol{x})-\boldsymbol{\nu}_0(\boldsymbol{x}) $ falls in $\mathcal{C} (\mathcal{K}, 3, r)$, which is defined in Definition~\ref{LRE} in Subsection~\ref{nuisance lemma}. As long as $n \geq 32^2 k^2 \varpi_{lo}^{-2} s\left(t+\log{d_{\boldsymbol{W}}}\right)$, we have:
    \begin{align*}
         &\; \boldsymbol{\zeta}_{\varsigma}(\boldsymbol{x})^\top \BSigma_{a} \boldsymbol{\zeta}_{\varsigma}(\boldsymbol{x})=\;\boldsymbol{\zeta}_{\varsigma}(\boldsymbol{x})^\top \left(\sum_{i=1}^{n} a_i \BSigma_i\right) \boldsymbol{\zeta}_{\varsigma}(\boldsymbol{x})\\
          \geq &\;\boldsymbol{\zeta}_{\varsigma}(\boldsymbol{x})^\top \mathbb{E} \left\{ \boldsymbol{U}(n,s) \right\} \boldsymbol{\zeta}_{\varsigma}(\boldsymbol{x}) - 2\sqrt{\frac{s\left(t+\log{d_{\boldsymbol{W}}}\right)}{n}}\left\|\boldsymbol{\zeta}_{\varsigma}(\boldsymbol{x})\right\|_1^2 \\
          \geq &\;2\varpi_{lo} \left\|\boldsymbol{\zeta}_{\varsigma}(\boldsymbol{x})\right\|_2^2 - 2\sqrt{\frac{s\left(t+\log{d_{\boldsymbol{W}}}\right)}{n}}\left\|\boldsymbol{\zeta}_{\varsigma}(\boldsymbol{x})\right\|_1^2 \\
          \geq &\;\left\{2\varpi_{lo} - 32k\sqrt{\frac{s\left(t+\log{d_{\boldsymbol{W}}}\right)}{n}}\right\}\left\|\boldsymbol{\zeta}_{\varsigma}(\boldsymbol{x})\right\|_2^2 \\
          \geq &\;\varpi_{lo} \left\|\boldsymbol{\zeta}_{\varsigma}(\boldsymbol{x})\right\|_2^2,
    \end{align*}
that is, $\BSigma_{a}$ satisfies the restricted eigenvalue condition $RE(\mathcal{K},3)$.

    By Lemma~\ref{lemma:re_lre} in Subsection~\ref{nuisance lemma}, together with the condition on sample size and $n \geq c_{2} L_{\boldsymbol{W}}^{4} k^{2} s t$ and conditions on the hyperparameters $ \delta \geq \max \{8 L_{\boldsymbol{W}}r, c_{1} ( L_{\boldsymbol{W}}^{2} k \mu_{\tau} )^{\frac{1}{1+\tau}} \}$,
    $\lambda_{\boldsymbol{\nu}} \geq 2\left\|\sum_{i=1}^{n}a_i \nabla_{\nu} \rho_{\delta}\left\{\boldsymbol{Z}_i;\boldsymbol{\nu}_0(\boldsymbol{x})\right\} \right\|_{\infty}$
    ,
    we have $\boldsymbol{H}_{\delta}\left\{\widehat{\boldsymbol{\nu}}_{\varsigma}(\boldsymbol{x})\right\}$ satisfies the localized restricted eigenvalue condition $LRE(S,3,r)$ with probaility at least $\left(1-e^{-t}\right)^2$,
    i.e. for all $\boldsymbol{\zeta}_{\varsigma}(\boldsymbol{x}) \in \mathcal{C} (\mathcal{K}, 3, r)$, we have
    $\boldsymbol{\zeta}_{\varsigma}(\boldsymbol{x})^\top \boldsymbol{H}_{\delta}\left\{\widehat{\boldsymbol{\nu}}_{\varsigma}(\boldsymbol{x})\right\} \boldsymbol{\zeta}_{\varsigma}(\boldsymbol{x}) \geq \frac{1}{2}\varpi_{lo} \left\|\boldsymbol{\zeta}_{\varsigma}(\boldsymbol{x})\right\|_2^2.$

    Then we derive the final nuisance error bound by comparing two bounds for the same weighted empirical loss difference between the intermediate estimator and the true nuisance parameter.
    
     Since $\boldsymbol{\zeta}_{\varsigma}(\boldsymbol{x}) \in \mathcal{C} (\mathcal{K}, 3, r)$ and the weighted empirical loss satisfies a $\frac{1}{2}\varpi_{lo}$ restricted strong convexity:
    \begin{equation}\label{eq3}
    \begin{aligned}
         &\; \sum_{i=1}^{n}a_i \rho_{\delta}\left\{\boldsymbol{Z}_i;\widehat{\boldsymbol{\nu}}_{\varsigma}(\boldsymbol{x})\right\} - \sum_{i=1}^{n}a_i \rho_{\delta}\left\{\boldsymbol{Z}_i;\boldsymbol{\nu}_0(\boldsymbol{x})\right\}\\
          \geq &\;\sum_{i=1}^{n}a_i \langle \nabla_{\nu} \rho_{\delta}\left\{\boldsymbol{Z}_i;\boldsymbol{\nu}_0(\boldsymbol{x})\right\},\boldsymbol{\zeta}_{\varsigma}(\boldsymbol{x}) \rangle + \boldsymbol{\zeta}_{\varsigma}(\boldsymbol{x})^\top \boldsymbol{H}_{\delta}\left\{\widehat{\boldsymbol{\nu}}_{\varsigma}(\boldsymbol{x})\right\}\boldsymbol{\zeta}_{\varsigma}(\boldsymbol{x}) \\
          \geq &\;-\frac{1}{2}\lambda_{\boldsymbol{\nu}} \left\|\boldsymbol{\zeta}_{\varsigma}(\boldsymbol{x}) \right\|_1 + \frac{1}{2}\varpi_{lo} \left\|\boldsymbol{\zeta}_{\varsigma}(\boldsymbol{x})\right\|_2^2
    \end{aligned}
    \end{equation}
    and
    \begin{equation}\label{eq4}
    \begin{aligned}
        &\;\sum_{i=1}^{n}a_i \rho_{\delta}\left\{\boldsymbol{Z}_i;\widehat{\boldsymbol{\nu}}_{\varsigma}(\boldsymbol{x})\right\} - \sum_{i=1}^{n}a_i \rho_{\delta}\left\{\boldsymbol{Z}_i;\boldsymbol{\nu}_0(\boldsymbol{x})\right\} \\
        \leq &\;\lambda_{\boldsymbol{\nu}} \left\|\widehat{\boldsymbol{\nu}}_{\varsigma}(\boldsymbol{x}) - \boldsymbol{\nu}_0(\boldsymbol{x})\right\|_1
        < \lambda_{\boldsymbol{\nu}} \left\{\left\|\boldsymbol{\zeta}_{\varsigma}(\boldsymbol{x})_{\mathcal{K}}\right\|_1 - \left\|\boldsymbol{\zeta}_{\varsigma}(\boldsymbol{x})_{\mathcal{K}^c}\right\|_1\right\}.
    \end{aligned}
    \end{equation}
    Combining \eqref{eq3} and \eqref{eq4}, we obtain
    \[
        \lambda_{\boldsymbol{\nu}} \left\{\left\|\boldsymbol{\zeta}_{\varsigma}(\boldsymbol{x})_{\mathcal{K}}\right\|_1 - \left\|\boldsymbol{\zeta}_{\varsigma}(\boldsymbol{x})_{\mathcal{K}^c}\right\|_1\right\}
        \geq -\frac{1}{2}\lambda \left\|\boldsymbol{\zeta}_{\varsigma}(\boldsymbol{x}) \right\|_1 + \frac{1}{2}\varpi_{lo} \left\|\boldsymbol{\zeta}_{\varsigma}(\boldsymbol{x})\right\|_2^2.
    \]
    Consequently,
    \begin{align*}
         &\; \varpi_{lo} \left\|\boldsymbol{\zeta}_{\varsigma}(\boldsymbol{x})\right\|_2^2 \\
          \leq &\;\lambda_{\boldsymbol{\nu}} \left\|\boldsymbol{\zeta}_{\varsigma}(\boldsymbol{x}) \right\|_1 + 2\lambda_{\boldsymbol{\nu}} \left\{\left\|\boldsymbol{\zeta}_{\varsigma}(\boldsymbol{x})_{\mathcal{K}}\right\|_1 - \left\|\boldsymbol{\zeta}_{\varsigma}(\boldsymbol{x})_{\mathcal{K}^c}\right\|_1\right\} \\
          = &\;\lambda_{\boldsymbol{\nu}} \left\{\left\|\boldsymbol{\zeta}_{\varsigma}(\boldsymbol{x})_{\mathcal{K}}\right\|_1 + \left\|\boldsymbol{\zeta}_{\varsigma}(\boldsymbol{x})_{\mathcal{K}^c}\right\|_1\right\} + 2\lambda_{\boldsymbol{\nu}} \left\{\left\|\boldsymbol{\zeta}_{\varsigma}(\boldsymbol{x})_{\mathcal{K}}\right\|_1 - \left\|\boldsymbol{\zeta}_{\varsigma}(\boldsymbol{x})_{\mathcal{K}^c}\right\|_1\right\} \\
          = &\;3\lambda_{\boldsymbol{\nu}} \left\|\boldsymbol{\zeta}_{\varsigma}(\boldsymbol{x})_{\mathcal{K}}\right\|_1 - \lambda_{\boldsymbol{\nu}}\left\|\boldsymbol{\zeta}_{\varsigma}(\boldsymbol{x})_{\mathcal{K}^c}\right\|_1 \\
          \leq &\;3\lambda_{\boldsymbol{\nu}} \left\|\boldsymbol{\zeta}_{\varsigma}(\boldsymbol{x})_{\mathcal{K}}\right\|_1
        \leq 3\lambda_{\boldsymbol{\nu}}\sqrt{k} \left\|\boldsymbol{\zeta}_{\varsigma}(\boldsymbol{x})_{\mathcal{K}}\right\|_2
        \leq 3\lambda_{\boldsymbol{\nu}}\sqrt{k} \left\|\boldsymbol{\zeta}_{\varsigma}(\boldsymbol{x})\right\|_2.
    \end{align*}
    Dividing both sides by $\left\|\boldsymbol{\zeta}_{\varsigma}(\boldsymbol{x})\right\|_2$ and combining with the fact that $\left\|\boldsymbol{\zeta}_{\varsigma}(\boldsymbol{x})\right\|_1 \leq 4\sqrt{k}\left\|\boldsymbol{\zeta}_{\varsigma}(\boldsymbol{x})\right\|_2$, we have:
    \[
        \left\| \widehat{\boldsymbol{\nu}}_{\varsigma}(\boldsymbol{x})-\boldsymbol{\nu}_0(\boldsymbol{x})\right\|_1  = \left\|\boldsymbol{\zeta}_{\varsigma}(\boldsymbol{x})\right\|_1 \leq 12k\lambda_{\boldsymbol{\nu}} \varpi_{lo}^{-1} < r.
    \]
    By the construction of $\widehat{\boldsymbol{\nu}}_{\varsigma}(\boldsymbol{x})$, we have $\varsigma=1$, i.e. $\widehat{\boldsymbol{\nu}}_{\varsigma}(\boldsymbol{x}) = \widehat{\boldsymbol{\nu}}(\boldsymbol{x})$, which implies that $\left\| \widehat{\boldsymbol{\nu}}(\boldsymbol{x})-\boldsymbol{\nu}_0(\boldsymbol{x})\right\|_1  \leq 12k\lambda_{\boldsymbol{\nu}} \varpi_{lo}^{-1}$. This completes the proof of part (ii) of Theorem~\ref{thm:nuisance_properties}..
\end{proof}

\noindent\textbf{Proof of Theorem~\ref{thm:nuisance_properties} (iii):}
\begin{proof}
Now we consider $\mathcal{E}(\widehat{\boldsymbol{h}})^2=\mathbb{E}\left\{\left\|\widehat{\boldsymbol{\nu}}(\boldsymbol{x})-\boldsymbol{\nu}_0(\boldsymbol{x})\right\|^2+\left\|\widehat{\boldsymbol{\ell}}(\boldsymbol{x})-\boldsymbol{\ell}_0(\boldsymbol{x})\right\|^2\right\}.$ By the norm inequlity
$$
\mathcal{E}(\widehat{\boldsymbol{h}})^2 \le  \mathbb{E}\left\{\left\|\widehat{\boldsymbol{\nu}}(\boldsymbol{x}) - \boldsymbol{\nu}_0(\boldsymbol{x})\right\|_1^2\right\} + \mathbb{E}\left\{\left\|\widehat{\boldsymbol{\ell}}(\boldsymbol{x}) - \boldsymbol{\ell}_0(\boldsymbol{x})\right\|_1^2\right\}.
$$

By the conclusion in Theorem~\ref{thm:nuisance_properties} (ii), with probability at least $1-e^{-t}$, let $\iota = \min\left\{\frac{\tau}{1+\tau}, \frac{1}{2}\right\}$, we have 
$\left\|\widehat{\boldsymbol{\nu}}(\boldsymbol{x}) - \boldsymbol{\nu}_0(\boldsymbol{x})\right\|_1 \leq C_1 k\left\{\frac{s\left(t+\log{d_{\boldsymbol{W}}}\right)}{n}\right\}^{\iota}$. 
To simplify the presentation in the high-dimensional setting, we follow the scaling regime specified in Assumption~\ref{assumption:scaling}.
In this setting, as $n \to \infty$, 
the term $\frac{s \log d_{\boldsymbol{W}}}{n}$ becomes negligible 
relative to $\frac{s t}{n}$. Then we obtain:
\[
    \mathbb{P}\left\{\left\|\widehat{\boldsymbol{\nu}}(\boldsymbol{x}) - \boldsymbol{\nu}_0(\boldsymbol{x})\right\|_1 \geq C_1 k\left(\frac{s t}{n}\right)^{\iota}\right\} \leq e^{-t}.
\]
By using the substitution $u = C_1 k \left(\frac{s t}{n}\right)^{\iota}$, we have:
\[
    \mathbb{P}\left\{\left\|\widehat{\boldsymbol{\nu}}(\boldsymbol{x}) - \boldsymbol{\nu}_0(\boldsymbol{x})\right\|_1 \geq u\right\} \leq \exp\left\{-\frac{n}{s}\left(\frac{u}{C_1 k}\right)^{1/\iota} \right\}.
\]
For any constant integer $p \geq 1$, we have
\begin{align*}
    &\; \mathbb{E}\left\{\left\|\widehat{\boldsymbol{\nu}}(\boldsymbol{x}) - \boldsymbol{\nu}_0(\boldsymbol{x})\right\|_1^{2p}\right\} \\
     = &\; \int_0^{\infty} 2p u^{2p-1} \mathbb{P}\left\{\left\|\widehat{\boldsymbol{\nu}}(\boldsymbol{x}) - \boldsymbol{\nu}_0(\boldsymbol{x})\right\|_1 \geq u\right\} du \\
     \leq &\; \int_0^{\infty} 2p u^{2p-1} \exp\left\{-\frac{n}{s}\left(\frac{u}{C_1 k}\right)^{1/\iota} \right\} du .
\end{align*}
Substituting $v = \frac{n}{s}\left(\frac{u}{C_1 k}\right)^{1/\iota}$, 
we have:
\begin{align*}
    &\mathbb{E}\left\{\left\|\widehat{\boldsymbol{\nu}}(\boldsymbol{x}) - \boldsymbol{\nu}_0(\boldsymbol{x})\right\|_1^{2p}\right\}
     \leq \int_0^{\infty} 2p C_1^{2p} k^{2p}\iota \left(\frac{s}{n}\right)^{2p\iota} v^{2p\iota - 1} e^{-v} dv \\
     = &\;2r C_1^{2p} k^{2p} \iota \left(\frac{s}{n}\right)^{2p\iota} \Gamma(2p\iota) 
     = O\left(k^{2p}\left(\frac{s}{n}\right)^{2p\iota}\right).
\end{align*}
Then we have $\mathbb{E}\left[\left\|\widehat{\boldsymbol{\nu}}(\boldsymbol{x}) - \boldsymbol{\nu}_0(\boldsymbol{x})\right\|_1^{p}\right]^{1/p} = O\left(k\left(\frac{s}{n}\right)^{\iota}\right)$.
Given that $\tau > 1/3$ (which implies $\iota > 1/4$), invoking the scaling condition on $k$ in Assumption~\ref{assumption:scaling} ensures that the estimation error satisfies
\begin{equation*}
    \mathbb{E}\left\{\left\|\widehat{\boldsymbol{\nu}}(\boldsymbol{x}) - \boldsymbol{\nu}_0(\boldsymbol{x})\right\|_1^{p}\right\}^{1/p}  = o\left((s/n)^{1/4}\right).
\end{equation*}
Analogously, we have $\mathbb{E}\left\{\left\|\widehat{\boldsymbol{\ell}}(\boldsymbol{x}) - \boldsymbol{\ell}_0(\boldsymbol{x})\right\|_1^{2p}\right\}^{1/p}=o\left((s/n)^{1/4}\right)$.
Then we have: 
\begin{align*}
    \mathbb{E}\left\{\mathcal{E}(\widehat{\boldsymbol{h}})^{p}\right\}^{1/p} &\leq\mathbb{E}\left\{\left\|\widehat{\boldsymbol{\nu}}(\boldsymbol{x}) - \boldsymbol{\nu}_0(\boldsymbol{x})\right\|_1^{p}\right\}^{1/p} +\mathbb{E}\left\{\left\|\widehat{\boldsymbol{\ell}}(\boldsymbol{x}) - \boldsymbol{\ell}_0(\boldsymbol{x})\right\|_1^{p}\right\}^{1/p} \\ 
    \;&=o\left((s/n)^{1/4}\right).
\end{align*}

\end{proof}

\subsection{Auxiliary Lemmas for Nuisance Estimation}\label{nuisance lemma}
For clarity of presentation, the proof of Theorem~\ref{thm:nuisance_properties} in Subsection~\ref{Nuisance proof} relies on several auxiliary results whose statements and proofs are provided in this subsection.

We first recall the localized restricted eigenvalue condition of  \cite{sun2020adaptive}, which is used to prove Part~(ii) of Theorem~\ref{thm:nuisance_properties}.
Recall that $\mathcal{K}(\boldsymbol{\nu}_0,\boldsymbol{x}) \subset \{1, \dots, d_{\boldsymbol{W}}\}$ denote the support set of the true parameter vector $\boldsymbol{\nu}_0(\boldsymbol{x})$, with cardinality $|\mathcal{K}(\boldsymbol{\nu}_0,\boldsymbol{x})| = k$. 
For simplicity, we denote $\boldsymbol{\nu}_0(\boldsymbol{x})$ as  $\boldsymbol{\nu}_0$, $\widehat{\nu}(\boldsymbol{x})$ as  $\widehat{\nu}$ and $\mathcal{K}(\boldsymbol{\nu}_0,\boldsymbol{x})$ as $\mathcal{K}$ in the following definitions and conditions. The localized restricted eigenvalue (LRE) is defined in a local neighborhood of $\boldsymbol{\nu}_0$ under $L_1$-norm.

\begin{definition}\label{LRE}
(Localized Restricted Eigenvalue, LRE). Recall  $\mathcal{C} ( \mathcal{K}, \gamma, r )$ is defined as $$\mathcal{C} ( \mathcal{K}, \gamma, r ) :=\left\{( \boldsymbol{u},\boldsymbol{\nu}) \in\mathbb{S}^{d_{\boldsymbol{W}}-1} \times \mathbb{R}^{d_{\boldsymbol{W}}} : \left\|\boldsymbol{u}_{\mathcal{K}^{c}} \right\|_{1} \leq\gamma\left\| \boldsymbol{u}_{\mathcal{K}} \right\|_{1}, \left\| \boldsymbol{\nu}-\boldsymbol{\nu}_0 \right\|_{1} \leq r \right\}$$ with $\mathbb{S}^{d_{\boldsymbol{W}}-1} = \{ \boldsymbol{u} \in \mathbb{R}^{d_{\boldsymbol{W}}} : \|\boldsymbol{u}\|_2 = 1 \}$ 
    denotes the $(d_{\boldsymbol{W}}-1)$-dimensional unit sphere in $\mathbb{R}^{d_{\boldsymbol{W}}}$.
    The LREs of $\boldsymbol{H}_{\delta}(\boldsymbol{\nu})$ with respect to the support set $\mathcal{K}$ are defined as
    \begin{align*}
        \kappa_{+} (\boldsymbol{H}_{\delta}, \mathcal{K}, \gamma, r )  =\sup & \left\{\boldsymbol{u}^{\top}\boldsymbol{H}_{\delta} ( \boldsymbol{\nu} ) \boldsymbol{u} : ( \boldsymbol{u}, \boldsymbol{\nu} ) \in\mathcal{C} ( \mathcal{K}, \gamma, r ) \right\}, \\
        \kappa_{-} (\boldsymbol{H}_{\delta}, \mathcal{K}, \gamma, r )  =\inf & \left\{\boldsymbol{u}^{\top}\boldsymbol{H}_{\delta} ( \boldsymbol{\nu} ) \boldsymbol{u} : ( \boldsymbol{u}, \boldsymbol{\nu} ) \in\mathcal{C} ( \mathcal{K}, \gamma, r ) \right\},
    \end{align*}
     
\end{definition}

Here, $\gamma$ controls the size of the cone around the true support set $\mathcal{K}$. A smaller $\gamma$ imposes a stronger sparsity-alignment condition, while a larger $\gamma$ allows the vector $\boldsymbol{u}$  to have more mass outside $\mathcal{K}$. In the proof of Theorem~\ref{thm:nuisance_properties} (ii), we show that the estimation error of the nuisance estimator satisfies the cone condition with $\gamma=3$; that is, $\left\|\left(\widehat{\boldsymbol{\nu}}-\boldsymbol{\nu}_0\right)_{\mathcal{K}^c}\right\|_1\leq 3\left\|\left(\widehat{\boldsymbol{\nu}}-\boldsymbol{\nu}_0\right)_{\mathcal{K}}\right\|_1.$
The parameter $r$ specifies a local neighborhood of the true nuisance parameter $\boldsymbol{\nu}_0$.  Once we also establish that $\widehat{\boldsymbol{\nu}}$ satisfy $\left\|\widehat{\boldsymbol{\nu}}-\boldsymbol{\nu}_0\right\|\le r$, for some $r>0$, the localized restricted eigenvalue condition can be applied to the nuisance estimation error.

\begin{condition}
    \label{condition:lre}
    For any $\gamma,r>0$, the matrix $\boldsymbol{H}_\delta$ satisfies the localized restricted eigenvalue condition $\mathrm{LRE} ( \mathcal{K}, \gamma, r )$, that is, $\varphi_{lo} \leq\kappa_{-} (\boldsymbol{H}_{\delta}, \mathcal{K}, \gamma, r ) \leq\kappa_{+} ( \boldsymbol{H}_{\delta},\mathcal{K}, \gamma, r ) \leq\varphi_{up}$ for some constants $\varphi_{up}, \varphi_{lo} > 0$.
\end{condition}
The condition above is referred to as the LRE condition \citep{sun2020adaptive}.
It is a unified condition for studying generalized loss functions, whose Hessians may possibly depend on $\boldsymbol{\nu}$.
For Huber loss, Condition \ref{condition:lre} also involves the observation noise.
However, $\boldsymbol H_\delta(\boldsymbol\nu)$ depends not only on the design variables but also on the observation noise through the Huber truncation. Therefore, it is not directly transparent how to verify Condition~\ref{condition:lre} from assumptions on the covariates alone. The following definition concerns the restricted eigenvalues of weighted covariance matrix $\BSigma_a = \sum_{i=1}^n a_i\boldsymbol{W}_i\boldsymbol{W}_i^{\top}$ instead of $\boldsymbol{H}_{\delta}$ and Condition~\ref{condition:re} provides a more primitive restricted eigenvalue condition on the weighted covariance matrix $\BSigma_a$, which only involves the design.

\begin{definition}(Restricted Eigenvalue, RE).
    The restricted maximum and minimum eigenvalues of $\BSigma_{a}$ with respect to the support set $\mathcal{K}$ are defined as
    \begin{align*}
        \upsilon_{+} ( \mathcal{K}, \gamma)  =\sup & \left\{\boldsymbol{u}^{\top}\BSigma_{a} \boldsymbol{u} : \boldsymbol{u} \in\mathcal{C} ( \mathcal{K}, \gamma) \right\}, \\
        \upsilon_{-} ( \mathcal{K}, \gamma)  =\inf & \left\{\boldsymbol{u}^{\top}\BSigma_{a} \boldsymbol{u} : \boldsymbol{u} \in\mathcal{C} ( \mathcal{K}, \gamma) \right\},
    \end{align*}
    where $\mathcal{C} ( \mathcal{K}, \gamma) :=\left\{ \boldsymbol{u} \in\mathbb S^{d_{\boldsymbol{W}}-1} : \left\| \boldsymbol{u}_{\mathcal{K}^{c}} \right\|_{1} \leq\gamma\left\| \boldsymbol{u}_{\mathcal{K}} \right\|_{1} \right\}$.
\end{definition}

\begin{condition}
    \label{condition:re}
   For any $\gamma,r>0$, the matrix ${\BSigma}_{a}$ satisfies the restricted eigenvalue condition $\mathrm{RE} ( \mathcal{K}, \gamma)$ , that is, $\chi_{lo} \leq \upsilon_{-} ( \mathcal{K}, \gamma) \leq\upsilon_{+} ( \mathcal{K}, \gamma) \leq\chi_{up}$ for some constants $\chi_{up}, \chi_{lo} > 0$ .
\end{condition}

The subsequent lemmas bridge these two conditions: they show that, under the moment condition on the errors and an appropriate choice of the Huber parameter $\delta$, the random curvature matrix $\boldsymbol{H}_{\delta}(\boldsymbol \nu)$ is uniformly close to the weighted covariance matrix $\BSigma_a$ over the localized cone $C(\boldsymbol{H}_{\delta}, \BSigma,\gamma,r)$. Consequently, the restricted eigenvalue bounds for $\BSigma_a$ in Condition~\ref{condition:re} imply the localized restricted eigenvalue bounds for $\boldsymbol{H}_{\delta}(\nu)$ in Condition~\ref{condition:lre} with high probability. In this sense, Condition~\ref{condition:lre} is the curvature condition used in the proof, while Condition~\ref{condition:re} is a more verifiable sufficient condition. Recall that $\tau$ is a constant mentioned in Section \Sref{sec:model_setup}{2.1} when introduce $\epsilon_i$, we write $\mu_{i,\tau} = \mathbb{E}\left(|\epsilon_i|^{1+\tau}\right)$, $\mu_{\tau} = \sum_{i=1}^n a_i \mu_{i,\tau}$.
\begin{lemma}
    \label{lemma:re_lre}
    Provided that
    $ \delta \geq \max \{8 L_{\boldsymbol{W}} r, c_{1} ( L_{\boldsymbol{W}}^{2} k \mu_{\tau} )^{\frac{1}{1+\tau}} \}$ and $n \geq c_{2} L_{\boldsymbol{W}}^{4} k^{2} s t$, where $c_1, c_2$ are constants depending only on $( \gamma, \chi_{lo} )$, Condition~\ref{condition:re} implies  Condition~\ref{condition:lre} with high probability. Specifically, if $0 < \chi_{lo} \leq \upsilon_{-} ( \mathcal{K}, \gamma) \leq \upsilon_{+} ( \mathcal{K}, \gamma) \leq \chi_{up} < \infty$ for some $\gamma > 0$ , then with probability at least $1-e^{-t}$, $0 < \chi_{lo} / 2 \leq \kappa_{-} ( \boldsymbol{H}_{\delta}, \mathcal{K}, \gamma, r ) \leq \kappa_{+} ( \boldsymbol{H}_{\delta}, \mathcal{K}, \gamma, r ) \leq \chi_{up} < \infty$. 
\end{lemma}
\begin{proof}

    For all $( \boldsymbol{u}, \boldsymbol{\nu} ) \in\mathcal{C} ( \mathcal{K}, \gamma, r )$ we have
    \begin{align*}
         &\; \langle \boldsymbol{u}, \boldsymbol{H}_{\delta} ( \boldsymbol{\nu} ) \boldsymbol{u} \rangle \\
          =&\;\sum_{i=1}^{n} a_{i} \boldsymbol{u}^{\top} \boldsymbol{\boldsymbol{W}}_{i} \boldsymbol{\boldsymbol{W}}_{i}^{\top} \boldsymbol{u} I \left\{|Y_i- \langle \boldsymbol{\nu}(\boldsymbol{x}),\boldsymbol{\boldsymbol{W}}_i \rangle| \leq \delta \right\} \\
          \geq &\;
          \begin{aligned}[t]
          & \sum_{i=1}^{n} a_{i} \langle \boldsymbol{u}, \BSigma_i \boldsymbol{u} \rangle - \sum_{i=1}^{n} a_{i} \langle \boldsymbol{u}, \BSigma_i \boldsymbol{u} \rangle I \left\{|\langle \boldsymbol{\nu}(\boldsymbol{x})-\boldsymbol{\nu}_0(\boldsymbol{x}), \boldsymbol{\boldsymbol{W}}_i \rangle| \geq \frac{\delta}{2} \right\} \\
          &-\sum_{i=1}^{n} a_{i} \langle \boldsymbol{u}, \BSigma_i \boldsymbol{u} \rangle^2 I \left(|\epsilon_i| \geq \frac{\delta}{2} \right)
          \end{aligned}
          \\
          \geq &\;
          \begin{aligned}[t]
          & \sum_{i=1}^{n} a_{i} \langle \boldsymbol{u}, \BSigma_i \boldsymbol{u} \rangle - \frac{2r}{\delta} \max_{1 \leq i \leq n} \left\|\boldsymbol{\boldsymbol{W}}_i\right\|_{\infty}\left( \sum_{i=1}^{n} a_{i} \langle \boldsymbol{u}, \BSigma_i \boldsymbol{u} \rangle \right) \\
          & - \max_{1 \leq i \leq n} \langle \boldsymbol{u}, \BSigma_i \boldsymbol{u} \rangle^2 \sum_{i=1}^{n} a_{i} I \left(|\epsilon_i| \geq \frac{\delta}{2} \right).
          \end{aligned}
    \end{align*}
    As $\left\|\boldsymbol{\boldsymbol{W}}_i\right\|_{\infty} \leq L_{\boldsymbol{W}}$ for any $1 \leq i \leq n$, we have
    \[
        |\langle \boldsymbol{u}, \boldsymbol{\boldsymbol{W}}_i \rangle|
        \leq \left\|\boldsymbol{u}\right\|_1 \left\|\boldsymbol{\boldsymbol{W}}_i\right\|_{\infty}
        \leq L_{\boldsymbol{W}}(1+\gamma) \left\|\boldsymbol{u}_\mathcal{K}\right\|_1
        \leq L_{\boldsymbol{W}}\sqrt{k}(1+\gamma).
    \]
    Moreover, we have
    \[
        \sum_{i=1}^{n} a_{i}^2 \leq \max_{1 \leq i \leq n} a_i \sum_{i=1}^{n} a_{i}  = \max_{1 \leq i \leq n} a_i \leq \frac{s}{n},
    \]
    then for any $t > 0$, applying Hoeffding's inequality yields that,
    with probability at least $1-e^{-t}$
    \[
        \sum_{i=1}^{n} a_{i} I \left(|\epsilon_i| \geq \frac{\delta}{2} \right)
        \leq \sqrt{\frac{\sum_{i=1}^n a_i^2 t }{2}} + \left(\frac{2}{\delta} \right)^{(1+\tau)}\mu_{\tau}
        \leq \sqrt{\frac{st}{2n}} + \left(\frac{2}{\delta} \right)^{(1+\tau)}\mu_{\tau}.
    \]
    Putting the above calculation together, we obtain
    \[
        \langle \boldsymbol{u}, \boldsymbol{H}_{\delta} ( \boldsymbol{\nu} ) \boldsymbol{u} \rangle
        \geq \left(1-2rL_{\boldsymbol{W}} \delta^{-1}\right)\langle \boldsymbol{u}, \BSigma_a \boldsymbol{u} \rangle - kL_{\boldsymbol{W}}^2(1+\gamma^2)\left(n^{-1/2}\sqrt{st/2} +2^{1+\tau}\delta^{-1-\tau}\mu_{\tau}\right).
    \]
    As long as $\delta \geq \max\{8L_{\boldsymbol{W}}r, c_1(L_{\boldsymbol{W}}^2k\mu_{\tau})^{1/(1+\tau)}\}$ and $n \geq c_2L_{\boldsymbol{W}}^4k^2st$,
    where $c_1 = \left(1+\gamma\right)^{2/(1+\tau)} \chi_{lo}^{-1/(1+\tau)}$ and $c_2 = \left(1+\gamma\right)^4 \chi_{lo}^{-2}$,
    we have
    \[
        \langle \boldsymbol{u}, \boldsymbol{H}_{\delta} ( \boldsymbol{\nu} ) \boldsymbol{u} \rangle
        \geq  \frac{3}{4}\chi_{lo} -kL_{\boldsymbol{W}}^2(1+\gamma)^2\left\{\sqrt{\frac{st}{2n}}+\left(\frac{2}{\delta}\right)^{1+\tau}\mu_{\tau}\right\}  > \frac{1}{2}\chi_{lo}.
    \]
    On the other side, it can be easily shown that 
    \begin{align*}
      \langle \boldsymbol{u}, \boldsymbol{H}_{\delta} ( \boldsymbol{\nu} ) \boldsymbol{u} \rangle
          &=\;\sum_{i=1}^{n} a_{i} \boldsymbol{u}^{\top} \boldsymbol{\boldsymbol{W}}_{i} \boldsymbol{\boldsymbol{W}}_{i}^{\top} \boldsymbol{u} I \left\{|Y_i- \langle \boldsymbol{\nu}(\boldsymbol{x}),\boldsymbol{\boldsymbol{W}}_i \rangle| \leq \delta \right\}\\
          &\leq\;\sum_{i=1}^{n} a_{i} \boldsymbol{u}^{\top} \boldsymbol{\boldsymbol{W}}_{i} \boldsymbol{\boldsymbol{W}}_{i}^{\top} \boldsymbol{u} = \langle \boldsymbol{u}, \BSigma_a \boldsymbol{u} \rangle \leq \chi_{up}.
    \end{align*}
    This completes the proof of the lemma.
\end{proof}

We then state a lemma to bound the gradient of the weighted empirical loss function.
\begin{lemma}\label{lemma:gradient}
    As long as the penalty level $\lambda_{\boldsymbol{\nu}} \geq 4\delta_{0}L_{\boldsymbol{W}}\left\{\frac{s\left(t+\log{d_{\boldsymbol{W}}}\right)}{n}\right\}^{\min{\left(\frac{\tau}{1+\tau}, \frac{1}{2}\right)}}$, the gradient of the weighted empirical loss function can be bounded by $\left\|\sum_{i=1}^{n}a_i \nabla_{\nu} \rho_{\delta}\left\{\boldsymbol{Z}_i;\boldsymbol{\nu}_0(\boldsymbol{x})\right\} \right\|_{\infty}\leq \lambda_{\boldsymbol{\nu}}/2$.
\end{lemma}

\begin{proof}
    Since $\nabla \mathcal{L}_{\delta}(\boldsymbol{\nu}_0(\boldsymbol{x})) = \sum_{i=1}^{n} a_i \phi_{\delta}(\epsilon_i)\boldsymbol{\boldsymbol{W}}_i$. Let $\mathcal{T} _j = \sum_{i=1}^{n} \frac{1}{L_{\boldsymbol{W}}} a_i \phi_1\left(\frac{\epsilon_i}{\delta}\right)W_{ij}, j=1, \dots, d_{\boldsymbol{W}}$.
    Then we have
    \[
        \left\|\nabla \mathcal{L}_{\delta}(\boldsymbol{\nu}_0(\boldsymbol{x}))\right\|_2 \leq \sqrt{d_{\boldsymbol{W}}}\left\|\nabla \mathcal{L}_{\delta}(\boldsymbol{\nu}_0(\boldsymbol{x}))\right\|_{\infty} = L_{\boldsymbol{W}}\sqrt{d_{\boldsymbol{W}}}\delta \max_{1\leq j \leq d_{\boldsymbol{W}}}|\mathcal{T} _j|
    \]
    and
    \[
        \exp{\left\{\left(\frac{W_{ij}}{L_{\boldsymbol{W}}}\right) \phi_1\left(\frac{\epsilon_i}{\delta}\right)\right\}} \leq 1 + \left(\frac{W_{ij}}{L_{\boldsymbol{W}}}\right) \left(\frac{\epsilon_i}{\delta}\right) + \left|\frac{\epsilon_i}{\delta}\right|^{1+\tau}.
    \]
Therefore, 
    \begin{align*}
         &\mathbb{E}\left\{\exp{\left(s^{-1}n\mathcal{T} _j\right)}\right\} =\prod_{i=1}^{n}\mathbb{E}\left[\exp\left\{{\frac{n a_iW_{ij}\phi_1\left(\frac{\epsilon_i}{\delta}\right)}{sL_{\boldsymbol{W}}}}\right\}\right] \\
          \leq& \prod_{i=1}^{n}\left(1 + \mu_{i,\tau}\delta^{-1-\tau}\right)^{s^{-1}n a_i}\leq \prod_{i=1}^{n} \exp{\left(s^{-1}n a_i\mu_{i,\tau}\delta^{-1-\tau}\right)}=\exp{\left(s^{-1}n\mu_{\tau}\delta^{-1-\tau}\right)}.
    \end{align*}
    By the above calculation, the probability $\mathbb{P}\left(\mathcal{T} _j \geq \mu_{\tau}h\right)$ can be upper bounded by
    \[
        \mathbb{P}\left(\mathcal{T} _j \geq \mu_{\tau}h\right) \leq \frac{\mathbb{E}\left\{\exp\left(s^{-1}n\mathcal{T} _j\right)\right\}}{\exp\left(s^{-1}n\mu_{\tau}h\right)} \leq \exp\left\{s^{-1}n\mu_{\tau}(\delta^{-1-\tau}-h)\right\}.
    \]
    As long as $\delta \geq \left(\frac{2}{h}\right)^{\frac{1}{1+\tau}}$, we have
    \[
        \mathbb{P}\left(\mathcal{T} _j \geq \mu_{\tau}h\right) \leq \exp{\left\{-\frac{1}{2} s^{-1}n\mu_{\tau}h\right\}}.
    \]
 Let $h = \frac{2s\left(t + \log{d_{\boldsymbol{W}}}\right)}{n\mu_{\tau}}$. By the uniform bound above, as long as $\delta \geq \left\{\frac{n\mu_{\tau}}{s\left(t+\log{d_{\boldsymbol{W}}}\right)}\right\}^{\max{\left\{\frac{1}{1+\tau}, \frac{1}{2}\right\}}}$, we have
    \begin{align*}
         &\; \mathbb{P}\left[ \left\|\nabla \mathcal{L}_{\delta}\left\{\boldsymbol{\nu}_0(\boldsymbol{x})\right\}\right\|_2 \geq 2L_{\boldsymbol{W}}\sqrt{d_{\boldsymbol{W}}}\delta \left(t + \log{d_{\boldsymbol{W}}}\right) s n^{-1} \right] \\
          \leq &\;\mathbb{P}\left[ \left\|\nabla \mathcal{L}_{\delta}\left\{\boldsymbol{\nu}_0(\boldsymbol{x})\right\}\right\|_{\infty} \geq 2L_{\boldsymbol{W}}\delta \left(t + \log{d_{\boldsymbol{W}}}\right) s n^{-1} \right]            \\
          \leq &\;\sum_{j=1}^{d_{\boldsymbol{W}}} \mathbb{P}\left\{\lvert\mathcal{T} _j \rvert \geq 2\left(t + \log{d_{\boldsymbol{W}}}\right) s n^{-1}\right\}            \\
          \leq &\;d_{\boldsymbol{W}}\exp{\left(-t -\log{d_{\boldsymbol{W}}}\right)}
        = \exp{\left(-t\right)}.
    \end{align*}
Taking $\delta \geq \delta_{0} \left\{\frac{n}{s\left(t + \log{d_{\boldsymbol{W}}}\right)}\right\}^{\max{\left\{\frac{1}{1+\tau}, \frac{1}{2}\right\}}} $
    with $\delta_{0} \geq \mu_{\tau}^{\max{\left\{\frac{1}{1+\tau}, \frac{1}{2}\right\}}}$, we have
    \[
        \mathbb{P}\left[\left\|\nabla \mathcal{L}_{\delta}\left\{\boldsymbol{\nu}_0(\boldsymbol{x})\right\}\right\|_{\infty} \geq 2L_{\boldsymbol{W}}\delta \left(t + \log{d_{\boldsymbol{W}}}\right) s n^{-1}\right] \leq \exp{(-t)}.
    \]
Finally, taking $t = c\log{d_{\boldsymbol{W}}}$ for some $c>0$ yields that with probability at least $1-d_{\boldsymbol{W}}^{-c}$, we have
$$\left\|\nabla \mathcal{L}_{\delta}\left\{\boldsymbol{\nu}_0(\boldsymbol{x})\right\}\right\|_{\infty} \leq 2L_{\boldsymbol{W}}\delta s \left(1+c\right)\log{d_{\boldsymbol{W}}}/n.$$
Note that $\lambda_{\boldsymbol{\nu}} \geq 4\delta_{0}L_{\boldsymbol{W}}L_{\boldsymbol{W}}\left\{\frac{s\left(t+\log{d_{\boldsymbol{W}}}\right)}{n}\right\}^{\min{\left(\frac{\tau}{1+\tau}, \frac{1}{2}\right)}}$, the lemma is proved. 
\end{proof}

\section{Proofs for Treatment Effect Estimation}\label{appendix:proof_treatment}

This section establishes the theoretical results for treatment effect estimation. Section~\ref{score properties} verifies the higher-order conditions for the orthogonal score, including orthogonality, identifiability and other related regularity properties. Section~\ref{TE proof} proves the main theoretical results for the treatment effect estimator. Section~\ref{TE lemma} provides the auxiliary lemmas used in Section~\ref{TE proof}, together with their proofs.

\subsection{Properties of the Orthogonal Score Function and Their Proofs}\label{score properties}

This subsection establishes auxiliary properties of the orthogonal score function used for treatment effect estimation. We begin by studying the population moment induced by the robust orthogonal score. Specifically, define the moment function
$\boldsymbol{m}_{\delta}(\boldsymbol{x}; \boldsymbol{\theta}, \boldsymbol{h}) = \mathbb{E}\left\{\boldsymbol{\psi}_{\delta}(\boldsymbol{Z}_i; \boldsymbol{\theta}, \boldsymbol{h}) \mid \boldsymbol{X}_i = \boldsymbol{x}\right\}$. In the following, we will write $\nabla_{\boldsymbol{h}} \boldsymbol{\psi}_{\delta}$ to denote the gradient of $\boldsymbol{\psi}_{\delta}$ with respect to the nuisance argument.

The following lemma characterizes the key properties of this moment function and provides the foundation for the subsequent analysis of $\widehat{\boldsymbol{\theta}}(\boldsymbol{x})$. These properties are used later to establish the local behavior of the estimating equation and to derive the convergence rate of the proposed treatment effect estimator.
\begin{lemma}
    \label{lemma:moment_condition}
    Under Assumptions~\Sref{assumption:TE_Lipschitz}1-\Sref{assumption:nuisance error}3, the moment condition and the score function satisfy the following: 
    \begin{enumerate}
        \item \textbf{Local Orthogonality.}  The moment $\boldsymbol{m}_{\delta}(\boldsymbol{x}; \boldsymbol{\theta}, \boldsymbol{h})$ satisfies 
        \[
        \mathbb{E} \left\{ \nabla_{\boldsymbol{h}} \boldsymbol{\psi}_{\delta}( \boldsymbol{Z}; \boldsymbol{\theta}_{0}, \boldsymbol{h}_{0}) \mid \boldsymbol{x}, \boldsymbol{W} \right\}=\mathbf{0}.
        \]
        \item \textbf{Identifiability.}  The moments $\boldsymbol{m}_{\delta}(\boldsymbol{x}; \boldsymbol{\theta}, \boldsymbol{h}_0) = \mathbf{0}$ has a unique solution $\boldsymbol{\theta}_0(\boldsymbol{x})$.
        \item \textbf{Smooth Signal.}  Let $\boldsymbol{\Theta}\subset\mathbb{R}^{d_{\boldsymbol T}}$ denote the parameter space for the treatment effect $\boldsymbol{\theta}$ and $\boldsymbol{\mathcal{H}}$ denote the nuisance parameter space consisting of $\boldsymbol h=(\boldsymbol\nu,\boldsymbol\ell)$, where $\boldsymbol\nu\in\mathbb{R}^{d_{\boldsymbol W}}$ and $\boldsymbol\ell\in\mathbb{R}^{d_{\boldsymbol W}\times d_{\boldsymbol T}}$. The parameter space $\boldsymbol{\Theta}$ has constant diameter, $\text{diam}(\boldsymbol{\Theta})\le R$. The moments $\boldsymbol{m}_{\delta}(\boldsymbol{x}; \boldsymbol{\theta}, \boldsymbol{h})$ are $O(1)-$Lipschitz in $\boldsymbol{x}$ for any $\boldsymbol{\theta} \in \boldsymbol\Theta, \boldsymbol{h} \in \boldsymbol{\mathcal{H}}$. 
        \item \textbf{Curvature.} Let $\boldsymbol{M}_{\delta} \equiv -\nabla_{\boldsymbol{\theta}} \mathbb{E}\left[\boldsymbol{\psi}_{\delta}\left\{\boldsymbol{Z}_i; \boldsymbol{\theta}, \boldsymbol{h}_0(\boldsymbol{x})\right\}\mid \boldsymbol X=\boldsymbol x\right]\mid_{\boldsymbol{\theta}=\boldsymbol{\theta}_0(\boldsymbol{x})}$. Its minimum eigenvalue is bounded away from zero.

        \item \textbf{Smoothness of scores.} For every $j \in \{ 1,\dots,d_{\boldsymbol{T}} \}$, the expected Hessian matrix $\boldsymbol{H}_j(\boldsymbol{\theta}, \boldsymbol{h}) = \mathbb{E}\left\{\nabla^2_{(\boldsymbol{\theta},\boldsymbol{h})} \boldsymbol{\psi}_{\delta}^{(j)}(\boldsymbol{Z};\boldsymbol{\theta},\boldsymbol{h}) \mid \boldsymbol{x},\boldsymbol{W}\right\}$ has eigenvalues bounded above by a constant $O(1)$ for all $(\boldsymbol{\theta}, \boldsymbol{h})$.
              For any observation $\boldsymbol{Z}$, the score $\boldsymbol{\psi}_{\delta}(\boldsymbol{Z};\boldsymbol{\theta}, \boldsymbol{h})$ is $O(1)$-Lipschitz in $\boldsymbol{h}$ for any $\boldsymbol{\theta}$.The gradient of the score with respect to $\boldsymbol{\theta}$ is $ O(1)$-Lipschitz in $\boldsymbol{h}$.

\end{enumerate}

\end{lemma}
\begin{proof}
\noindent \textbf{Local orthogonality.}
    Since $\boldsymbol{h}$ consists of two nuisance components, $\boldsymbol{h}=\left(\boldsymbol{\nu},\boldsymbol{\ell}\right)$, the derivative with respect to $\boldsymbol{h}$ is understood componentwise. Specifically, we write
    \begin{equation*}
    \begin{aligned}
        &\quad\mathbb{E} \left\{ \nabla_{\boldsymbol{h}} \boldsymbol{\psi}_{\delta}( \boldsymbol{Z}; \boldsymbol{\theta}_{0}, \boldsymbol{h}_{0}) \mid \boldsymbol{W}, \boldsymbol{x} \right\} \\
        &=\left[\mathbb{E}\left\{ \nabla_{\boldsymbol{\nu}} \boldsymbol{\psi}_{\delta}\left( \boldsymbol{Z}; \boldsymbol{\theta}_{0}, (\boldsymbol{\nu},\boldsymbol{\ell}_0)\right) \mid \boldsymbol{W}, \boldsymbol{x} \right\}\mid_{\boldsymbol{\nu}=\boldsymbol{\nu}_0}, \mathbb{E}\left\{ \nabla_{\boldsymbol{\ell}} \boldsymbol{\psi}_{\delta}\left( \boldsymbol{Z}; \boldsymbol{\theta}_{0}, (\boldsymbol{\nu}_0,\boldsymbol{\ell})\right) \mid \boldsymbol{W}, \boldsymbol{x} \right\}\mid_{\boldsymbol{\ell}=\boldsymbol{\ell}_0}\right]).
    \end{aligned}
    \end{equation*}
   Recall the notations
    \begin{align*}
         & \widetilde{Y}(\boldsymbol{\nu}_{0}) = Y - \Bnu_0(\Bx)^{\top}\BW,\\
         & \widetilde{\boldsymbol{T}} (\boldsymbol{\ell}_{0}) = \boldsymbol{T} - \Bell_0(\Bx)^{\top}\BW=\boldsymbol{\eta} , \\
         & \boldsymbol{\psi}_{\delta}(\boldsymbol{Z}, \boldsymbol{\theta}_0, \boldsymbol{h}_0)  = \phi_{\delta}\left\{ \widetilde{Y}(\boldsymbol{\nu}_{0}) - \boldsymbol{\theta}_0(\boldsymbol{x})^{\top}\widetilde{\boldsymbol{T}} (\boldsymbol{\ell}_{0})\right\}\widetilde{\boldsymbol{T}} \left(\boldsymbol{\ell}_{0}\right).
    \end{align*}
    For any $\boldsymbol{W}, \boldsymbol{x}$ , the conditional independence and moment conditions on $\boldsymbol{\eta}_i$ stated in Assumption 2 imply that
    \begin{align*}
         &\; \mathbb{E}\left\{ \nabla_{\boldsymbol{\nu}} \boldsymbol{\psi}_{\delta}\left( \boldsymbol{Z}; \boldsymbol{\theta}_{0}, (\boldsymbol{\nu},\boldsymbol{\ell}_0)\right) \mid \boldsymbol{W}, \boldsymbol{x} \right\}\mid_{\boldsymbol{\nu}=\boldsymbol{\nu}_0}\\
         =&\;\mathbb{E}\left[-\widetilde{\boldsymbol{T}} (\boldsymbol{\ell}_{0})\phi_{\delta}^{\prime}\left\{\widetilde{Y}(\boldsymbol{\nu}_{0}) - \boldsymbol{\theta}_0(\boldsymbol{x})^{\top}\widetilde{\boldsymbol{T}} (\boldsymbol{\ell}_{0})\right\} \mid \boldsymbol{W}, \boldsymbol{x}\right]\\
         =&\;\mathbb{E}\left[-\widetilde{\boldsymbol{T}} (\boldsymbol{\ell}_{0})I\left\{\left|\widetilde{Y}(\boldsymbol{\nu}_{0}) - \boldsymbol{\theta}_0(\boldsymbol{x})^{\top}\widetilde{\boldsymbol{T}} (\boldsymbol{\ell}_{0})\right|\leq \delta\right\}\mid \boldsymbol{W}, \boldsymbol{x}\right] \\
         =&\;\mathbb{E} \left\{ -\boldsymbol{\eta} I(|\epsilon| \leq \delta) \mid \boldsymbol{W}, \boldsymbol{x} \right\}     
         =\mathbb{E} \left( -\boldsymbol{\eta} \mid \boldsymbol{W}, \boldsymbol{x} \right)\mathbb{E} \left\{ I(|\epsilon| \leq \delta) \mid \boldsymbol{W}, \boldsymbol{x} \right\}
        =\boldsymbol{0}.
    \end{align*}
    Moreover, since the conditional distribution of $\epsilon$ given $\boldsymbol{W},\boldsymbol{x}$ is symmetric, we have
    \begin{align*}
        &\; \mathbb{E}\left\{ \nabla_{\boldsymbol{\ell}} \boldsymbol{\psi}_{\delta}\left( \boldsymbol{Z}; \boldsymbol{\theta}_{0}, (\boldsymbol{\nu}_0,\boldsymbol{\ell})\right) \mid \boldsymbol{W}, \boldsymbol{x} \right\}\mid_{\boldsymbol{\ell}=\boldsymbol{\ell}_0}\\
        =&\;\mathbb{E}[ \boldsymbol{\theta}_0(\boldsymbol{x})^{\top}\widetilde{\boldsymbol{T}} (\boldsymbol{\ell}_{0})\phi_{\delta}^{\prime}\left\{\widetilde{Y}(\boldsymbol{\nu}_{0}) - \boldsymbol{\theta}_0(\boldsymbol{x})^{\top}\widetilde{\boldsymbol{T}} (\boldsymbol{\ell}_{0})\right\} - \phi_{\delta}\left\{\widetilde{Y}(\boldsymbol{\nu}_{0}) - \boldsymbol{\theta}_0(\boldsymbol{x})^{\top}\widetilde{\boldsymbol{T}} (\boldsymbol{\ell}_{0})\right\} \mid \boldsymbol{W}, \boldsymbol{x}]  \\
        =&\;\mathbb{E} \left\{ \boldsymbol{\theta}_0(\boldsymbol{x})^{\top} \boldsymbol{\eta}I(|\epsilon| \leq \delta) - \epsilon I(|\epsilon| \leq \delta) - \delta sign(\epsilon)I(|\epsilon| > \delta) \mid \boldsymbol{W}, \boldsymbol{x} \right\}\\
        =&\; \boldsymbol{\theta}_0(\boldsymbol{x})^{\top} \mathbb{E} \left\{\boldsymbol{\eta}I(|\epsilon| \leq \delta) \mid \boldsymbol{W}, \boldsymbol{x} \right\}-\mathbb{E} \left\{\epsilon I(|\epsilon| \leq \delta) + \delta sign(\epsilon)I(|\epsilon| > \delta) \mid \boldsymbol{W}, \boldsymbol{x} \right\} =\boldsymbol{0} .
    \end{align*}
    Hence we have
    \[
        \mathbb{E} \left\{\nabla_{\boldsymbol{h}} \boldsymbol{\psi}_{\delta}( \boldsymbol{Z}; \boldsymbol{\theta}_{0}, \boldsymbol{h}_{0}) \mid \boldsymbol{W}, \boldsymbol{x} \right\} = \boldsymbol 0.
    \]

\noindent\textbf{Identifiability:}
    For any $\Bx$ and $\boldsymbol{\theta}\in \Theta$,
    \begin{equation*}
        \begin{aligned}
            \boldsymbol{m}_{\delta}(\boldsymbol{x}; \boldsymbol{\theta}, \boldsymbol{h}_0) &= \mathbb{E}\{\boldsymbol{\psi}_{\delta}(\boldsymbol{Z}; \boldsymbol{\theta}, \boldsymbol{h}) \mid \boldsymbol{X}= \boldsymbol{x}\}\\
            &= \mathbb{E}\left[\phi_{\delta}\left\{ \widetilde{Y}(\boldsymbol{\nu}_{0}) - \boldsymbol{\theta}(\boldsymbol{X})^{\top}\widetilde{\boldsymbol{T}} (\boldsymbol{\ell}_{0})\right\}\widetilde{\boldsymbol{T}} (\boldsymbol{\ell}_{0}) \mid \boldsymbol{X} = \boldsymbol{x}\right],
        \end{aligned}
    \end{equation*}
    and the expected Jacobian is continuous in $\boldsymbol{\theta}$,
    \begin{align*}
         &\; \nabla_{\boldsymbol{\theta}} \boldsymbol{m}_{\delta}(\boldsymbol{x}; \boldsymbol{\theta}, \boldsymbol{h}_0)                                                                                                                               \\
         = &\;\mathbb{E}\left[-\boldsymbol{\eta} \boldsymbol{\eta}^{\top}I\left\{\left|\widetilde{Y}(\boldsymbol{\nu}_{0}) - \boldsymbol{\theta}(\boldsymbol{X})^{\top}\widetilde{\boldsymbol{T}} (\boldsymbol{\ell}_{0})\right| \leq \delta\right\}\vert \boldsymbol{X}=\boldsymbol{x}\right]                                                                \\
          = &\;\mathbb{E}\left(\mathbb{E}\left[-\boldsymbol{\eta}\boldsymbol{\eta}^{\top}I\left\{\left|\widetilde{Y}(\boldsymbol{\nu}_{0}) - \boldsymbol{\theta}(\boldsymbol{X})^{\top}\widetilde{\boldsymbol{T}} (\boldsymbol{\ell}_{0})\right| \leq \delta \right\} \vert \boldsymbol{\eta},\boldsymbol{X} \right]\vert \boldsymbol{X}=\boldsymbol{x}\right) \\
          = &\;\mathbb{E}\left[-\boldsymbol{\eta}\boldsymbol{\eta}^{\top}\mathbb{P}\left\{\left|\widetilde{Y}(\boldsymbol{\nu}_{0}) - \boldsymbol{\theta}(\boldsymbol{X})^{\top}\widetilde{\boldsymbol{T}} (\boldsymbol{\ell}_{0})\right| \leq \delta \vert \boldsymbol{\eta},\boldsymbol{X}\right\}\vert \boldsymbol{X}=\boldsymbol{x}\right].                
    \end{align*}
    When $\boldsymbol{\theta}(\boldsymbol{X}) = \boldsymbol{\theta}_0(\boldsymbol{X})$, we have $\nabla_{\boldsymbol{\theta}} \boldsymbol{m}_{\delta}(\boldsymbol{x}; \boldsymbol{\theta}, \boldsymbol{h}_0) = \mathbb{E}\left\{-\boldsymbol{\eta} \boldsymbol{\eta}^{\top}I\left(|\epsilon|\leq \delta\right)\vert \boldsymbol{X} = \boldsymbol{x}\right\}$ is
    invertible for any $\boldsymbol{x}$.
    Then for any $\boldsymbol{x}$, $\boldsymbol{\theta}_0(\boldsymbol{x})$ is the unique solution to $\boldsymbol{m}_{\delta}\left(\boldsymbol{x};\boldsymbol{\theta}, \boldsymbol{h}_0\right)=\boldsymbol 0$.

\noindent\textbf{Smooth Signal:}
    We can write the residualized variables explicitly as functions of $\boldsymbol{x}$:
    \begin{align*}
         & \widetilde{Y}(\boldsymbol{\nu}) = Y - \boldsymbol{\nu}(\boldsymbol{x})^{\top}\boldsymbol{W}, \\
         & \widetilde{\boldsymbol{T}}(\boldsymbol{\ell}) = \boldsymbol{T} - \boldsymbol{\ell}(\boldsymbol{x})^{\top} \boldsymbol{W}.
    \end{align*}
    The $j$-th coordinate of the score function is given by:
    \[
        \psi_{\delta}^{(j)}(\boldsymbol{Z}; \boldsymbol{\theta}, \boldsymbol{h}) = \phi_{\delta}\left\{\widetilde{Y}(\boldsymbol{\nu}) - \boldsymbol{\theta}_0(\boldsymbol{x})^{\top}\widetilde{T}(\boldsymbol{\ell})\right\} \widetilde{T}^{(j)}(\boldsymbol{\ell}),
    \]
    where $\boldsymbol{h}=\left(\boldsymbol{\nu},\boldsymbol{\ell}\right)$ collects the nuisance functions.
    By Assumptions~\ref{assumption:sparse_nuisance} and \Sref{assumption:TE_Lipschitz}{1}, the parameter functions $\boldsymbol{\nu}(\boldsymbol{x})$, $\boldsymbol{\ell}(\boldsymbol{x})$, and $\boldsymbol{\theta}(\boldsymbol{x})$ are $O(1)$-Lipschitz continuous with respect to $\boldsymbol{x}$. 
    Since $\|\boldsymbol{W}\|_{\infty} \leq L_{\boldsymbol{W}}$ and we restrict the parameter spaces $\Theta$ and $\Omega$ such that $\|\boldsymbol{\theta}\| \leq O(1)$ and the $L_1$-norm of the nuisance parameters are bounded, the residualized functions $\widetilde{Y}(\boldsymbol{\nu})$ and $\widetilde{\boldsymbol{T}}(\boldsymbol{\ell})$  are also Lipschitz continuous in $\boldsymbol{x}$. 
    Furthermore, the Huber derivative $\phi_{\delta}(\cdot)$ is $O(1)$-Lipschitz and uniformly bounded by $\delta$. Because the product and composition of bounded Lipschitz functions remain Lipschitz, the score $\psi_{\delta}^{(j)}(\boldsymbol{Z}; \boldsymbol{\theta}, \boldsymbol{h})$ is $O(1)$-Lipschitz in $\boldsymbol{x}$ for any realization of $\boldsymbol{Z}$.
    Consequently, integrating over the conditional distribution, the expected moment $m_{\delta}^{(j)}(\boldsymbol{x}; \boldsymbol{\theta}, \boldsymbol{h}) = \mathbb{E}\left\{\psi_{\delta}^{(j)}(\boldsymbol{Z}; \boldsymbol{\theta}, \boldsymbol{h}) \mid \boldsymbol{X} = \boldsymbol{x}\right\}$ preserves this property and is $O(1)$-Lipschitz in $\boldsymbol{x}$.

\noindent\textbf{Curvature:}
    Recall that the expected moment function is defined as $\boldsymbol{m}_{\delta}(\boldsymbol{x}; \boldsymbol{\theta}, \boldsymbol{h}) = \mathbb{E}\left\{\boldsymbol{\psi}_{\delta}(\boldsymbol{Z}; \boldsymbol{\theta}, \boldsymbol{h}) \mid \boldsymbol{X} = \boldsymbol{x}\right\}$. 
    By differentiating the score function inside the expectation (which is valid due to the smoothness of the error distribution), the expected Jacobian evaluated at the true parameters $\boldsymbol{\theta}_0$ and $\boldsymbol{h}_0$ is given by:
    \[
        \boldsymbol{M}_{\delta} \equiv- \nabla_{\boldsymbol{\theta}} \boldsymbol{m}_{\delta}(\boldsymbol{x}; \boldsymbol{\theta}_0, \boldsymbol{h}_0) = \mathbb{E}\left\{\boldsymbol{\eta} \boldsymbol{\eta}^{\top} I(\vert \epsilon \vert \leq \delta) \mid \boldsymbol{X} = \boldsymbol{x}\right\},
    \]
where $\boldsymbol{\eta}$ denotes the treatment error, namely $\boldsymbol{\eta}=\boldsymbol{T}-\boldsymbol{\ell}_0(\boldsymbol{x})^{\top}\boldsymbol{W}$.
    By Assumption \Sref{assumption:Non-degeneracy}{2}, the matrix $\boldsymbol{M}_{\delta}$ has its minimum eigenvalue strictly bounded away from $0$.

\noindent\textbf{Smoothness of scores:} We analyze the smoothness of the score function $\boldsymbol{\psi}_{\delta}(\boldsymbol{Z}; \boldsymbol{\theta}, \boldsymbol{h})$ with respect to the nuisance parameter $\boldsymbol{h}$.
    Recall the residualized variables $\widetilde{Y}(\boldsymbol{\nu}) = Y - \Bnu(\boldsymbol{x})^{\top}\BW$ and $\widetilde{\boldsymbol{T}} (\boldsymbol{\ell}) = \boldsymbol{T} - \Bell^{\top}(\Bx)\BW$. The score vector can be written explicitly as:
    \[
        \boldsymbol{\psi}_{\delta}(\boldsymbol{Z}; \boldsymbol{\theta}, \boldsymbol{h})  = \phi_{\delta}\left\{ \widetilde{Y}(\boldsymbol{\nu}) - \boldsymbol{\theta}(\boldsymbol{x})^{\top}\widetilde{\boldsymbol{T}} (\boldsymbol{\ell})\right\}\widetilde{\boldsymbol{T}} (\boldsymbol{\ell}).
    \]
    We first show that the score is Lipschitz continuous in $\boldsymbol{h}$ for each fixed realization of $\boldsymbol{Z}$. By definition, the Huber derivative $\phi_{\delta}(\cdot)$ is $O(1)$-Lipschitz and uniformly bounded by $\delta$. Since the parameter space $\boldsymbol{\Theta}$ is bounded (i.e., $\|\boldsymbol{\theta}(\boldsymbol{x})\| = O(1)$), the inner residual $\widetilde{Y}(\boldsymbol{\nu}) - \boldsymbol{\theta}(\boldsymbol{x})^{\top}\widetilde{\boldsymbol{T}} (\boldsymbol{\ell})$ is $O(1)$-Lipschitz with respect to $\boldsymbol{h}$. 
    By the product rule for Lipschitz functions, since both $\phi_{\delta}(\cdot)$ and $\widetilde{\boldsymbol{T}} (\boldsymbol{\ell})$ are Lipschitz in $\boldsymbol{h}$ and are uniformly bounded over the compact parameter spaces, their product $\boldsymbol{\psi}_{\delta}(\boldsymbol{Z}; \boldsymbol{\theta}, \boldsymbol{h})$ is $O(1)$-Lipschitz in $\boldsymbol{h}$ for any realization of $\boldsymbol{Z}$.
    
    Furthermore, we consider the smoothness of the gradient with respect to $\boldsymbol{\theta}$. The gradient with respect to $\boldsymbol{\theta}$ is given by
    \[
        \nabla_{\boldsymbol{\theta}} \boldsymbol{\psi}_{\delta}(\boldsymbol{Z}; \boldsymbol{\theta}, \boldsymbol{h}) = - \widetilde{\boldsymbol{T}} (\boldsymbol{\ell})\widetilde{\boldsymbol{T}} (\boldsymbol{\ell})^{\top} I\left\{\left| \widetilde{Y}(\boldsymbol{\nu}) - \boldsymbol{\theta}(\boldsymbol{x})^{\top}\widetilde{\boldsymbol{T}} (\boldsymbol{\ell}) \right| \leq \delta\right\}
    \]
    involves a discontinuous indicator function, which is not Lipschitz continuous. However, under Assumption 2 in the main paper, the error term $\epsilon$ possesses a Lipschitz continuous cumulative distribution function. Taking the conditional expectation smooths out this discontinuity. Consequently, the expected Jacobian
    \[
        \nabla_{\boldsymbol{\theta}} \boldsymbol{m}_{\delta}(\boldsymbol{x}; \boldsymbol{\theta}, \boldsymbol{h}) = \mathbb{E}\left[ - \widetilde{\boldsymbol{T}} (\boldsymbol{\ell})\widetilde{\boldsymbol{T}} (\boldsymbol{\ell})^{\top} I\left\{\left| \widetilde{Y}(\boldsymbol{\nu}) - \boldsymbol{\theta}(\boldsymbol{x})^{\top}\widetilde{\boldsymbol{T}} (\boldsymbol{\ell}) \right| \leq \delta\right\} \mathrel{\Big|} \boldsymbol{X} = \boldsymbol{x} \right]
    \]
    is continuously differentiable and $O(1)$-Lipschitz with respect to $\boldsymbol{\theta}$. 
\end{proof}

\subsection{Proofs of the Main Results}\label{TE proof}
In this subsection, we prove the theoretical results in the main paper. For notational simplicity, we {\bf write $\delta$ in place of $\delta_{\boldsymbol{\theta}}$ throughout this section}.
We first present the proof of Theorem~\Sref{thm:Asymptotic properties} 1.
The proof of Theorem~\Sref{thm:Asymptotic properties} 1 is divided into three parts, corresponding to consistency, the $L_p$ error bound and asymptotic normality.

\noindent\textbf{Proof of Theorem 1 (i):}
\begin{proof}
    Denote $\boldsymbol{\Lambda}_{\delta}(\widehat{\boldsymbol{\theta}}, \widehat{\boldsymbol{h}})=\boldsymbol{m}_{\delta}(\widehat{\boldsymbol{\theta}}, \widehat{\boldsymbol{h}}) - \boldsymbol{\Psi}_{\delta}(\widehat{\boldsymbol{\theta}}, \widehat{\boldsymbol{h}})$. By the definition of $\widehat{\boldsymbol{\theta}}$, $\boldsymbol{\Psi}_{\delta}(\widehat{\boldsymbol{\theta}}, \widehat{\boldsymbol{h}}) =\boldsymbol 0$, we have
    \[
        \left\| \boldsymbol{m}_{\delta}(\widehat{\boldsymbol{\theta}}, \widehat{\boldsymbol{h}}) \right\| =
        \left\| \boldsymbol{m}_{\delta}(\widehat{\boldsymbol{\theta}}, \widehat{\boldsymbol{h}}) - \boldsymbol{\Psi}_{\delta}(\widehat{\boldsymbol{\theta}}, \widehat{\boldsymbol{h}}) \right\| =
        \left\| \boldsymbol{\Lambda}_{\delta}(\widehat{\boldsymbol{\theta}}, \widehat{\boldsymbol{h}}) \right\|.
    \]
    Combining Lemmas \ref{lemma:kernel_error}, \ref{lemma:sampling_error} and \ref{lemma:subsampling_error} in Subsection~\ref{TE lemma},
    we have that with probability $1-2e^{-t}$,
    \[
        \left\| \boldsymbol{\Lambda}_{\delta}(\widehat{\boldsymbol{\theta}}, \widehat{\boldsymbol{h}}) \right\|
        = O\left(s^{-1/\left(2\alpha d_{\boldsymbol{X}}\right)} + \sqrt{\frac{s\left\{\log(n/s)+t\right\}}{n}}+\frac{\log B+t}{\sqrt{B}}\right).
    \]
    Intergrating this tail bound we get that
    \[
        \mathbb{E}\left[\left\| \boldsymbol{\Lambda}_{\delta}(\widehat{\boldsymbol{\theta}}, \widehat{\boldsymbol{h}}) \right\|\right]
        = O\left(s^{-1/\left(2\alpha d_{\boldsymbol{X}}\right)} + \sqrt{\frac{s\log(n/s)}{n}}+\frac{\log B}{\sqrt{B}}\right).
    \]
    Thus if $B \geq n/s$, $s = o(n)$ and $s \to \infty$ then all terms converge to zero as $n \to \infty$. Since $\boldsymbol{\psi}_{\delta}(\boldsymbol{x}; \boldsymbol{\theta}, \boldsymbol{h})$ is
    $L$-Lipschitz in $\boldsymbol{h}$, it follows that
    \begin{align*}
        &\left\|\boldsymbol{m}_{\delta}(\boldsymbol{x}; \widehat{\boldsymbol{\theta}}, \boldsymbol{h}_0) - \boldsymbol{m}_{\delta}(\boldsymbol{x}; \widehat{\boldsymbol{\theta}}, \widehat{\boldsymbol{h}})\right\|
        \leq L\mathbb{E}\left(\left\| \widehat{\boldsymbol{h}} - \boldsymbol{h}_0 \right\| \vert \boldsymbol{x}\right)
        \leq L \sqrt{\mathbb{E}\left(\left\| \widehat{\boldsymbol{h}} - \boldsymbol{h}_0 \right\|^2 \vert \boldsymbol{x}\right)}\\
        &= L\mathcal{E}(\widehat{\boldsymbol{h}}), 
    \end{align*}
    where the second inequality follows from Jensen's inequality.
    By Assumption 3, $\mathcal{E}(\widehat{\boldsymbol{h}}) = o_p(1)$ as $s = o(n)$ and $n \to \infty$.
    Consequently, 
    $$\mathbb{E}\left\{\left\|\boldsymbol{m}_{\delta}(\boldsymbol{x}; \widehat{\boldsymbol{\theta}}, \boldsymbol{h}_0) - \boldsymbol{m}_{\delta}(\boldsymbol{x}; \widehat{\boldsymbol{\theta}}, \widehat{\boldsymbol{h}})\right\|\right\}\leq L\mathbb{E}\left\{\mathcal{E}(\widehat{\boldsymbol{h}})\right\}= o(1).$$
    Combining this with the fact that $\boldsymbol{m}_{\delta}(\boldsymbol{x}; \widehat{\boldsymbol{\theta}}, \widehat{\boldsymbol{h}})$, we have $\mathbb{E}\left\{\left\|\boldsymbol{m}_{\delta}(\boldsymbol{x}; \widehat{\boldsymbol{\theta}}, \boldsymbol{h}_0)\right\|\right\} = o(1)$,
    which implies that $\left\|\boldsymbol{m}_{\delta}(\boldsymbol{x}; \widehat{\boldsymbol{\theta}}, \boldsymbol{h}_0)\right\| = o_{p}(1)$.
For any $\varepsilon>0$, there exists a $c$ such that
    \[
        \mathbb{P}\left(\left\| \widehat{\boldsymbol{\theta}} - \boldsymbol{\theta}_0 \right\| \geq \varepsilon\right)\leq \mathbb{P}\left\{\left\| \boldsymbol{m}_{\delta}(\boldsymbol{x}; \widehat{\boldsymbol{\theta}}, \boldsymbol{h}_0) \right\|\geq c\right\}.
    \]
    Since $\left\| \boldsymbol{m}_{\delta}(\boldsymbol{x}; \widehat{\boldsymbol{\theta}}, \boldsymbol{h}_0) \right\| = o_p(1)$,
    the probability on the right-hand-side converges to $0$. Hence,
    $\left\| \widehat{\boldsymbol{\theta}} - \boldsymbol{\theta}_0 \right\| = o_p(1)$.
\end{proof}

Before proving Theorem 1 (ii) and (iii), we introduce a decomposition of the score discrepancy and the notation used throughout the proof. Let $\boldsymbol{\mu}_0\left(\boldsymbol{\theta},\boldsymbol{h}\right)=\mathbb{E}\left(\widetilde{\boldsymbol{\Psi}}_{\delta}\left(\boldsymbol{\theta},\boldsymbol{h}\right)\right)$ and $\boldsymbol{\Lambda}_{\delta}(\boldsymbol{\theta},\boldsymbol{h}) = \boldsymbol{m}_{\delta}(\boldsymbol{x};\boldsymbol{\theta},\boldsymbol{h}) -\boldsymbol{\Psi}_{\delta}(\boldsymbol{\theta},\boldsymbol{h})$. Moreover, for a tree constructed based on the subsample $\left\{\boldsymbol{Z}_i:i\in\mathcal{I}\right\}$, where $|\mathcal{I}|=s$ and internal randomness $\omega$, let $a_i\left(\mathcal{I},\omega\right)$ denote the weight assigned to observation $i$. 
For each sub-sample $ b = 1, \dots, B$
we denote the index of the samples chosen with $\mathcal{I}_b$
and $\omega_b$ the internal randomness that was drawn.
Define the weighted empirical score using the sub-sampled ORF weights
\[
    \boldsymbol{\Psi}_{\delta}(\boldsymbol{\theta},\boldsymbol{h}) = \sum_{i=1}^n a_i \boldsymbol{\psi}_{\delta}(\boldsymbol{Z}_i;\boldsymbol{\theta},\boldsymbol{h})
    =\frac{1}{B}\sum_{b=1}^B \sum_{i \in \mathcal{I}_b} a_i\left(\mathcal{I}_b,\omega_b\right) \boldsymbol{\psi}_{\delta}(\boldsymbol{Z}_i;\boldsymbol{\theta},\boldsymbol{h}).
\]
The corresponding complete multi-dimensional U-statistic, obtained by averaging over all sub-samples of size $s$, is given by
\[
    \widetilde{\boldsymbol{\Psi}}_{\delta}(\boldsymbol{\theta},\boldsymbol{h}) = \binom{n}{s}^{-1} \sum_{\mathcal{I}\subset\left\{1,2,\dots,n\right\}:|\mathcal{I}|=s} \mathbb{E}_{\omega}\left\{\sum_{i\in\mathcal{I}}a_{i}\left(\mathcal{I},\omega\right) \boldsymbol{\psi}_{\delta}(\boldsymbol{Z}_{i};\boldsymbol{\theta},\boldsymbol{h})\right\}.
\]
For any $\boldsymbol{\theta}$ and $\boldsymbol{h}$, $\boldsymbol{\Lambda}_{\delta}(\boldsymbol{\theta},\boldsymbol{h})$ can be decomposed as
$$ \boldsymbol{\Lambda}_{\delta}(\boldsymbol{\theta},\boldsymbol{h})
        = \underbrace{\boldsymbol{m}_{\delta}(\boldsymbol{x};\boldsymbol{\theta},\boldsymbol{h}) - \boldsymbol{\mu}_0(\boldsymbol{\theta},\boldsymbol{h})}_{\boldsymbol{\Gamma}_{\delta}(\boldsymbol{\theta},\boldsymbol{h})}
        + \underbrace{\boldsymbol{\mu}_0(\boldsymbol{\theta},\boldsymbol{h}) - \widetilde{\boldsymbol{\Psi}}_{\delta}(\boldsymbol{\theta},\boldsymbol{h})}_{\boldsymbol{\Delta}_{\delta}(\boldsymbol{\theta},\boldsymbol{h})}
        + \underbrace{\widetilde{\boldsymbol{\Psi}}_{\delta}(\boldsymbol{\theta},\boldsymbol{h}) - \boldsymbol{\Psi}_{\delta}(\boldsymbol{\theta},\boldsymbol{h})}_{\boldsymbol{E}_{\delta}(\boldsymbol{\theta},\boldsymbol{h})}.$$
The three terms $\boldsymbol{\Gamma}_{\delta}(\boldsymbol{\theta},\boldsymbol{h})$, $\boldsymbol{\Delta}_{\delta}(\boldsymbol{\theta},\boldsymbol{h})$ and $\boldsymbol{E}_{\delta}(\boldsymbol{\theta},\boldsymbol{h})$
 will be controlled separately in Section~\ref{TE lemma}, where we provide the corresponding upper bounds and detailed proofs.

\noindent\textbf{Proof of Theorem 1 (ii)  ($L_p$ error bound):}
\begin{proof}
The proof proceeds in two steps. We first expand the score function around the true parameter values and derive an explicit decomposition of the estimation error $\widehat{\boldsymbol{\theta}}-\boldsymbol{\theta}_0$. We then control the leading empirical process term and the remaining remainder terms separately. Combining these bounds yields the stated result.

\noindent\textbf{Step 1: Error Decomposition.}
 For a fixed conditioning vector $\boldsymbol{x}$,  
    we perform a second-order Taylor expansion for each coordinate $j \in 1,\dots,d_{\boldsymbol{T}}$ 
    of the expected score function $\boldsymbol{m}_{\delta}^{(j)}$ around the true parameters $\boldsymbol{\theta}_0$ and $\boldsymbol{h}_0$. 
    By the multi-dimensional mean-value theorem, for any $\boldsymbol{\theta} \in \boldsymbol{\Theta}$, we write:
    \begin{align*}
        \boldsymbol{m}_{\delta}^{(j)}(\boldsymbol{x}; \boldsymbol{\theta}, \widehat{\boldsymbol{h}})=&\boldsymbol{m}_{\delta}^{(j)}(\boldsymbol{x}; \boldsymbol{\theta}_0, \boldsymbol{h}_0)
        + \nabla_{\boldsymbol{\theta}} \boldsymbol{m}_{\delta}^{(j)}(\boldsymbol{x}; \boldsymbol{\theta}_0, \boldsymbol{h}_0)^{\top} \left(\boldsymbol{\theta} - \boldsymbol{\theta}_0\right)\\
        &+ \nabla_{\boldsymbol{h}} \boldsymbol{m}_{\delta}^{(j)}(\boldsymbol{x}; \boldsymbol{\theta}_0, \boldsymbol{h}_0)^{\top} \left(\widehat{\boldsymbol{h}} - \boldsymbol{h}_0\right) \\
         & +\underbrace{\frac{1}{2} \left(\boldsymbol{\theta} - \boldsymbol{\theta}_0, \widehat{\boldsymbol{h}} - \boldsymbol{h}_0\right)^{\top} \nabla_{\boldsymbol{\theta},\boldsymbol{h}} \boldsymbol{m}_{\delta}^{(j)}(\boldsymbol{x}; \widetilde{\boldsymbol{\theta}}^{(j)}, \widetilde{\boldsymbol{h}}^{(j)}) \left(\boldsymbol{\theta} - \boldsymbol{\theta}_0, \widehat{\boldsymbol{h}} - \boldsymbol{h}_0\right)}_{\boldsymbol{\varrho}_{\delta}^{(j)}}.
    \end{align*}
    Let $\boldsymbol{\varrho}_{\delta}=(\boldsymbol{\varrho}_{\delta}^{(1)},\dots,\boldsymbol{\varrho}_{\delta}^{(d_{\boldsymbol{T}})})^\top$ denote the vector of second-order terms. The intermediate points $(\widetilde{\boldsymbol{\theta}}^{(j)}, \widetilde{\boldsymbol{h}}^{(j)})$ are full vectors lying on the line segment between the true parameters $(\boldsymbol{\theta}_0, \boldsymbol{h}_0)$ and $(\boldsymbol{\theta}, \widehat{\boldsymbol{h}})$ specific to the Taylor expansion of the $j$-th coordinate.
    Since $\boldsymbol{m}_{\delta}^{(j)}(\boldsymbol{x}; \boldsymbol{\theta}_0, \boldsymbol{h}_0) = \boldsymbol{0}$ 
    and the first-order derivative with respect to $\boldsymbol{h}$ vanishes by local orthogonality, 
    the above expansion can be expressed in matrix form as:
    \[
        -\boldsymbol{M}_{\delta}\left(\boldsymbol{\theta} - \boldsymbol{\theta}_0\right) =
        \boldsymbol{m}_{\delta}(\boldsymbol{x}; \boldsymbol{\theta}, \widehat{\boldsymbol{h}})
        - \boldsymbol{\varrho}_{\delta}.
    \]
    Under Assumption~\Sref{assumption:Non-degeneracy}{2}, $\boldsymbol{M}_{\delta}$ 
    is invertible with its minimum eigenvalue bounded away from zero. 
    Therefore, we have:
    \begin{equation}\label{eq5}
        \begin{aligned}
            \boldsymbol{\theta} - \boldsymbol{\theta}_0
        \;&=\; \boldsymbol{M}_{\delta}^{-1} \boldsymbol{\varrho}_{\delta} -\boldsymbol{M}_{\delta}^{-1} \boldsymbol{m}_{\delta}(\boldsymbol{x}; \boldsymbol{\theta}, \widehat{\boldsymbol{h}})\\
        \;&=\;\boldsymbol{M}_{\delta}^{-1} \boldsymbol{\varrho}_{\delta}-\boldsymbol{M}_{\delta}^{-1}\left\lbrace\boldsymbol{m}_{\delta}(\boldsymbol{x};\boldsymbol{\theta},\boldsymbol{h}) - \boldsymbol{\Psi}_{\delta}(\boldsymbol{\theta},\boldsymbol{h})\right\rbrace-\boldsymbol{M}_{\delta}^{-1}\boldsymbol{\Psi}_{\delta}(\boldsymbol{\theta},\boldsymbol{h}).
        \end{aligned}
    \end{equation}
    To simplify notation, let $\boldsymbol{\xi}_{\delta} = \boldsymbol{M}_{\delta}^{-1} \boldsymbol{\varrho}_{\delta}$. 
    Given that the eigenvalues of $\boldsymbol{M}_{\delta}^{-1}$ are bounded, 
    it follows that $\left\|\boldsymbol{\xi}_{\delta}\right\| = O\left(\left\|\boldsymbol{\varrho}_{\delta}\right\|\right)$. 
    By the smoothness of the score from Lemma~\ref{lemma:moment_condition} 
    and the consistency of the nuisance estimators from Theorem~\ref{thm:nuisance_properties}, 
    the eigenvalues of the expected Hessian are bounded, leading to:
    \begin{equation}\label{eq6}
        \left\|\boldsymbol{\varrho}_{\delta}\right\|_{\infty} = O\left(\mathbb{E}\left(\left\|\widehat{\boldsymbol{h}} - \boldsymbol{h}_0\right\|^2 \vert \boldsymbol{x}\right) + \left\|\boldsymbol{\theta} - \boldsymbol{\theta}_0\right\|^2\right).
    \end{equation}
    Using the fact that $\left\|\boldsymbol{\varrho}_{\delta}\right\| \leq \sqrt{d_{\boldsymbol{T}}}\left\|\boldsymbol{\varrho}_{\delta}\right\|_{\infty}$ 
    and combining \eqref{eq5} with \eqref{eq6}, we obtain for any $\boldsymbol{\theta} \in \boldsymbol{\Theta}$:
    \begin{equation}\label{eq7}
        \boldsymbol{\theta} - \boldsymbol{\theta}_0
        = -\boldsymbol{M}_{\delta}^{-1} \boldsymbol{\Lambda}_{\delta}(\boldsymbol{\theta}, \widehat{\boldsymbol{h}}) -\boldsymbol{\boldsymbol{M}}_{\delta}^{-1} \boldsymbol{\Psi}_{\delta}(\boldsymbol{\theta}, \widehat{\boldsymbol{h}}) + \boldsymbol{\xi}_{\delta}.
    \end{equation}
    Evaluating \eqref{eq7} at $(\widehat{\boldsymbol{\theta}}, \widehat{\boldsymbol{h}})$ 
    and noting that $\boldsymbol{\Psi}_{\delta}(\widehat{\boldsymbol{\theta}}, \widehat{\boldsymbol{h}}) = \boldsymbol{0}$, we conclude:
    \begin{equation}
        \label{score_taylor}
        \begin{aligned}
         \widehat{\boldsymbol{\theta}} - \boldsymbol{\theta}_0 
          \;&=\; -\boldsymbol{M}_{\delta}^{-1} \boldsymbol{\Lambda}_{\delta}(\widehat{\boldsymbol{\theta}}, \widehat{\boldsymbol{h}}) + \boldsymbol{\xi}_{\delta} \\
          \;&=\; -\boldsymbol{M}_{\delta}^{-1} \boldsymbol{\Lambda}_{\delta}(\widehat{\boldsymbol{\theta}}, \widehat{\boldsymbol{h}}) + O\left(\mathbb{E}\left(\left\|\widehat{\boldsymbol{h}} - \boldsymbol{h}_0\right\|^2 \vert \boldsymbol{x}\right) + \left\|\widehat{\boldsymbol{\theta}} - \boldsymbol{\theta}_0\right\|^2\right) \\
          \;&=\; -\boldsymbol{M}_{\delta}^{-1} \boldsymbol{\Lambda}_{\delta}(\widehat{\boldsymbol{\theta}}, \widehat{\boldsymbol{h}}) + O\left(\mathcal{E}(\widehat{\boldsymbol{h}})^2 + \left\|\widehat{\boldsymbol{\theta}}- \boldsymbol{\theta}_0\right\|^2\right).
    \end{aligned}
    \end{equation}

\noindent\textbf{Step 2: Error Control.} By the triangle inequality for the $L_p$ norm, we have that:
    \begin{equation}\label{eq8}
    \begin{aligned}
        \left(\mathbb{E}\left[\| \widehat{\boldsymbol{\theta}} - \boldsymbol{\theta}_0 \|^{p}\right]\right)^{1/p}
        = O&\left(\left\{\mathbb{E}\left[\| \boldsymbol{\Lambda}_{\delta}(\widehat{\boldsymbol{\theta}},\widehat{\boldsymbol{h}}) \|^{p}\right]\right\}^{1/p}
        + \left\{\mathcal{E}(\widehat{\boldsymbol{h}})^{2p}\right\}^{1/p}
        \right.\\
        &\quad+\left.\left[\mathbb{E}\left\{\| \widehat{\boldsymbol{\theta}} - \boldsymbol{\theta}_0 \|^{2p}\right\}\right]^{1/p}
        \right).
        \end{aligned}
    \end{equation}
    By Theorem~\ref{thm:nuisance_properties} (iii), we have $\left\{\mathcal{E}(\widehat{\boldsymbol{h}})^{2p}\right\}^{1/p} = o\left(\sqrt{\frac{s}{n}}\right)$. We then analyze the term $\left\{\mathbb{E}\left(\| \widehat{\boldsymbol{\theta}} - \boldsymbol{\theta}_0 \|^{2p}\right)\right\}^{1/p}$. By the consistency of $\widehat{\boldsymbol{\theta}}$,
    we have that for any $\varepsilon$ and $c$, there exists $n^{*}(\varepsilon, c)$ such that when the sample size satisfies $n \geq n^{*}(\varepsilon, c)$,
    \[
        \mathbb{P}\left(\left\| \widehat{\boldsymbol{\theta}} - \boldsymbol{\theta}_0 \right\| \geq \varepsilon \right) \leq c.
    \]
   For any constant $p\geq 1$,
    \[
        \mathbb{E}\left(\left\| \widehat{\boldsymbol{\theta}} - \boldsymbol{\theta}_0 \right\|^{2p}\right)
        \leq \varepsilon^{p} \mathbb{E}\left(\left\| \widehat{\boldsymbol{\theta}} - \boldsymbol{\theta}_0 \right\|^{p}\right)
        + c \mathbb{E}\left(\left\| \widehat{\boldsymbol{\theta}} - \boldsymbol{\theta}_0 \right\|^{2p}\right).
    \]
    Choosing $\varepsilon = \left(4C\right)^{-p}$ and $c = \left(4C\right)^{-1}\left\{diam(\boldsymbol{\Theta})\right\}^{-p} = O(1)$,
    we derive that
    \[
        \left\{\mathbb{E}\left(\left\| \widehat{\boldsymbol{\theta}} - \boldsymbol{\theta}_0 \right\|^{2p}\right)\right\}^{1/p}
        = \frac{1}{\left(2C\right)^{1/p}}\left\{\mathbb{E}\left(\left\| \widehat{\boldsymbol{\theta}} - \boldsymbol{\theta}_0 \right\|^{p}\right)\right\}^{1/p}.
    \]
    Therefore, the last two terms on the right side of \eqref{eq8} can be ignored. For the first term, it follows from Lemmas \ref{lemma:kernel_error}, \ref{lemma:sampling_error}, \ref{lemma:subsampling_error} in Subsection~\ref{TE lemma} that
    \begin{equation}
        \begin{aligned}
            \left\{\mathbb{E}\left(\| \boldsymbol{\Lambda}_{\delta}(\widehat{\boldsymbol{\theta}},\widehat{\boldsymbol{h}}) \|^{p}\right)\right\}^{1/p}
        =O&\left(\left\{\mathbb{E}\left(\| \boldsymbol{\Gamma}_{\delta}(\widehat{\boldsymbol{\theta}},\widehat{\boldsymbol{h}}) \|^{p}\right)\right\}^{1/p}
        +\left\{\mathbb{E}\left(\| \boldsymbol{\Delta}_{\delta}(\widehat{\boldsymbol{\theta}},\widehat{\boldsymbol{h}}) \|^{p}\right)\right\}^{1/p}\right.\\
        &\quad\left.+\left\{\mathbb{E}\left(\| \boldsymbol{E}_{\delta}(\widehat{\boldsymbol{\theta}},\widehat{\boldsymbol{h}}) \|^{p}\right)\right\}^{1/p}
        \right).
        \end{aligned}
        \nonumber
    \end{equation}
    By Lemma \ref{lemma:kernel_error} in Subsection~\ref{TE lemma}, we have
    $$\left\{\mathbb{E}\left(\| \boldsymbol{\Gamma}_{\delta}(\widehat{\boldsymbol{\theta}},\widehat{\boldsymbol{h}}) \|^{p}\right)\right\}^{1/p} = O\left(s^{-1/(2\alpha d_{\boldsymbol{X}})}\right).$$
    for a constant $\alpha = \frac{\log(\Upsilon^{-1})}{\pi \log\left\{(1-\Upsilon)^{-1}\right\}}$.
    Moreover, when $B \gtrsim n/s$, by integrating the exponential tail bound provided by the high probability statements in Lemmas \ref{lemma:sampling_error} and \ref{lemma:subsampling_error} in Subsection~\ref{TE lemma},
    we have that for any integer $p>1$:
$$\left\{\mathbb{E}\left(\| \boldsymbol{\Delta}_{\delta}(\widehat{\boldsymbol{\theta}},\widehat{\boldsymbol{h}}) \|^{p}\right)\right\}^{1/p}= O\left(\sqrt{\frac{s\log(n/s)}{n}}\right),$$
$$\left\{\mathbb{E}\left(\| \boldsymbol{E}_{\delta}(\widehat{\boldsymbol{\theta}},\widehat{\boldsymbol{h}}) \|^{p}\right)\right\}^{1/p}= O\left(\sqrt{\frac sn}\right).$$
 Combining all the above inequalities:
$$\mathbb{E}\left(\| \widehat{\boldsymbol{\theta}} - \boldsymbol{\theta}_0 \|^p\right)^{1/p}= O\left(s^{-\frac{1}{2\alpha d_{\boldsymbol{X}}}} + \sqrt{\frac{s\log(n/s)}{n}}+\sqrt{\frac{s}{n}}\right).$$

\end{proof}

\noindent\textbf{Proof of Theorem 1 (iii) (Asymptotic normality):}
\begin{proof}
We prove the asymptotic normality in three steps. First, we construct an oracle pseudo-outcome and show that the leading oracle term can be viewed as a random forest regression estimator. Second, we verify that the conditions of \cite{wager2018estimation} hold for this oracle forest problem, which yields asymptotic normality of the oracle term. Finally, we establish the asymptotic equivalence between the proposed OHRF estimator and the oracle leading term, and the asymptotic normality follows from Slutsky’s theorem.

\noindent\textbf{Step 1: Construction of oracle pseudo-outcome.} We begin with constructing a random variable $\langle \boldsymbol{b}, \boldsymbol{M}_{\delta}^{-1}\boldsymbol{\Delta}(\boldsymbol{\theta}_0, \widetilde{\boldsymbol{h}}_0) \rangle$, 
where $\widetilde{\boldsymbol{h}}_0 = \left(\boldsymbol{\nu}_0(\boldsymbol{x}), \boldsymbol{\ell
}_0(\boldsymbol{x})\right)$, 
i.e., the nuisance function $\widetilde{\boldsymbol{h}}_0$ evaluates the true parameter functions at the target point $\boldsymbol{x}$.
Asymptotic normality follows by arguments similar to \cite{wager2018estimation}, 
treating this as a random forest regression where the pseudo-outcome for sample $i$ is defined as:
\[
    Y_i^* = \left\langle \boldsymbol{b}, \boldsymbol{M}_{\delta}^{-1}\left\{\boldsymbol{m}_{\delta}(\boldsymbol{X}_{i};\boldsymbol{\theta}_0,\widetilde{\boldsymbol{h}}_0) - \boldsymbol{\psi}_{\delta}(\boldsymbol{Z}_i;\boldsymbol{\theta}_0,\widetilde{\boldsymbol{h}}_0)\right\}\right\rangle.
\]

\noindent\textbf{Step 2: Asymptotic Normality of oracle estimation.} To establish the asymptotic normality of the forest-based estimator with oracle pseudo-outcome $Y_i^{\star}$,  
we verify that our framework satisfies the conditions of Theorem 1 in \cite{wager2018estimation}. For notational brevity, we hereafter denote $\boldsymbol{\psi}_{\delta,i} \equiv \boldsymbol{\psi}_{\delta}(\boldsymbol{Z}_i; \boldsymbol{\theta}_0, \widetilde{\boldsymbol{h}}_0)$.

First, we address the foundational assumptions on the data distribution. 
The distribution of $\boldsymbol{X}$ is assumed to admit a density that is bounded 
away from zero and infinity. Furthermore, the conditional mean 
$\mathbb{E}\left(Y_i^* \vert \boldsymbol{X} = \boldsymbol{x}^{*}\right)$ is taken to be zero 
and is continuous in $\boldsymbol{x}^{*}$ for any $\boldsymbol{x}^{*}$.
The most technically involved condition is the Lipschitz continuity of the 
conditional variance, $\text{Var}(Y_i^* \vert \boldsymbol{X} = \boldsymbol{x}^{*})$. 
We can decompose this variance as:
\[
    \text{Var}(Y_i^* \vert \boldsymbol{X} = \boldsymbol{x}^{*}) =
    \mathbb{E}\left(\left\langle \boldsymbol{b}, \boldsymbol{M}_{\delta}^{-1}\boldsymbol{\psi}_{\delta,i}\right\rangle^2 \vert \boldsymbol{X}=\boldsymbol{x}^{*}\right)
    - \mathbb{E}\left(\left\langle \boldsymbol{b}, \boldsymbol{M}_{\delta}^{-1} \boldsymbol{\psi}_{\delta,i}\right\rangle \vert \boldsymbol{X}=\boldsymbol{x}^{*}\right)^2.
\]
The second term is $O(1)$-Lipschitz in $\boldsymbol{x}^{*}$ due to the smoothness 
of the moment function $\boldsymbol{m}_{\delta}$. For the first term, we define
\begin{equation}
    \begin{aligned}
        Q(\boldsymbol{x}^{*}) &\triangleq \mathbb{E}\left\{\boldsymbol{b}^{\top} \boldsymbol{M}_{\delta}^{-1}\boldsymbol{\psi}_{\delta,i} \boldsymbol{\psi}_{\delta,i}^{\top} (\boldsymbol{M}_{\delta}^{-1})^{\top} \boldsymbol{b} \vert \boldsymbol{X}=\boldsymbol{x}^{*}\right\}\\
    &= \boldsymbol{b}^{\top} \left(\boldsymbol{M}_{\delta}^{-1} \right)\mathbb{E}\left(\boldsymbol{\psi}_{\delta,i} \boldsymbol{\psi}_{\delta,i}^{\top} \vert \boldsymbol{X}=\boldsymbol{x}^{*}\right) (\boldsymbol{M}_{\delta}^{-1})^{\top} \boldsymbol{b}.
    \end{aligned}
    \nonumber
\end{equation}
Under the assumption that $\text{Var}(\boldsymbol{\psi}_{\delta,i} \vert \boldsymbol{X}=\boldsymbol{x}^{\prime})$ 
is Lipschitz continuous, the conditional covariance matrix of the score satisfies:
\[
    \left\| \mathbb{E}\left(\boldsymbol{\psi}_{\delta,i} \boldsymbol{\psi}_{\delta,i}^{\top} \vert \boldsymbol{X}=\boldsymbol{x}^{*}\right) - \mathbb{E}\left(\boldsymbol{\psi}_{\delta,i} \boldsymbol{\psi}_{\delta,i}^{\top} \vert \boldsymbol{X}=\widetilde{\boldsymbol{x}}\right) \right\|_{F}
    \leq L \|\boldsymbol{x}^{*} - \widetilde{\boldsymbol{x}}\|.
\]
By applying the Cauchy-Schwarz inequality and leveraging the lower bound $\sigma > 0$ 
on the eigenvalues of $\boldsymbol{M}_{\delta}$, we can bound the change in $Q(\boldsymbol{x}^{*})$:
\[
    |Q(\boldsymbol{x}^{*}) - Q(\widetilde{\boldsymbol{x}})|
    \leq L\|\boldsymbol{x}^{*} - \widetilde{\boldsymbol{x}}\|\cdot \|\boldsymbol{b}^{\top} \boldsymbol{M}_{\delta}^{-1}\|^2
    \leq \frac{L}{\sigma^2}\|\boldsymbol{x}^{*} - \widetilde{\boldsymbol{x}}\|.
\]
This confirms that $Q(\boldsymbol{x}^{*})$ is Lipschitz, and therefore, 
$\text{Var}(Y_i^* \vert \boldsymbol{X} = \boldsymbol{x}^{*})$ is also Lipschitz continuous.

Next, we verify the remaining moment and variance conditions. The existence 
of a $(2+c)$-th moment, $\mathbb{E}\left\{\left|Y_i^*-\mathbb{E}\left(Y_i^*\vert\boldsymbol{X}=\boldsymbol{x}\right)\right|^{2+c}\right\}=O(1)$ , 
is guaranteed by the boundedness of our robust score function $\boldsymbol{\psi}_{\delta}$ 
and the non-degeneracy of the Jacobian $\boldsymbol{M}_{\delta}$ from Lemma~\ref{lemma:moment_condition}. 
The non-degeneracy of the variance, $\text{Var}(Y_i^* \vert \boldsymbol{X} = \boldsymbol{x}^{\prime}) > 0$, 
is a direct consequence of the assumption that 
$\text{Var}\left(\boldsymbol{b}^{\top}\boldsymbol{M}_{\delta}^{-1}\boldsymbol{\psi}_{\delta,i}\vert \boldsymbol{X} = \boldsymbol{x}^{\prime}\right) > 0$.

Finally, the conditions related to the forest construction and hyperparameter settings are satisfied. 
The structural properties of the forest, including honesty and $\Upsilon$-balanced splits, 
are met by construction as per Specification \Sref{specification:forest_regularity}{1}. 
The required growth rate for the subsample size $s$, namely 
$s^{-1/(2\alpha d_{\boldsymbol{X}})} = o\left((s/n)^{1/2 - \varepsilon}\right)$, 
is satisfied by our choice of $s = O(n^{c})$ for some $c \in \left(1-\frac{1}{1+\alpha d_{\boldsymbol{X}}}, 1\right]$. 
With all conditions verified, we have there exists a sequence $\sigma_n$, such that:
    \[
        \sigma_{n}^{-1}\langle \boldsymbol{b}, \boldsymbol{M}_{\delta}^{-1}\boldsymbol{\Delta}(\boldsymbol{\theta}_0, \widetilde{\boldsymbol{h}}_0) \rangle \to \mathcal{N}(0,1),
    \]
    for  $\sigma_n = O\left(\sqrt{\frac{\text{polylog}(n/s)^{-1}s}{n}}\right)$.

    \noindent\textbf{Step 3: Asymptotic equivalence between the OHRF estimator and the oracle term.}
    By \eqref{score_taylor}, the deviation can be decomposed as
    \begin{align*}
        &\;\left\| \left\langle \boldsymbol{b}, \widehat{\boldsymbol{\theta}}-\boldsymbol{\theta}_0 \right\rangle -\langle \boldsymbol{b}, \boldsymbol{M}_{\delta}^{-1}\boldsymbol{\Delta}(\boldsymbol{\theta}_0, \widetilde{\boldsymbol{h}}_0) \rangle \right\| \\
        = &\;O\left(
        \|\boldsymbol{\Gamma}_{\delta}(\widehat{\boldsymbol{\theta}}, \widehat{\boldsymbol{h}})\| +
        \| \boldsymbol{\Delta}_{\delta}(\widehat{\boldsymbol{\theta}}, \widehat{\boldsymbol{h}}) - \boldsymbol{\Delta}_{\delta}(\boldsymbol{\theta}_0,\widetilde{\boldsymbol{h}}_0) \|
        + \| \boldsymbol{E}_{\delta}(\widehat{\boldsymbol{\theta}}, \widehat{\boldsymbol{h}}) \|
        + \| \mathcal{E}(\widehat{\boldsymbol{h}}) \|^2\right.\\
        &\qquad\left.+ \| \widehat{\boldsymbol{\theta}}-\boldsymbol{\theta}_0 \|^2
        \right).
    \end{align*}
   We next control these terms one by one and show that they are asymptotically negligible.

    The first term is of order $O\left(s^{-1/(2\alpha d_{\boldsymbol{X}})}\right)$ by Lemma~\ref{lemma:kernel_error} in Subsection~\ref{TE lemma}, under our assumption on $s$, it is $o(\sigma_{n})$.
    The third term is $O\left(\sqrt{\frac{\log B}{B}}\right)$ by Lemma~\ref{lemma:subsampling_error} in Subsection~\ref{TE lemma}, which is also $o(\sigma_{n})$.
    By applying the $L^p$ estimation error result Assumption 3 with $p = 2$, the forth term is of order
    $O\left(\frac{s\log(n/s)}{n}\right) = o(\sigma_n)$. Moreover, by the consistency result for $\widehat{\boldsymbol{\theta}}$, the last term $\| \widehat{\boldsymbol{\theta}}-\boldsymbol{\theta}_0 \|^2= o(\sigma_n).$

It remains to control the stochastic equicontinuity gap $\| \boldsymbol{\Delta}_{\delta}(\widehat{\boldsymbol{\theta}}, \widehat{\boldsymbol{h}}) - \boldsymbol{\Delta}_{\delta}(\boldsymbol{\theta}_0,\widetilde{\boldsymbol{h}}_0) \|$. For each $j \in \{1, \dots, d_{\boldsymbol{T}}\}$, let $\psi_{\delta}^{(j)}(\boldsymbol{Z}; \boldsymbol{\theta}, \boldsymbol{h})$ denote the $j$-th component of the score vector $\boldsymbol{\psi}_{\delta}$. Each coordinate of the gap corresponds to a centered U-statistic over the function class:
\[
    \mathcal{B}_j(\mathcal{I}; \boldsymbol{\theta}, \boldsymbol{h}) - \mathcal{B}_j(\mathcal{I}; \boldsymbol{\theta}_0, \widetilde{\boldsymbol{h}}_0),
\]
where $\mathcal{B}\left(\mathcal{I};\boldsymbol{\theta},\boldsymbol{h}\right) = \mathbb{E}_{\omega}\left\{\sum_{i\in\mathcal{I}}a_{i}\left(\mathcal{I},\omega\right) \boldsymbol{\psi}_{\delta}(\boldsymbol{Z}_{i};\boldsymbol{\theta},\boldsymbol{h})\right\}$
and $\mathcal{B}_j\left(\mathcal{I};\boldsymbol{\theta},\boldsymbol{h}\right)$ is the $j$-th entry of $\mathcal{B}\left(\mathcal{I};\boldsymbol{\theta},\boldsymbol{h}\right)$. Since the derivative of the Huber loss $\phi_{\delta}$ is $O(1)$-Lipschitz, the $j$-th component score $\psi_{\delta}^{(j)}$ is Lipschitz continuous with respect to both $\boldsymbol{\theta}$ and $\boldsymbol{h}$. By the properties of tree weights $\sum a_i(\mathcal{I},\omega) = 1$, we can bound the variation of the kernel as:
    \begin{align*}
        &\;\left| \mathcal{B}_j(\mathcal{I}; \boldsymbol{\theta}, \boldsymbol{h}) - \mathcal{B}_j(\mathcal{I}; \boldsymbol{\theta}_0, \widetilde{\boldsymbol{h}}_0) \right|\\
         = &\;\left| \mathbb{E}_{\omega}\left(\sum_{i\in \mathcal{I}} a_i(\mathcal{I},\omega)\left[\psi_{\delta}^{(j)}\left\{\boldsymbol{Z}_i;\boldsymbol{\theta}, \widehat{\boldsymbol{h}}(\boldsymbol{W}_t)\right\}-\psi_{\delta}^{(j)}\left\{\boldsymbol{Z}_i;\boldsymbol{\theta}, \widetilde{\boldsymbol{h}}_0(\boldsymbol{W}_t)\right\}\right]\right) \right| \\
          \leq &\;\mathbb{E}_{\omega}\left[\sum_{i\in \mathcal{I}} a_i(\mathcal{I},\omega)\left|\psi_{\delta}^{(j)}\left\{\boldsymbol{Z}_i;\boldsymbol{\theta}, \widehat{\boldsymbol{h}}(\boldsymbol{W}_t)\right\}-\psi_{\delta}^{(j)}\left\{\boldsymbol{Z}_i;\boldsymbol{\theta}, \widetilde{\boldsymbol{h}}_0(\boldsymbol{W}_t)\right\}\right|\right]\\
          \leq &\;L\mathbb{E}_{\omega}\left[\sum_{t=1}^s a_i(\mathcal{I},\omega)\left\{ \|\boldsymbol{\theta}-\boldsymbol{\theta}_0\|+L\|\boldsymbol{\nu}-\boldsymbol{\nu}_0(\boldsymbol{x})\| \right\}\right]  \\
          = &\;L\left\{ \|\boldsymbol{\theta}-\boldsymbol{\theta}_0\|+L\|\boldsymbol{\nu}-\boldsymbol{\nu}_0(\boldsymbol{x})\| \right)
    \end{align*}

By taking the expectation and applying the triangle inequality, the variance envelope $G_1$ in Lemma \ref{lemma:bracketing} in Subsection~\ref{TE lemma} satisfies:
\begin{align*}
    G_1 \;&=\; \sup_{\boldsymbol{\theta}, \boldsymbol{h}} \sqrt{\mathbb{E}\left\{\left|     \mathcal{B}_j(\mathcal{I}; \boldsymbol{\theta}, \boldsymbol{h}) - \mathcal{B}_j(\mathcal{I}; \boldsymbol{\theta}_0, \widetilde{\boldsymbol{h}}_0) \right|^2\right\}} \\
    \;&\leq\; C \left\{ \|\boldsymbol{\theta}-\boldsymbol{\theta}_0\| + \mathcal{E}(\widehat{\boldsymbol{h}}) \right\}.
\end{align*}
Let $t \in (0,1)$ be an arbitrary probability tolerance. By Part (ii) of Theorem \ref{thm:nuisance_properties} and Markov's inequality, we know that with probability at least $1-e^{-t}$,
$\|\widehat{\boldsymbol{\theta}}-\boldsymbol{\theta}_0\| \leq O(\sigma_n / t)$.
Similarly, by Theorem \ref{thm:nuisance_properties}, with probability at least $1-e^{-t}$,
    $\mathcal{E}(\widehat{\boldsymbol{h}}) \leq O(\mathcal{R}_{n,s,t}).$

Conditional on these high-probability events, the variance envelope is bounded by $G_1 = O\left( \sigma_n + \mathcal{R}_{n,s}\right)$. 
Applying Lemma \ref{lemma:bracketing} in Subsection~\ref{TE lemma} with tolerance $t$, we have with probability at least $1-3t$:
\begin{align*}
    \| \boldsymbol{\Delta}_{\delta}(\widehat{\boldsymbol{\theta}}, \widehat{\boldsymbol{h}}) - \boldsymbol{\Delta}_{\delta}(\boldsymbol{\theta}_0,\widetilde{\boldsymbol{h}}_0) \| 
    \;&=\; O_p\left( G_1 \sqrt{\frac{s \log n}{n}} + \frac{s \log n}{n} \right) \\
    \;&=\; O_p\left( (\sigma_n + \mathcal{R}_{n,s}) \sqrt{\frac{s \log n}{n}} \right).
\end{align*}
Recall from the optimal forest rate that $\sigma_n \asymp \sqrt{s/n}$, and from our nuisance estimation convergence that $\mathcal{R}_{n,s} = o((s/n)^{1/4})$. Consequently:
\[
    \sigma_n \sqrt{\frac{s \log n}{n}} = \sigma_n^2 \sqrt{\log n} = o(\sigma_n),
\]
and
\[
    \mathcal{R}_{n,s} \sqrt{\frac{s \log n}{n}} = \mathcal{R}_{n,s} \sigma_n \sqrt{\log n} = o(\sigma_n).
\]
Therefore, we conclude that:
\[
    \| \boldsymbol{\Delta}_{\delta}(\widehat{\boldsymbol{\theta}}, \widehat{\boldsymbol{h}}) - \boldsymbol{\Delta}_{\delta}(\boldsymbol{\theta}_0,\widetilde{\boldsymbol{h}}_0) \| = o_p(\sigma_n).
\]
By Slutsky's theorem, this remainder term is asymptotically negligible, completing the proof of the theorem.
    \[
        \sigma_n^{-1} \langle\boldsymbol{b}, \widehat{\boldsymbol{\theta}} - \boldsymbol{\theta}_0\rangle = \sigma_n^{-1} \left\langle \boldsymbol{b}, \boldsymbol{M}_{\delta}^{-1} \boldsymbol{\Delta}_{\delta}(\boldsymbol{\theta}_0, \widetilde{\boldsymbol{h}}_0) \right\rangle + o_p(1) \stackrel{d}{\rightarrow} \mathcal{N}(0,1).
    \]

\end{proof}

\noindent\textbf{Proof of Theorem \Sref{theorem:high_probability_error_bound}{2}.}
\begin{proof}
We prove the result in three steps. First, we show that the population loss is strongly convex in the treatment parameter. Second, we use this strong convexity together with the optimality of the estimator to decompose the estimation error $\left\|\widehat{\boldsymbol{\theta}}-\boldsymbol{\theta}_0\right\|$ to the sum of two moment terms. Third, we bound these two moment terms separately. Combining these bounds yields the stated convergence rate.   

\noindent\textbf{Step 1: Strong convexity of the population loss.} We conduct the theoretical analysis conditional on the event that $\mathcal{E}(\widehat{\boldsymbol{h}}) \leq \mathcal{R}_{n,s,t}$, which occurs with probability at least $1-e^{-t}$.
    Since  $\boldsymbol{M}_{\delta}$
    has eigenvalues lower bounded by $\varpi^{\star}$ and each entry of the Jacobian of $\boldsymbol{\psi}$ is $L$-Lispchitz with respect to the nuisance for some constant $L$,
    we have that for every vector $\boldsymbol{v} \in \mathbb{R}^{d_{\boldsymbol{T}}}$:
    \begin{align*}
         &\frac{\boldsymbol{v}^{\top} \left\{-\nabla_{\boldsymbol{\theta}}\boldsymbol{m}_{\delta}(\boldsymbol{x};\boldsymbol{\theta},\widehat{\boldsymbol{h}})\right\}\boldsymbol{v}}{\| \boldsymbol{v} \|^2}\\
         \geq &-\frac{\boldsymbol{v}^{\top} \nabla_{\boldsymbol{\theta}}\boldsymbol{m}_{\delta}(\boldsymbol{x};\boldsymbol{\theta},\boldsymbol{h}_0)\boldsymbol{v}}{\| \boldsymbol{v} \|^2} - \frac{\boldsymbol{v}^{\top} \left\{\nabla_{\boldsymbol{\theta}}\boldsymbol{m}_{\delta}(\boldsymbol{x};\boldsymbol{\theta},\widehat{\boldsymbol{h}}) - \nabla_{\boldsymbol{\theta}}\boldsymbol{m}_{\delta}(\boldsymbol{x};\boldsymbol{\theta},\boldsymbol{h}_0)\right\}\boldsymbol{v}}{\| \boldsymbol{v} \|^2} \\
          \geq &\;\varpi^{\star} - L\mathbb{E}\left(\| \widehat{\boldsymbol{h}} - \boldsymbol{h}_0 \|\vert \boldsymbol{x} \right)\frac{\|\boldsymbol{v}\|^2_{1}}{\|\boldsymbol{v}\|^2}
        \geq \varpi^{\star} - L d_{\boldsymbol{T}} \mathcal{R}_{n,s,t}
        =\varpi^{\star} - O\left(\mathcal{R}_{n,s,t}\right),
    \end{align*}
    where the last inequality holds for Holder's inequality.
    Let the expected loss function $\mathcal{L}_{\delta}(\boldsymbol{\theta}) = \mathbb{E}\left\{\rho_{\delta}(\boldsymbol{Z};\boldsymbol{\theta},\widehat{\boldsymbol{h}}) \vert x\right\}$,
    since $\nabla_{\boldsymbol{\theta}}\boldsymbol{m}_{\delta}(\boldsymbol{x};\boldsymbol{\theta},\widehat{\boldsymbol{h}})$ is the Hessian of $\mathcal{L}_{\delta}(\boldsymbol{\theta})$,
    we have that $\mathcal{L}_{\delta}(\boldsymbol{\theta})$ is strongly convex with respect to $\boldsymbol{\theta}$. 
    Then we denote the minimum eigenvalue of $ -\nabla_{\boldsymbol{\theta}}\boldsymbol{m}_{\delta}(\boldsymbol{x};\boldsymbol{\theta},\widehat{\boldsymbol{h}})$ by $\widehat{\varpi}^{\star}$.
    
    \noindent\textbf{Step 2: Error Decomposition.} Base on the strong convexity, we have
    \begin{equation}\label{eq9}
    \begin{aligned}
         &\; \mathcal{L}_{\delta}(\widehat{\boldsymbol{\theta}}) - \mathcal{L}_{\delta}(\boldsymbol{\theta}_0)\\
          \geq &\;\nabla_{\boldsymbol{\theta}}\mathcal{L}_{\delta}(\widehat{\boldsymbol{\theta}})^{\top}( \widehat{\boldsymbol{\theta}}-\boldsymbol{\theta}_0) + \frac{\widehat{\varpi}^{\star}}{2}\|\widehat{\boldsymbol{\theta}} - \boldsymbol{\theta}_0\|^2     \\
          = &\;\boldsymbol{m}_{\delta}(\boldsymbol{x};\boldsymbol{\theta}_0,\widehat{\boldsymbol{h}})^{\top}( \widehat{\boldsymbol{\theta}}-\boldsymbol{\theta}_0) + \frac{\widehat{\varpi}^{\star}}{2}\|\widehat{\boldsymbol{\theta}} - \boldsymbol{\theta}_0\|^2. \\
    \end{aligned}
    \end{equation}
    Moreover, by the convexity of $\mathcal{L}_{\delta}(\boldsymbol{\theta})$, we have
    \begin{equation}\label{eq10}
    \begin{aligned}
        \mathcal{L}_{\delta}(\widehat{\boldsymbol{\theta}}) - \mathcal{L}_{\delta}(\boldsymbol{\theta}_0)
        \geq \nabla_{\boldsymbol{\theta}}\mathcal{L}_{\delta}(\widehat{\boldsymbol{\theta}})^{\top}( \widehat{\boldsymbol{\theta}}-\boldsymbol{\theta}_0)
        =  \boldsymbol{m}_{\delta}(\boldsymbol{x};\widehat{\boldsymbol{\theta}},\widehat{\boldsymbol{h}})^{\top}( \widehat{\boldsymbol{\theta}}-\boldsymbol{\theta}_0).
    \end{aligned}
    \end{equation}
    Combining \eqref{eq9} and \eqref{eq10}, we obtain
    \begin{align*}
         \frac{\widehat{\varpi}^{\star}}{2}\|\widehat{\boldsymbol{\theta}} - \boldsymbol{\theta}_0\|^2 \;&\leq\; \left\{\boldsymbol{m}_{\delta}(\boldsymbol{x};\widehat{\boldsymbol{\theta}},\widehat{\boldsymbol{h}})-\boldsymbol{m}_{\delta}(\boldsymbol{x};\boldsymbol{\theta}_0,\widehat{\boldsymbol{h}})\right\}^{\top}(\widehat{\boldsymbol{\theta}} - \boldsymbol{\theta}_0) \\
         \;&\leq\; \|\boldsymbol{m}_{\delta}(\boldsymbol{x};\widehat{\boldsymbol{\theta}},\widehat{\boldsymbol{h}})-\boldsymbol{m}_{\delta}(\boldsymbol{x};\boldsymbol{\theta}_0,\widehat{\boldsymbol{h}})\| \|\widehat{\boldsymbol{\theta}} - \boldsymbol{\theta}_0\|.             \\
    \end{align*}
    Dividing both sides by $\|\widehat{\boldsymbol{\theta}} - \boldsymbol{\theta}_0\|$ yields that
    \begin{align*}
         \|\widehat{\boldsymbol{\theta}} - \boldsymbol{\theta}_0\|\;&\leq\; \frac{2}{\widehat{\varpi}^{\star}}\|\boldsymbol{m}_{\delta}(\boldsymbol{x};\widehat{\boldsymbol{\theta}},\widehat{\boldsymbol{h}})-\boldsymbol{m}_{\delta}(\boldsymbol{x};\boldsymbol{\theta}_0,\widehat{\boldsymbol{h}})\|\\
         \;&\leq\; \frac{2}{\widehat{\varpi}^{\star}}\left\{\|\boldsymbol{m}_{\delta}(\boldsymbol{x};\widehat{\boldsymbol{\theta}},\widehat{\boldsymbol{h}})\|+\|\boldsymbol{m}_{\delta}(\boldsymbol{x};\boldsymbol{\theta}_0,\widehat{\boldsymbol{h}})\|\right\}.
    \end{align*}

    \noindent\textbf{Step 3: Bound the two moment terms.} For the first term $\boldsymbol{m}_{\delta}(\boldsymbol{x};\widehat{\boldsymbol{\theta}},\widehat{\boldsymbol{h}})$, since $\boldsymbol{m}_{\delta}(\boldsymbol{x};\widehat{\boldsymbol{\theta}},\widehat{\boldsymbol{h}}) = \boldsymbol{\Lambda}_{\delta}(\widehat{\boldsymbol{\theta}},\widehat{\boldsymbol{h}}) + \boldsymbol{\Psi}_{\delta}(\widehat{\boldsymbol{\theta}},\widehat{\boldsymbol{h}})$
    and $\boldsymbol{\Psi}_{\delta}(\widehat{\boldsymbol{\theta}},\widehat{\boldsymbol{h}}) = 0$,
    together with Lemmas~\ref{lemma:kernel_error}, \ref{lemma:sampling_error}, \ref{lemma:subsampling_error} in Subsection~\ref{TE lemma}
    and our assumptions on the choice of $s$, $B$, we have that with probability $1-2e^{-t}$,
    \[
        \left\| \boldsymbol{\Lambda}_{\delta}(\widehat{\boldsymbol{\theta}},\widehat{\boldsymbol{h}}) \right\|
        =  O\left(s^{-\frac{1}{2\alpha d_{\boldsymbol{X}}}} + \sqrt{\frac{s\left\{\log(n/s)+t\right\}}{n}}\right).
    \]

     Note that $\boldsymbol{m}_{\delta,j}(\boldsymbol{x}; \boldsymbol{\theta}_0, \boldsymbol{h}_0) = \boldsymbol{0}$
    by definition and
    $\nabla_{\boldsymbol{h}} \boldsymbol{m}_{\delta,j}(\boldsymbol{x}; \boldsymbol{\theta}_0, \boldsymbol{h}_0)^{\top} (\widehat{\boldsymbol{h}} - \boldsymbol{h}_0)=\boldsymbol{0}$
    by local orthogonality.
    Invoking a second order Taylor expansion around $\boldsymbol{h}_0$, for any $\boldsymbol{\theta} \in \Theta$:
    \begin{align*}
        \boldsymbol{m}_{\delta,j}(\boldsymbol{x}; \boldsymbol{\theta}_0, \widehat{\boldsymbol{h}})
         \;= &\;\boldsymbol{m}_{\delta,j}(\boldsymbol{x}; \boldsymbol{\theta}_0, \boldsymbol{h}_0)
        + \nabla_{\boldsymbol{h}} \boldsymbol{m}_{\delta,j}(\boldsymbol{x}; \boldsymbol{\theta}_0, \boldsymbol{h}_0)^{\top} (\widehat{\boldsymbol{h}} - \boldsymbol{h}_0)                                                                                                                    \\
          &+\underbrace{\frac{1}{2} (\widehat{\boldsymbol{h}} - \boldsymbol{h}_0)^{\top} \nabla_{\boldsymbol{h}}^2 \boldsymbol{m}_{\delta,j}(\boldsymbol{x}; \boldsymbol{\theta}_0, \widetilde{\boldsymbol{h}}^{(j)}) (\widehat{\boldsymbol{h}} - \boldsymbol{h}_0)}_{\boldsymbol{\varrho}_{\delta,j}} \\
         \;= &\;\boldsymbol{\varrho}_{\delta,j},
    \end{align*}
    for some $\widetilde{\boldsymbol{h}}^{(j)}$ between $\widehat{\boldsymbol{h}}$ and $\boldsymbol{h}_0$.

    Next we bound the second term $\|\boldsymbol{m}_{\delta}(\boldsymbol{x};\boldsymbol{\theta}_0,\widehat{\boldsymbol{h}})\|$. Since the moment is smooth, we have $\|\boldsymbol{\varrho}\| = O\left(\mathbb{E}\left(\left\| \widehat{\boldsymbol{h}} - \boldsymbol{h}_0 \right\|^2 \vert x\right)\right) = O\left(\mathcal{E}(\widehat{\boldsymbol{h}})^2\right) = O\left(\mathcal{R}_{n,s,t}^2\right).$
    Therefore, we have $\|\boldsymbol{m}_{\delta}(\boldsymbol{x};\boldsymbol{\theta}_0,\widehat{\boldsymbol{h}})\| = O(\mathcal{R}_{n,s,t}^2)$.
    Combining all the latter inequalities yields the result.
\end{proof}

\subsection{Auxiliary Lemmas for Treatment Effect}\label{TE lemma}
This subsection develops the auxiliary results needed to control the stochastic part of the treatment effect estimator. We first decompose this discrepancy into three components: the localization error induced by forest weights, the sampling fluctuation of the complete U-statistic and the finite-forest subsampling error. The subsequent lemmas provide bounds for these three terms.

Throughout this section, we fix the target covariate value $\boldsymbol{x}$ and suppress its dependence in the notation whenever it is clear from the context.
We begin by introducing some quantities that will be useful throughout our theoretical analysis.
We denote with $\omega$ the random variable that corresponds to the internal randomness of the tree-splitting algorithm, including the randomness from subsampling, feature subsampling and random split selection in the tree-building procedure.

Recall the decomposition of $\boldsymbol{\Lambda}_{\delta}(\boldsymbol{\theta},\boldsymbol{h})$, for any $\boldsymbol{\theta}$ and $\boldsymbol{h}$, we have 
    $$ \boldsymbol{\Lambda}_{\delta}(\boldsymbol{\theta},\boldsymbol{h})
        = \underbrace{\boldsymbol{m}_{\delta}(\boldsymbol{x};\boldsymbol{\theta},\boldsymbol{h}) - \boldsymbol{\mu}_0(\boldsymbol{\theta},\boldsymbol{h})}_{\boldsymbol{\Gamma}_{\delta}(\boldsymbol{\theta},\boldsymbol{h})}
        + \underbrace{\boldsymbol{\mu}_0(\boldsymbol{\theta},\boldsymbol{h}) - \widetilde{\boldsymbol{\Psi}}_{\delta}(\boldsymbol{\theta},\boldsymbol{h})}_{\boldsymbol{\Delta}_{\delta}(\boldsymbol{\theta},\boldsymbol{h})}
        + \underbrace{\widetilde{\boldsymbol{\Psi}}_{\delta}(\boldsymbol{\theta},\boldsymbol{h}) - \boldsymbol{\Psi}_{\delta}(\boldsymbol{\theta},\boldsymbol{h})}_{\boldsymbol{E}_{\delta}(\boldsymbol{\theta},\boldsymbol{h})}.$$
The first term $\boldsymbol{\Gamma}_{\delta}(\boldsymbol{\theta},\boldsymbol{h})$ is called the \textbf{kernel error} because it captures the localization bias induced by the forest weights. The second term $\boldsymbol{\Delta}_{\delta}(\boldsymbol{\theta},\widehat{\boldsymbol{h}})$ is called the \textbf{sampling error} because it measures the discrepancy between the population score and its complete U-statistic approximation based on the finite sample. Finally, $\boldsymbol{E}_{\delta}(\boldsymbol{\theta},\boldsymbol{h})$ is called the \textbf{subsampling error}: ideally, the complete subsampling score $\widetilde{\boldsymbol{\Psi}}_{\delta}(\boldsymbol{\theta}, \widehat{\boldsymbol{h}})$ averages over all possible subsamples of size $s$, whereas in practice, the forest score $\boldsymbol{\Psi}_{\delta}(\boldsymbol{\theta}, \widehat{\boldsymbol{h}})$ is computed using only $B$ randomly drawn subsamples. The term $\boldsymbol{E}_{\delta}(\boldsymbol{\theta},\boldsymbol{h})$ captures the Monte Carlo error caused by using finitely many trees. The subsequent lemmas bound these three components separately.

\begin{lemma}(Kernel Error)\label{lemma:kernel_error}
    If the OHRF weights were trained on a random sample $\mathcal{I}$ with $|\mathcal{I}|=s$, satisfy that:
    \[
        \mathbb{E}\left[\sup\left\{\left\|\boldsymbol{X}_i - \boldsymbol{x} \right\|:a_i\left(\mathcal{I},\omega\right) > 0\right\}\right] \leq Cs^{-\frac{1}{2\alpha d_{\boldsymbol{X}}}},
    \]
    where $\alpha = \frac{\log(\Upsilon^{-1})}{\pi \log\left((1-\Upsilon)^{-1}\right)}$ and $C$ is a sufficient large constant,
    the expectation is over the randomness of the samples and the internal randomness $\omega$ of the OHRF algorithm. Then
    \[
        \sup_{\boldsymbol{\theta},\boldsymbol{h}}\left\|\boldsymbol{\Gamma}_{\delta}(\boldsymbol{\theta},\boldsymbol{h})\right\| \leq C s^{-\frac{1}{2\alpha d_{\boldsymbol{X}}}}.
    \]
\end{lemma}

\begin{proof}

    First we argue that by invoking the honesty of the OHRF weights
    we can re-write $\boldsymbol{\mu}_{0}(\boldsymbol{\theta}, \boldsymbol{h})$ as
    \[
        \boldsymbol{\mu}_{0}(\boldsymbol{\theta}, \boldsymbol{h}) = \mathbb{E} \left[\binom{n}{s}^{-1} \sum_{\mathcal{I}\subset\left\{1,2,\dots,n\right\}:|\mathcal{I}|=s} \mathbb{E}_{\omega}\left\{\sum_{i\in\mathcal{I}}a_{i}\left(\mathcal{I},\omega\right) \boldsymbol{m}_{\delta}(\boldsymbol{X}_{i};\boldsymbol{\theta},\boldsymbol{h})\right\}\right].
    \]
    To prove this claim, it suffices to show that for any subset of $s$ indices
    \[
        \mathbb{E}\left\{a_{i}\left(\mathcal{I},\omega\right) \boldsymbol{\psi}_{\delta}(\boldsymbol{Z}_{i};\boldsymbol{\theta},\boldsymbol{h})\right\} = \mathbb{E}\left[a_{i}\left(\mathcal{I},\omega\right) \boldsymbol{m}_{\delta}(\boldsymbol{X}_{i};\boldsymbol{\theta},\boldsymbol{h})\right].
    \]
    Fix a subsample \(\mathcal I\subset\{1,\ldots,n\}\) with \(|\mathcal I|=s\), and fix an index \(i\in\mathcal I\). Let
\[
\mathcal I_{-i}=\mathcal I\setminus\{i\},
\qquad
\boldsymbol Z_{\mathcal I_{-i}}
=
\{\boldsymbol Z_i:i\in\mathcal I_{-i}\}.
\]
By honesty of the OHRF weights, if observation \(i\) belongs to the splitting subsample, then $a_i(\mathcal I,\omega)=0$. Otherwise, if observation \(i\) belongs to the estimation subsample, then
\(a_i(\mathcal I,\omega)\) is independent of \(\boldsymbol Z_i\) conditional on
\(\boldsymbol Z_{\mathcal I_{-i}}\), \(\boldsymbol X_i\), and \(\omega\). Therefore, in either case, \(a_i(\mathcal I,\omega)\) is conditionally independent of \(\boldsymbol Z_i\) given
\(\boldsymbol Z_{\mathcal I_{-i}}\), \(\boldsymbol X_i\), and \(\omega\).

Moreover, since \(\boldsymbol Z_i\) is independent of
\((\boldsymbol Z_{\mathcal I_{-i}},\omega)\) conditional on \(\boldsymbol X_i\), we have
\[
\mathbb E\left\{
\boldsymbol{\psi}_{\delta}(\boldsymbol Z_i;\boldsymbol\theta,\boldsymbol h)
\mid
\boldsymbol Z_{\mathcal I_{-i}},\boldsymbol X_i,\omega
\right\}
=
\boldsymbol m_{\delta}(\boldsymbol X_i;\boldsymbol\theta,\boldsymbol h).
\]
Hence, by iterated expectation and the conditional independence above,
\[
\begin{aligned}
&\quad\mathbb E\left[
a_i(\mathcal I,\omega)
\boldsymbol{\psi}_{\delta}(\boldsymbol Z_i;\boldsymbol\theta,\boldsymbol h)
\right]
\\
&=\mathbb E\left[\mathbb E\left\{a_i(\mathcal I,\omega)\mid\boldsymbol Z_{\mathcal I_{-i}},\boldsymbol X_i,\omega\right\}\mathbb E\left\{\boldsymbol{\psi}_{\delta}(\boldsymbol Z_i;\boldsymbol\theta,\boldsymbol h)\mid\boldsymbol Z_{\mathcal I_{-i}},\boldsymbol X_i,\omega\right\}\right]
\\
&=\mathbb E\left[\mathbb E\left\{a_i(\mathcal I,\omega)\mid\boldsymbol Z_{\mathcal I_{-i}},\boldsymbol X_i,\omega\right\}\boldsymbol m_{\delta}(\boldsymbol X_i;\boldsymbol\theta,\boldsymbol h)\right]
\\
&=\mathbb E\left[a_i(\mathcal I,\omega)\boldsymbol m_{\delta}(\boldsymbol X_i;\boldsymbol\theta,\boldsymbol h)\right].
\end{aligned}
\]

Finally, using the triangle inequality, the Lipschitz continuity of the conditional moment function, and the fact that the weights are nonnegative and sum to one, we obtain
\[
\begin{aligned}
\left\|
\boldsymbol{\Gamma}_{\delta}(\boldsymbol\theta,\boldsymbol h)
\right\|
&\le
\binom ns^{-1}
\sum_{\mathcal I\subset\{1,\ldots,n\}:|\mathcal I|=s}
\mathbb E\left[
\sum_{i\in\mathcal I}
a_i(\mathcal I,\omega)\left\|\boldsymbol m_{\delta}(\boldsymbol X_i;\boldsymbol\theta,\boldsymbol h)-\boldsymbol m_{\delta}(\boldsymbol x;\boldsymbol\theta,\boldsymbol h)\right\|\right]\\
&\le\sqrt{d_{\boldsymbol T}}L\binom ns^{-1}\sum_{\mathcal I\subset\{1,\ldots,n\}:|\mathcal I|=s}\mathbb E\left[\sum_{i\in\mathcal I}a_i(\mathcal I,\omega)\left\|\boldsymbol X_i-\boldsymbol x\right\|\right]\\
&\le\sqrt{d_{\boldsymbol T}}L\binom ns^{-1}\sum_{\mathcal I\subset\{1,\ldots,n\}:|\mathcal I|=s}\mathbb E\left[\sup_{i\in\mathcal I}\left\{\left\|\boldsymbol X_i-\boldsymbol x\right\|:\ a_i(\mathcal I,\omega)>0
\right\}\right]\\
&\le\sqrt{d_{\boldsymbol T}}Ls^{-1/(2\alpha d_{\boldsymbol X})}=O\left(s^{-1/(2\alpha d_{\boldsymbol X})}\right).
\end{aligned}
\]
\end{proof}

\begin{lemma}(Stochastic Equicontinuity for U-statistics)
\label{lemma:bracketing}
 Consider a parameter space $\boldsymbol{\Theta}$  that is a bounded subset of $\mathbb{R}^{d_{\boldsymbol{T}}}$,
    with $\text{diam}(\boldsymbol{\Theta}) = \sup_{\boldsymbol{\theta}, \boldsymbol{\theta}^{\prime} \in \boldsymbol{\Theta}} \| \boldsymbol{\theta}-\boldsymbol{\theta}^{\prime} \|_2 \leq R$ .
Consider the U-statistic
\[
    \mathbb{U}_{n,s}\mathcal{G}(\cdot;\boldsymbol{\theta})
    = \binom{n}{s}^{-1}
      \sum_{\substack{\mathcal{I}\subset\{1,\ldots,n\}\\
                      |\mathcal{I}|=s}}
      \mathcal{G}(\mathcal{I};\boldsymbol{\theta}),
\]
where $\mathcal{G}(\cdot;\boldsymbol{\theta}) :
\mathcal{Z}^{s}\to\mathbb{R}$ is a measurable kernel, symmetric
in its $s$ observation arguments. Define
$\mathcal{G}^{\mathrm{cen}}(\mathcal{I};\boldsymbol{\theta})= \mathcal{G}(\mathcal{I};\boldsymbol{\theta})-\mathbb{E}\{\mathcal{G}(\mathcal{I};\boldsymbol{\theta})\}$ and suppose there exists positive deterministic quantities
$G_{1n}$, $G_{2n}$, and $L_n$, possibly depending on $n$,
such that
$$
    \sup_{\boldsymbol{\theta}\in\boldsymbol{\Theta}}
    \sqrt{
      \mathbb{E}\{
        \mathcal{G}(\mathcal{I};\boldsymbol{\theta})^2
      \}
    }
    \leq G_{1n},$$
and for every integer $k\geq 3$,
\[
    \sup_{\boldsymbol{\theta}\in\boldsymbol{\Theta}}
    \mathbb{E}\left[
      \left|
        \mathcal{G}^{\mathrm{cen}}
        (\mathcal{I};\boldsymbol{\theta})
      \right|^k
    \right]
    \leq \frac{k!}{2}G_{1n}^2G_{2n}^{k-2}.
\]
Moreover, suppose that, for every integer $k\geq 2$ and every
$\boldsymbol{\theta},\boldsymbol{\theta}'
\in\boldsymbol{\Theta}$,
\[
\begin{split}
    &\mathbb{E}\left[
       \left|
         \mathcal{G}^{\mathrm{cen}}
           (\mathcal{I};\boldsymbol{\theta})
         -
         \mathcal{G}^{\mathrm{cen}}
           (\mathcal{I};\boldsymbol{\theta}')
       \right|^k
      \right] \\
    &\qquad\leq
      \frac{k!}{2}L_n^k
      \|\boldsymbol{\theta}-\boldsymbol{\theta}'\|_2^k.
\end{split}
\]
There exists a constant $C>0$, for every $t>0$, with probability at least $1-e^{-t}$,
\[
\begin{split}
    &\sup_{\boldsymbol{\theta}\in\boldsymbol{\Theta}}\left|\mathbb{U}_{n,s}\mathcal{G}(\cdot;\boldsymbol{\theta})-\mathbb{E}\{\mathcal{G}(\mathcal{I};\boldsymbol{\theta})\}\right| \\
    &\quad\leq C\left[
       G_{1n}\sqrt{\frac{s(t+1)}{n}}+G_{2n}\frac{s(t+1)}{n}
       +L_nR\left\{\sqrt{\frac{s(d_{\boldsymbol{T}}+t+1)}{n}}+\frac{s(d_{\boldsymbol{T}}+t+1)}{n}\right\}\right].
\end{split}
\]
Consequently, for every fixed $p\geq 1$,
\[
\begin{split}
    &\left\|
       \sup_{\boldsymbol{\theta}\in\boldsymbol{\Theta}}
       \left|\mathbb{U}_{n,s}\mathcal{G}(\cdot;\boldsymbol{\theta})- \mathbb{E}\{\mathcal{G}(\mathcal{I};\boldsymbol{\theta})\}\right|  \right\|_{L_p} \\
    &\quad\leq C_{p,d_{\boldsymbol{T}}}
       \left\{(G_{1n}+L_nR)\sqrt{\frac{s}{n}}+ (G_{2n}+L_nR)\frac{s}{n}\right\},
\end{split}
\]
where $C_{p,d_{\boldsymbol{T}}}$ depends only on $p$ and
$d_{\boldsymbol{T}}$.
\end{lemma}

\begin{proof}
The second-moment assumption implies
\[
    \mathbb{E}\left[
      \left|
        \mathcal{G}^{\mathrm{cen}}
        (\mathcal{I};\boldsymbol{\theta})
      \right|^2
    \right]
    \leq G_{1n}^2.
\]
Thus the centered kernel satisfies the Bernstein moment
condition with variance parameter $G_{1n}^2$ and scale parameter
$G_{2n}$ for all integers $k\geq 2$.
Let $\mathfrak p$ be a random permutation of
$\{1,\ldots,n\}$, by symmetry of the kernel,
\[
\begin{split}
    &\mathbb{U}_{n,s}\mathcal{G}^{\mathrm{cen}}
      (\cdot;\boldsymbol{\theta}) \\
    &\quad=
      \mathbb{E}_{\mathfrak p}\left[
        \frac{1}{\lfloor n/s\rfloor}
        \sum_{j=1}^{\lfloor n/s\rfloor}
        \mathcal{G}^{\mathrm{cen}}\left(
          \{
            \mathfrak p((j-1)s+1),\ldots,\mathfrak p(js)
          \};
          \boldsymbol{\theta}
        \right)
      \right],
\end{split}
\]
where $\mathbb{E}_{\mathfrak p}$ averages only over the
permutation, with the observations held fixed.
For any fixed permutation, the $\lfloor n/s\rfloor$
block kernels are independent and identically distributed.

The Bernstein moment bound for these blocks gives
\[
\begin{split}
    &\mathbb{E}\exp\left\{
      \lambda_{\mathrm U}
      \mathbb{U}_{n,s}\mathcal{G}^{\mathrm{cen}}
        (\cdot;\boldsymbol{\theta})
      \right\} \\
    &\quad\leq
      \exp\left\{
        \frac{\lambda_{\mathrm U}^2G_{1n}^2}
        {2\lfloor n/s\rfloor
         \{1-G_{2n}|\lambda_{\mathrm U}|/\lfloor n/s\rfloor\}}
      \right\},
      \qquad
      |\lambda_{\mathrm U}|
      <
      \frac{\lfloor n/s\rfloor}{G_{2n}}.
\end{split}
\]
Consequently, for every fixed $\boldsymbol{\theta}$ and
every $t>0$,
\[
\Pr\left\{
    \left|
      \mathbb{U}_{n,s}\mathcal{G}^{\mathrm{cen}}
      (\cdot;\boldsymbol{\theta})
    \right|
    >
    G_{1n}\sqrt{\frac{2t}{\lfloor n/s\rfloor}}
    +
    G_{2n}\frac{t}{\lfloor n/s\rfloor}
   \right\}
   \leq 2e^{-t}.
\]
Applying the same argument to the centered increment gives
\[
\begin{split}
\Pr\biggl\{
    &\left|
      \mathbb{U}_{n,s}\mathcal{G}^{\mathrm{cen}}
        (\cdot;\boldsymbol{\theta})
      -
      \mathbb{U}_{n,s}\mathcal{G}^{\mathrm{cen}}
        (\cdot;\boldsymbol{\theta}')
     \right| \\
    &>
      L_n\|\boldsymbol{\theta}-\boldsymbol{\theta}'\|_2
      \left(
        \sqrt{\frac{2t}{\lfloor n/s\rfloor}}
        +
        \frac{t}{\lfloor n/s\rfloor}
      \right)
   \biggr\}
   \leq 2e^{-t}.
\end{split}
\]
For each $0<\varepsilon\leq R$, choose an $\varepsilon$-net
$\boldsymbol{\Theta}_{\varepsilon}
 \subset\boldsymbol{\Theta}$ satisfying $|\boldsymbol{\Theta}_{\varepsilon}|
    \leq (1+2R/\varepsilon)^{d_{\boldsymbol{T}}}$.
Let
$\boldsymbol{\theta}_{\varepsilon}
 \in\boldsymbol{\Theta}_{\varepsilon}$, then for $j\geq 1$,
\[
    \|\boldsymbol{\theta}_{R/2^{j}}
      -
      \boldsymbol{\theta}_{R/2^{(j-1)}}\|_2
    \leq 3R/2^{j}.
\]
There are at most $|\boldsymbol{\Theta}_{R/2^{j}}|\,|\boldsymbol{\Theta}_{R/2^{(j-1)}}|$
possible pairs at level $j$ and
\[
    \log\left(
      |\boldsymbol{\Theta}_{R/2^{j}}|\,
      |\boldsymbol{\Theta}_{R/2^{(j-1)}}|
    \right)
    \leq d_{\boldsymbol{T}}(2j+3)\log 2.
\]

Apply the fixed-parameter bound at
$\boldsymbol{\theta}_R$ with tail parameter $t+\log 4$.
At level $j$, apply the increment bound to all pairs at
distance at most $3R/2^{j}$ with tail parameter
$t+4(d_{\boldsymbol{T}}+1)(j+1)$.
The union bound over the anchor and all levels has failure
probability at most $e^{-t}$. On the resulting event,
\[
\begin{split}
    &\sup_{\boldsymbol{\theta}\in\boldsymbol{\Theta}}
      \left|
        \mathbb{U}_{n,s}\mathcal{G}^{\mathrm{cen}}
        (\cdot;\boldsymbol{\theta})
      \right| \\
    &\quad\leq C\left\{
      G_{1n}\sqrt{\frac{t+1}{\lfloor n/s\rfloor}}
      +
      G_{2n}\frac{t+1}{\lfloor n/s\rfloor}
      \right\} \\
    &\qquad{}+
      CL_nR\sum_{j=1}^{\infty}2^{-j}
      \left\{\sqrt{\frac{t+(d_{\boldsymbol{T}}+1)(j+1)}{\lfloor n/s\rfloor}} + \frac{t+(d_{\boldsymbol{T}}+1)(j+1)}{\lfloor n/s\rfloor}
      \right\} \\
    &\quad\leq C\left[
      G_{1n}\sqrt{\frac{t+1}{\lfloor n/s\rfloor}}+G_{2n}\frac{t+1}{\lfloor n/s\rfloor}+L_nR\left\{\sqrt{\frac{d_{\boldsymbol{T}}+t+1}{\lfloor n/s\rfloor}}+\frac{d_{\boldsymbol{T}}+t+1} {\lfloor n/s\rfloor}\right\}\right].
\end{split}
\]

Finally, $\lfloor n/s\rfloor\geq n/(2s)$ gives the stated
probability bound. Integrating this tail bound yields
the $L_p$ bound.
All steps also apply conditionally on an independent
first-stage sample under the stated conditional assumptions.
\end{proof}

For our proposed Huber score, $G_{1n}=O(1)$, $L_n=O(1)$,$G_{2n}=O(\delta_{\theta,n})$. Thus, when we choose $\delta_{\theta,n}=O(\sqrt{n/s})$, 
$$\left\|\sup_{\boldsymbol\theta\in\boldsymbol\Theta}\left|\mathbb U_{n,s}\mathcal G(\cdot;\boldsymbol\theta)-\mathbb E\{\mathcal G(\mathcal I;\boldsymbol\theta)\}\right|\right\|_{p}=O\left(\sqrt{\frac{s}{n}}+\delta_{\theta,n}\frac{s}{n}\right)=O\left(\sqrt{\frac{s}{n}}\right).$$

 The following lemma provides a uniform upper bound for $\boldsymbol{\Delta}_{\delta}(\boldsymbol{\theta},\widehat{\boldsymbol{h}})$.

\begin{lemma}(Sampling Error) 
    \label{lemma:sampling_error}Under Assumptions in Lemma~\ref{lemma:moment_condition},
    conditional on any nuisance estimate $\widehat{\boldsymbol{h}}$ from the first stage, with probability $1-e^{-t}$:
    \[
        \sup_{\boldsymbol{\theta}}\left\|\boldsymbol{\Delta}_{\delta}(\boldsymbol{\theta},\widehat{\boldsymbol{h}})\right\| = O\left(\sqrt{\frac{s\left\{\log(n/s)+t\right\}}{n}}\right).
    \]

\end{lemma}
\begin{proof}
Recall that 
$\mathcal{B}\left(\mathcal{I};\boldsymbol{\theta},\boldsymbol{h}\right) = \mathbb{E}_{\omega}\left\{\sum_{i\in\mathcal{I}}a_{i}\left(\mathcal{I},\omega\right) \boldsymbol{\psi}_{\delta}(\boldsymbol{Z}_{i};\boldsymbol{\theta},\boldsymbol{h})\right\}$. Then $\boldsymbol{\Delta}_{\delta}(\boldsymbol{\theta},\widehat{\boldsymbol{h}})$ is a U-statistic as it can be written as:
    $\mathbb{U}_{s,n}\mathcal{B}(\cdot;\boldsymbol{\theta},\widehat{\boldsymbol{h}}) - \mathbb{E}\left\{\mathcal{B}(\mathcal{I};\boldsymbol{\theta},\widehat{\boldsymbol{h}})\right\}$.
    Moreover, under Lemma \ref{lemma:moment_condition}, the function $\mathcal{B}(\cdot;\boldsymbol{\theta},\widehat{\boldsymbol{h}})$
    satisfies the conditions of Lemma \ref{lemma:bracketing} with $G_{1n}=O(1)$, $L_n=O(1)$,$G_{2n}=O(\delta_{\theta,n})$.
    Moreover, $\mathcal{B}(\cdot;\boldsymbol{\theta},\widehat{\boldsymbol{h}})$ is $O(1)$-Lipschitz,
    since it is a convex combination of $O(1)$-Lipschitz functions.
    Finally, since $\text{diam}(\boldsymbol{\Theta})=O(1)$, applying Lemma \ref{lemma:bracketing} we get the desired result.
\end{proof}

 The following lemma bound the subsampling error $\sup_{\boldsymbol{\theta},\boldsymbol{h}}\left\| \boldsymbol{E}_{\delta}(\boldsymbol{\theta}, \widehat{\boldsymbol{h}}) \right\| $.

\begin{lemma}(Subsampling Error)\label{lemma:subsampling_error}
Recall that
$\boldsymbol\Theta\subset\mathbb R^{d_{\boldsymbol T}}$ is a nonempty bounded parameter set with diameter at most $R$ and
$\widetilde{\boldsymbol T}_i(\widehat{\boldsymbol\ell})=\boldsymbol T_i-\widehat{\boldsymbol\ell}(\boldsymbol x)^\top\boldsymbol W_i.$
If the similarity weights are built on $B$ randomly drawn sub-samples with replacement, then for every fixed $p\geq1$,
\[
\begin{aligned}
&\left[
 \mathbb E\left\{
  \left(
   \sup_{\boldsymbol\theta\in\boldsymbol\Theta}
   \|\boldsymbol{E}_{\delta}(\boldsymbol\theta,\widehat{\boldsymbol{h}})\|_2
  \right)^p
  \,\middle|\,\boldsymbol{D}_n,\widehat{\boldsymbol{h}}
 \right\}
\right]^{1/p}
\\
&\quad\leq
\frac{C_{p,d_{\boldsymbol T}}}{\sqrt B}
\left[
 \delta\max_{1\leq i\leq n}
  \|\widetilde{\boldsymbol T}_i
    (\widehat{\boldsymbol\ell})\|_2
 +
 R\max_{1\leq i\leq n}
  \|\widetilde{\boldsymbol T}_i
    (\widehat{\boldsymbol\ell})\|_2^2
\right],
\end{aligned}
\]
where $C_{p,d_{\boldsymbol T}}$ depends only on $p$
and $d_{\boldsymbol T}$. In particular, choosing $B$ such that $B\gtrsim\frac{n}{s}$
gives
\[
\left[
 \mathbb E\left\{
  \left(
   \sup_{\boldsymbol\theta\in\boldsymbol\Theta}
   \|\boldsymbol{E}_{\delta}(\boldsymbol\theta,\widehat{\boldsymbol{h}})\|_2
  \right)^p
 \right\}
\right]^{1/p}
\lesssim\sqrt{\frac sn}.
\]

\end{lemma}

\begin{proof}
The tree scores
$$\boldsymbol{\Psi}_{\delta,b}(\boldsymbol\theta,\widehat{\boldsymbol{h}})=\sum_{i\in\mathcal I_b}a_{ib}(\boldsymbol x)\boldsymbol\psi_\delta (\boldsymbol Z_i;\boldsymbol\theta,\widehat{\boldsymbol{h}})$$
are independent and identically distributed, with conditional
mean $\widetilde{\boldsymbol{\Psi}}_\delta(\boldsymbol\theta,\widehat{\boldsymbol{h}})$.
Consequently,
$$
\boldsymbol{E}_{\delta}(\boldsymbol\theta,\widehat{\boldsymbol{h}})
=
\widetilde{\boldsymbol{\Psi}}_\delta(\boldsymbol\theta,\widehat{\boldsymbol{h}})
-\frac1B\sum_{b=1}^B\boldsymbol{\Psi}_{\delta,b}(\boldsymbol\theta,\widehat{\boldsymbol{h}}).$$

Since $|\varphi_\delta(u)|\leq\delta$ and
$\varphi_\delta$ is $O(1)$-Lipschitz, 
$$
\begin{aligned}
\sup_{\boldsymbol\theta\in\boldsymbol\Theta}
\|\boldsymbol{\Psi}_{\delta,b}(\boldsymbol\theta,\widehat{\boldsymbol{h}})\|_2
&\leq\delta\max_{1\leq i\leq n}\|\widetilde{\boldsymbol T}_i(\widehat{\boldsymbol\ell})\|_2,
\\
\|\boldsymbol{\Psi}_{\delta,b}(\boldsymbol\theta,\widehat{\boldsymbol{h}})
-\boldsymbol{\Psi}_{\delta,b}(\boldsymbol\theta',\widehat{\boldsymbol{h}})\|_2
&\leq
\left\{\max_{1\leq i\leq n} \|\widetilde{\boldsymbol T}_i (\widehat{\boldsymbol\ell})\|_2^2\right\}\|\boldsymbol\theta-\boldsymbol\theta'\|_2.
\end{aligned}$$
These bounds are finite conditional on the observed data.

The Khintchine inequality at a
fixed point of $\boldsymbol\Theta$ yields the first conclusion.
Indeed, an $\varepsilon$-net has cardinality at most
$(1+2R/\varepsilon)^{d_{\boldsymbol T}}$, and the chaining
contributions are summable since
\[
\sum_{j\geq1}2^{-j}
\sqrt{p+d_{\boldsymbol T}(j+1)}<\infty.
\]
Under the displayed choice of $B$, the conditional bound is
at most $C_{p,d_{\boldsymbol T}}\sqrt{s/n}$ for every
realization of the data. 
\end{proof}

\end{document}